\documentclass[
  journal=pasa,
  manuscript=research-paper, 
  year=2021,
  volume=37,
]{cup-journal}

\usepackage{graphicx}
\usepackage{xcolor}
\usepackage{microtype,siunitx}
\usepackage{amssymb}
\usepackage{amsmath}
\usepackage[backref=page]{hyperref}
\definecolor{dodgerblue}{RGB}{30, 144, 255}
\definecolor{crimson}{RGB}{220, 20, 60}
\definecolor{darkerblue}{RGB}{0, 0, 139}
\hypersetup{colorlinks,citecolor=dodgerblue,linkcolor=blue,urlcolor=darkerblue}
\usepackage{enumitem, xcolor}
\let\svitem\item

\usepackage{booktabs} 
\usepackage{threeparttable,longtable}  
\usepackage{makecell}

\usepackage[skip=0.5ex]{subcaption}
\usepackage{cprotect}
\usepackage{multirow}

\title{The Rapid ASKAP Continuum Survey IV: continuum imaging at 1367.5\,MHz and the first data release of RACS-mid}

\author{S.~W.~Duchesne}
\affiliation{CSIRO Space and Astronomy, PO Box 1130, Bentley WA 6102, Australia}
\email[S.~W.~Duchesne]{Stefan.Duchesne@csiro.au}

\author{A.~J.~M.~Thomson}
\affiliation{CSIRO Space and Astronomy, PO Box 1130, Bentley WA 6102, Australia}

\author{J.~Pritchard}
\affiliation{Sydney Institute for Astronomy, School of Physics, University of Sydney, NSW 2006, Australia}
\alsoaffiliation{CSIRO Space and Astronomy, PO Box 76, Epping, NSW, 1710, Australia}

\author{E.~Lenc}
\affiliation{CSIRO Space and Astronomy, PO Box 76, Epping, NSW, 1710, Australia}

\author{V.~A.~Moss}
\affiliation{CSIRO Space and Astronomy, PO Box 76, Epping, NSW, 1710, Australia}
\alsoaffiliation{Sydney Institute for Astronomy, School of Physics, University of Sydney, NSW 2006, Australia}

\author{D.~McConnell}
\affiliation{CSIRO Space and Astronomy, PO Box 76, Epping, NSW, 1710, Australia}

\author{M.~H.~Wieringa}
\affiliation{CSIRO Space and Astronomy, PO Box 76, Epping, NSW, 1710, Australia}

\author{M.~T.~Whiting}
\affiliation{CSIRO Space and Astronomy, PO Box 76, Epping, NSW, 1710, Australia}

\author{Z.~Wang}
\affiliation{Sydney Institute for Astronomy, School of Physics, University of Sydney, NSW 2006, Australia}
\alsoaffiliation{CSIRO Space and Astronomy, PO Box 76, Epping, NSW, 1710, Australia}

\author{Y.~Wang}
\affiliation{Sydney Institute for Astronomy, School of Physics, University of Sydney, NSW 2006, Australia}
\alsoaffiliation{CSIRO Space and Astronomy, PO Box 76, Epping, NSW, 1710, Australia}

\author{K.~Rose}
\affiliation{Sydney Institute for Astronomy, School of Physics, University of Sydney, NSW 2006, Australia}

\author{W.~Raja}
\affiliation{CSIRO Space and Astronomy, PO Box 76, Epping, NSW, 1710, Australia}

\author{Tara~Murphy}
\affiliation{Sydney Institute for Astronomy, School of Physics, University of Sydney, NSW 2006, Australia}

\author{J.~K.~Leung}
\affiliation{Sydney Institute for Astronomy, School of Physics, University of Sydney, NSW 2006, Australia}
\alsoaffiliation{CSIRO Space and Astronomy, PO Box 76, Epping, NSW, 1710, Australia}

\author{M.~T.~Huynh}
\affiliation{CSIRO Space and Astronomy, PO Box 1130, Bentley WA 6102, Australia}

\author{A.~W.~Hotan}
\affiliation{CSIRO Space and Astronomy, PO Box 1130, Bentley WA 6102, Australia}

\author{T.~Hodgson}
\affiliation{International Centre for Radio Astronomy Research (ICRAR), Curtin University, 1 Turner Ave, Bentley, WA 6102, Australia}
\alsoaffiliation{Curtin Institute for Computation, Curtin
University, GPO Box U1987, Perth, WA 6845, Australia}

\author{G.~H.~Heald}
\affiliation{CSIRO Space and Astronomy, PO Box 1130, Bentley WA 6102, Australia}

\doi{10.1017/pasa.20XX.XX}

\received {dd Mmm YYYY}
\revised  {dd Mmm YYYY}
\accepted {dd Mmm YYYY}
\published{22 September 2020}


\definecolor{dodgerblue}{RGB}{30, 144, 255}

\newcommand{\corrs}[1]{{\color{black} #1}}
\newcommand{\CORRS}[1]{{\color{black} #1}}
\newcommand{\CORRRS}[1]{{\color{black} #1}}

\def\rmsmedian{198}
\def\rmsmedianv{165}
\def\rmserror{$\rmsmedian_{-18}^{+20}$}
\def\rmserrorv{$\rmsmedianv_{-10}^{+11}$}
\def\rmsmin{140}
\def\rmsminerror{$\rmsmin_{-9}^{+8}$}
\def\maxpsfmajor{47.5}
\def\minpsfmajor{8.1}
\def\medpsfmajor{10.1}
\def\maxpsfminor{18.3}
\def\minpsfminor{6.8}

\begin{document}

\begin{abstract}
The Australian SKA Pathfinder (ASKAP) is being used to undertake a campaign to rapidly survey the sky in three frequency bands across its operational spectral range. The first pass of the Rapid ASKAP Continuum Survey (RACS) at 887.5\,MHz in the low band has already been completed, with images, visibility datasets, and catalogues made available to the wider astronomical community through the CSIRO ASKAP Science Data Archive (CASDA). This work presents details of the second observing pass in the mid band at 1367.5\,MHz, RACS-mid, and associated data release comprising images and visibility datasets covering the whole sky south of $\delta_{\text{J2000}}=+49^\circ$. This data release incorporates selective peeling to reduce artefacts around bright sources, as well as accurately modelled primary beam responses. The Stokes I \corrs{images} reach a median noise of $\rmsmedian$\,\textmu Jy\,PSF$^{-1}$ with a declination-dependent angular resolution of \minpsfmajor--\maxpsfmajor\,arcsec that fills a niche in the existing ecosystem of large-area astronomical surveys. We also supply Stokes V \CORRS{images after application of a widefield leakage correction}, with \corrs{a median noise of $\rmsmedianv$\,\textmu Jy\,PSF$^{-1}$.} \CORRS{We find the residual leakage of Stokes I into V to be} \corrs{$\lesssim 0.9$--$2.4$\% over the survey}. This initial RACS-mid data release will be complemented by a future release comprising catalogues of the survey region. As with other RACS data releases, data products from this release will be made available through CASDA.
\end{abstract}


\section{Introduction}
\label{sec:int}

\defcitealias{racs1}{Paper~I}
\defcitealias{racs2}{Paper~II}

\begin{table*}[t]
\begin{threeparttable}
\caption{
\label{tab:surveys} An update to table~1 in \citet{racs1} of representative properties of comparable  completed and on-going large-area surveys.}
\centering
\begin{tabular}{lccccccc}
\toprule
Survey & Frequency & Bandwidth & Resolution & Sky coverage & Sensitivity & Polarization & N$_\mathrm{sources}$ \tnote{a} \\
& (MHz) & (MHz) & (arcsec) & (deg$^2$) & (mJy\,PSF$^{-1}$) &  & ($\times10^6$) \\
\midrule%
VLSSr & 73.8 & 3.12 & 75 & 30\,793 & 100 & I & 0.93 \\
GLEAM & 87, 118, 154, 185, 215 & 30.72 & $\sim 140$--$196$ \tnote{b} & 27\,691 & $\sim 10$--$28$ \tnote{b} & I,Q,U,V & 0.33 \\
GLEAM-X \tnote{c} & 87, 118, 154,185, 215  & 30.72 & $\sim 75$--110 \tnote{a} & 30\,954 & $\gtrsim 1.2$ \tnote{b} & I,Q,U,V & $\sim 1.5$ \\
LoTSS \& V-LoTSS \tnote{d} & 144 & 48 & 6 & 5\,634 & 0.095 & I,Q,U,V & 4.4 \\
TGSS & 150 & 16.7& 25 & 36\,900 & 2--5 & I & 0.62 \\
RACS-low & 887.5 & 288 & 15--25 & 34\,240 & 0.2--0.4 & I & 2.1 \\
RACS-mid \tnote{e} & 1\,367.5 & 144 & $\gtrsim 8$ & 36\,449 & $\sim 0.15$--0.4 & I,V & $\sim 3.0$ \\
RACS-high \tnote{f} & 1\,667.5 & 288 & $\gtrsim 8$ & $\sim$35\,955 & 0.2--0.4 & I,V & $\sim 3.0$ \\
SUMSS \& MGPS-2 & 843 & 3 & 45 & 10\,300 & 1.5 & RC & 0.2 \\
NVSS & 1\,346, 1\,435 & 42 & 45 & 33\,800 & 0.45 & I,Q,U & 2 \\
FIRST & 1\,346, 1\,435 \& 1\,335 & 42 \& 128 & 5 & 10\,575 & 0.13 & I & 0.9 \\
VLASS & 3\,000 & 2\,000 & 2.5 & 33\,885 & 0.07 & I,Q,U & 5.3 \\
AWES \& AMES \tnote{g} & 1361.25 & 137.5 & $\geq 11$ & $\sim 1000$ & $\sim 0.04$ & I,V & $\sim 0.25$ \\
\bottomrule
\end{tabular}
\begin{tablenotes}[flushleft]

{\footnotesize \item[] \textit{Surveys, references, and notes.}  \item[] VLSSr: Very Large Array (VLA) Low-frequency Sky Survey Redux \citep{lcv+14} \item[] GLEAM: the GaLactic and Extragalactic MWA survey \citep[][]{wlb+15,gleamegc,HurleyWalker2019a,Franzen2021}. \item[] GLEAM-X: GLEAM-eXtended \citep{HurleyWalker2022}. \item[] LOFAR: LOw-Frequency ARray (LOFAR) Two-metre Sky Survey  \citep[LoTSS;][]{lotss-dr1,Tasse2021,lotss-dr2} \corrs{and the circularly polarized LoTSS \citep[V-LoTSS;][]{Callingham2022}}. \item[] TGSS: Tata Institute for Fundamental Research Giant Metrewave Radio Telescope Sky Survey \citep[alternate data release 1;][]{ijmf16}. \item[] SUMSS: Sydney University Molonglo Sky Survey \citep{bls99,mmb+03}. \item[] MGPS-2: The second epoch Molonglo Galactic Plane Survey \citep{mmg+07}. \item[] NVSS: National Radio Astronomy Observatory VLA Sky Survey \citep{ccg+98}. \item[] FIRST: Faint Images of the Radio Sky at Twenty centimetres \citep[][]{first,wbhg97,hwb15}. \item[] VLASS: VLA Sky Survey \citep[][]{vlass}. \item[] AWES and AMES: Apertif \citep{vanCappellen2022} Wide-area/Medium-deep Extragalactic Surveys \citep[][]{Adams2022}. \item[a] \corrs{Stokes I sources.} \item[b] Values reported for the 200-MHz wideband data, declination-dependent. \item[c] Projected for full release \citep{HurleyWalker2022}. \item[d] Based on data release 2 \citep{lotss-dr2}, which overlaps the sky coverage of data release 1 \citep{lotss-dr1}. \item[e] This work. \item[f] Projected based on this work and first pass processing/observing. \item[g] Continuum data products, based on the first data release \citep{Adams2022,Kutkin2022}. 
}
\end{tablenotes}
\end{threeparttable}
\end{table*}

The Australian SKA Pathfinder \citep[ASKAP;][]{askap1,askap3,Hotan2021} is a 36-antenna interferometer located \corrs{within Inyarrimanha Ilgari Bundara, the CSIRO \footnote{Commonwealth Scientific and Industrial Research Organisation.} Murchison Radio-astronomy Observatory in Western Australia.} ASKAP operates from 700--1800\,MHz with 288-MHz of instantaneous bandwidth and features 12-m diameter dishes. The array is arranged in a dense core with a small number of outlying antennas to achieve high angular resolution and good surface brightness sensitivity with baselines ranging from 22\,m to 6\,km.

ASKAP was designed as a survey instrument and has been an in-field test for Phased Array Feed (PAF) technology \citep[][]{Hotan2014,McConnell2016}. The PAF digitally forms 36 primary beams that can be arranged within a tile (hereafter this arrangement is referred to as the PAF footprint) which allows ASKAP to observe a frequency-dependent $\sim$15--31\,deg$^2$ area instantaneously \citep{McConnell2017}. Although the beams are largely independent, adjacent beams share some of their contributing PAF elements and so their noise is correlated by up to 20\% \citep{Serra2015}. ASKAP is working towards a range of large-area surveys, including deep Stokes I total intensity mapping \citep[the Evolutionary Map of the Universe, EMU;][]{emu2}, polarized intensity and rotation measure mapping \citep[the Polarisation Sky Survey of the Universe's Magnetism, POSSUM;][]{possum1}, as well as spectral line studies of Galactic and extragalactic radio sources \citep[e.g.][]{Rhee2022:dingo,gaskap1,wallaby1,Allison2022} and studies of variability and transient sources \citep[e.g.][]{craft1,Murphy2021}. These deep, large-area surveys will complement and expand on the existing ecosystem of multi-wavelength surveys covering the sky with a range of frequencies, sensitivities, and angular resolutions. Table~\ref{tab:surveys} summarises many of these completed and in-progress surveys.

\begin{table*}[t]
\begin{threeparttable}
\caption{\corrs{RACS-low and RACS-mid observing parameters.}\label{tab:observing}}
\begin{tabular}{lcc} \toprule
     & RACS-low & RACS-mid \\\midrule
     Observed central frequency  (MHz) & 887.5 & {1295.5} \tnote{a} \\
     Effective central frequency (MHz) & 887.5 & 1367.5 \tnote{a} \\
     Observed bandwidth (MHz) & 288 & 288\tnote{a} \\
     Effective bandwidth (MHz) & 288 & 144 \tnote{a} \\
     Integration per tile (min) & 15 & 15 \\
     PAF footprint & \texttt{square\_6x6} (Figure~\ref{fig:footprint}) & \texttt{closepack36} (Figure~\ref{fig:footprint}) \\
     Beam spacing (deg) & $1.05$ & $0.9$ \\
     Tiles & 903 & 1493 (Figure~\ref{fig:tiling}) \\
     Sky coverage & $\delta_\text{J2000} \lesssim +41^\circ$ & $\delta_\text{J2000} \lesssim +49^\circ$ \\
    Surveyed area (deg$^2$) & 34\,240 & 36\,449  \\\bottomrule
\end{tabular}
\begin{tablenotes}[flushleft]
{\footnotesize \item[a] Note \corrs{that for RACS-mid} half of the band is flagged due to RFI (Section~\ref{sec:spectralcoverage}).
}
\end{tablenotes}
\end{threeparttable}
\end{table*}

\begin{figure}[t]
    \centering
    \includegraphics[width=1\linewidth]{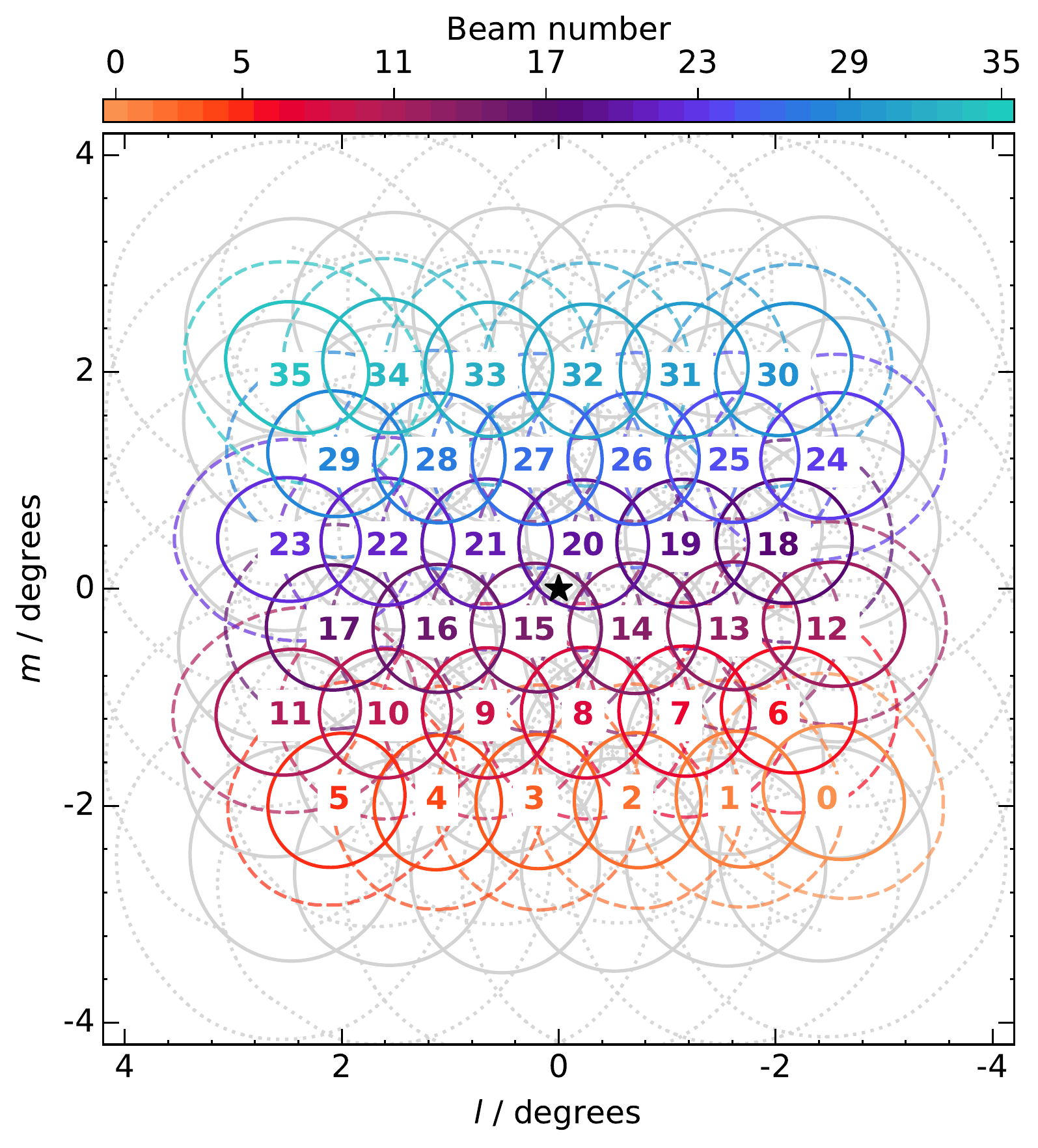}
    \caption{\label{fig:footprint} \corrs{Representative RACS-mid \texttt{closepack36} PAF footprint layout and shape. The coloured solid contours indicate 50\% attenuation for a particular beam, and the faint, dashed contour indicates 12\% (i.e.~apparent brightness is attenuated to 12\% of the sky brightness). The light grey contours correspond to the RACS-low \texttt{square\_6x6} footprint for comparison (solid, 50\%; dotted, 12\%). The black star indicates the centre of the footprint.}}
\end{figure}

\begin{figure}[t]
\centering
\includegraphics[width=1\linewidth]{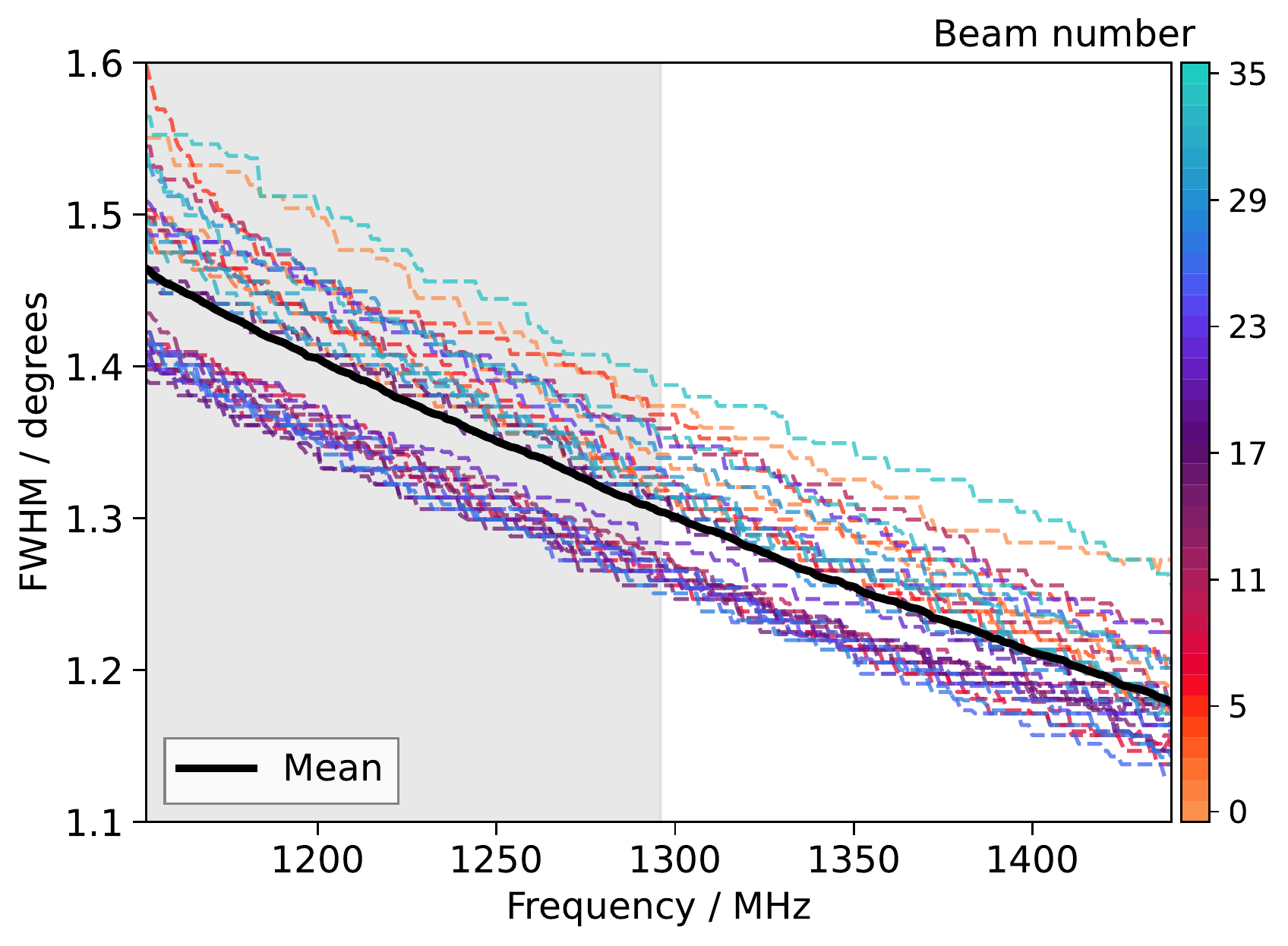}
\caption{\label{fig:fwhm} \corrs{FWHM of the primary beam response as a function of frequency across the full RACS-mid band for each beam in the PAF footprint. The black, solid line indicates the beam-averaged FWHM, \CORRS{and the grey, shaded region is flagged (see Section~\ref{sec:spectralcoverage})}}.}
\end{figure}

\begin{figure}[t]
    \centering
    \includegraphics[width=1\linewidth]{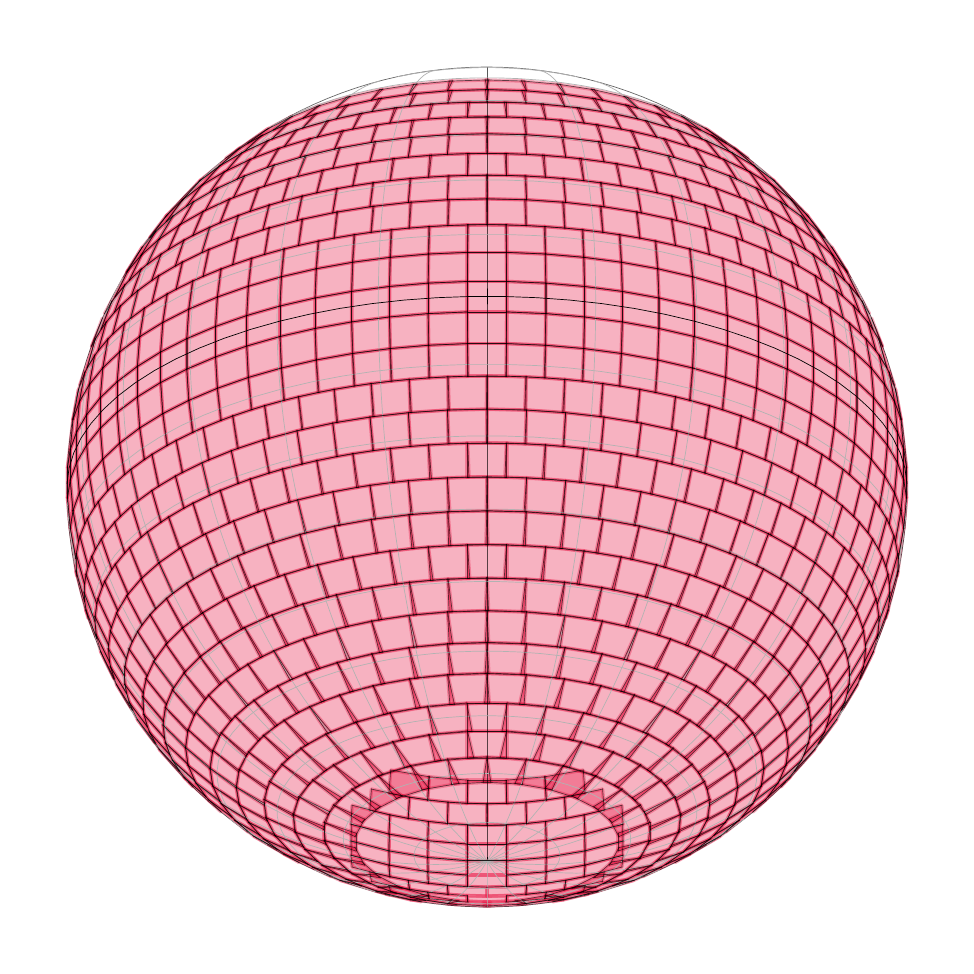}
    \caption{\label{fig:tiling}\corrs{ Tiling of the celestial sphere for RACS-mid, with a view centered on $(\alpha_\text{J2000},\delta_\text{J2000}) = (0, -27)^\circ$.}}
\end{figure}

The Rapid ASKAP Continuum Survey (RACS) was started as a CSIRO-led Observatory Project \citep[][hereinafter, \citetalias{racs1}]{racs1}~\footnote{\url{https://research.csiro.au/racs/}.} 
with the goal of creating a global sky model for calibration of the ASKAP surveys. RACS will cover the sky available to ASKAP to a moderate sensitivity across ASKAP's observing frequency range in three bands. The first pass of the survey in the low band at 887.5\,MHz (hereinafter RACS-low) was released in 2020  and a set of catalogues and uniform sensitivity images were released shortly after to provide one of the deepest large-area surveys to date \citep[][hereinafter, \citetalias{racs2}]{racs2}.

While a global sky model has been the primary motivation for RACS, numerous scientific works in Galactic, extragalactic, and cosmological contexts have benefited from the first epoch of RACS-low. \citet{Darling2022} has performed a cosmological study combining RACS-low with VLASS to provide the most sensitive all-sky source counts. Variability of active galactic nuclei (AGN) \citep{Ross2022}, stars \citep{Wang2021,Driessen2022}, and other sources \citep{Murphy2021} have made use of the first epoch of RACS-low, which provides a unique epoch for flux density measurements at 887.5\,MHz. High-redshift galaxies/AGN have been both newly discovered \citep{Ighina2021} and characterised \citep{Drouart2021,Ighina2022,Cai2022,Broderick2022} thanks to the sensitivity and angular resolution of the first epoch of RACS-low. Extended radio sources have also featured in work using images from the first epoch of RACS-low, including stellar bow shocks \citep{VandenEijnden2022}, radio emission associated with galaxy cluster mergers \citep[e.g.][]{Duchesne2021c,Duchesne2022}, searches for giant radio galaxies \citep[e.g.][]{Andernach2021}, and characterisation of nearby star-forming galaxies \citep{Kornecki2022}. In addition to these science results, RACS-low has been instrumental in providing lessons in data processing, autonomising ASKAP science operations (Moss et al., in prep), and general understanding of the performance of ASKAP \citepalias{racs1}. This knowledge has been absorbed by the observatory and the various ASKAP survey teams and applied during the pilot survey phase of ASKAP \citep[e.g.][]{For2021,Allison2022}. This highlights the utility of the comparatively shallow RACS project even in the upcoming era of deep ASKAP survey science.

The present work details efforts to survey the sky in A\-S\-K\-A\-P's mid-frequency band 2 (hereinafter RACS-mid) and represents the third data release from the RACS project. Future releases will include \corrs{creation of complementary, curated catalogues} in the mid band {(Duchesne et al., in prep)}, observations and catalogues in the high-frequency band 3, RACS-high, as well as a second epoch of RACS-low to benefit from instrument and data-processing improvements since the initial RACS-low observations and data release. Alongside the continuum data releases, a complementary project to extract polarized spectra is underway. Spectra and Polarization In Cutouts of Extragalactic sources from RACS (SPICE-RACS, Thomson et al., submitted) will initially make use of the RACS-low data products, and will extend to all bands later.

The following sections will describe RACS-mid and its initial data release.

\section{RACS-mid survey design and execution}

Generally, RACS-mid and subsequent RACS epochs follow a similar observing strategy to RACS-low with many $\sim$900\,s observations covering the sky. In this section, key differences from the survey description provided in \citetalias{racs1} will be highlighted and a brief overview provided. Specific observing parameters for RACS-mid are collected in Table~\ref{tab:observing}. A database~\footnote{\url{https://bitbucket.csiro.au/projects/ASKAP_SURVEYS/repos/racs/browse}.} is available that summarises observations for all of the RACS epochs, including observed field details and cross-matched source-lists for resulting images. RACS-mid is collected under \texttt{epoch\_1}~\footnote{The first epoch of RACS-low corresponds to \texttt{epoch\_0}.}. \corrs{Note that validation files produced as part of this database are not intended for scientific use.} Hereinafter references to a `database' are for that repository.

\subsection{Field-of-view, tiling, and scheduling the observations}\label{sec:observations}

\begin{figure*}
\centering
\includegraphics[width=0.8\linewidth]{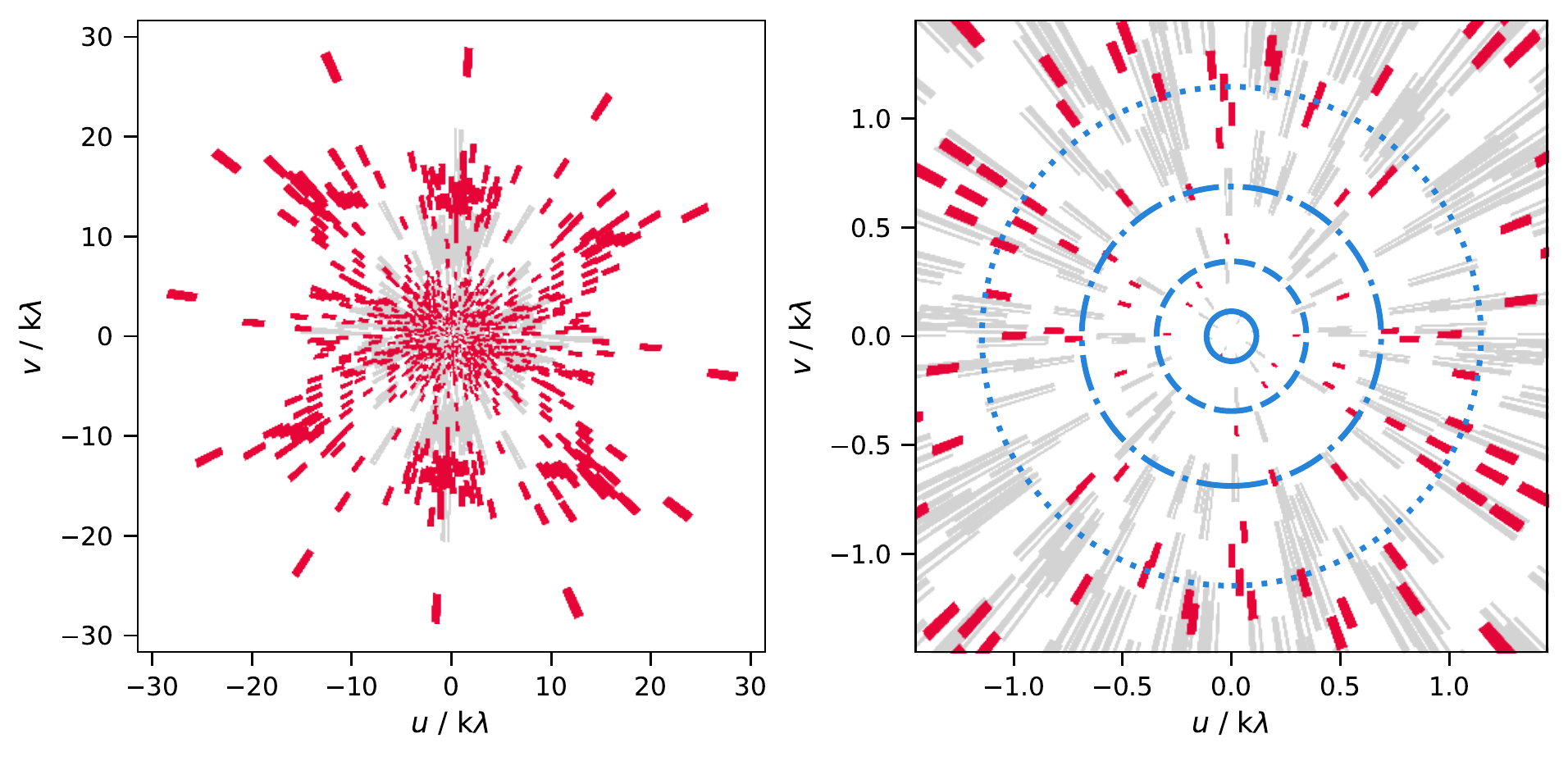}
\caption[]{\corrs{\label{fig:uvcover} Example $(u,v)$ coverage for a single central beam of the RACS-mid observation (SB21663, field \texttt{RACS\_0812-28}, beam 15, red) close to zenith, compared to a similar individual beam from the first epoch of RACS-low (SB8576, field \texttt{RACS\_1618-25A}, beam 0, light-grey). The left panel shows the full $(u,v)$ coverage, and the right panel shows the inner $\sim 1.5$\,k$\lambda$. The blue circles on the right panel enclose the $(u,v)$ range corresponding to angular scales of 3 (dotted), 5 (dot-dash), 10 (dashed), and 30 (solid)\,arcmin.}}
\end{figure*}

\begin{figure*}[t]
\centering
\includegraphics[width=1\linewidth]{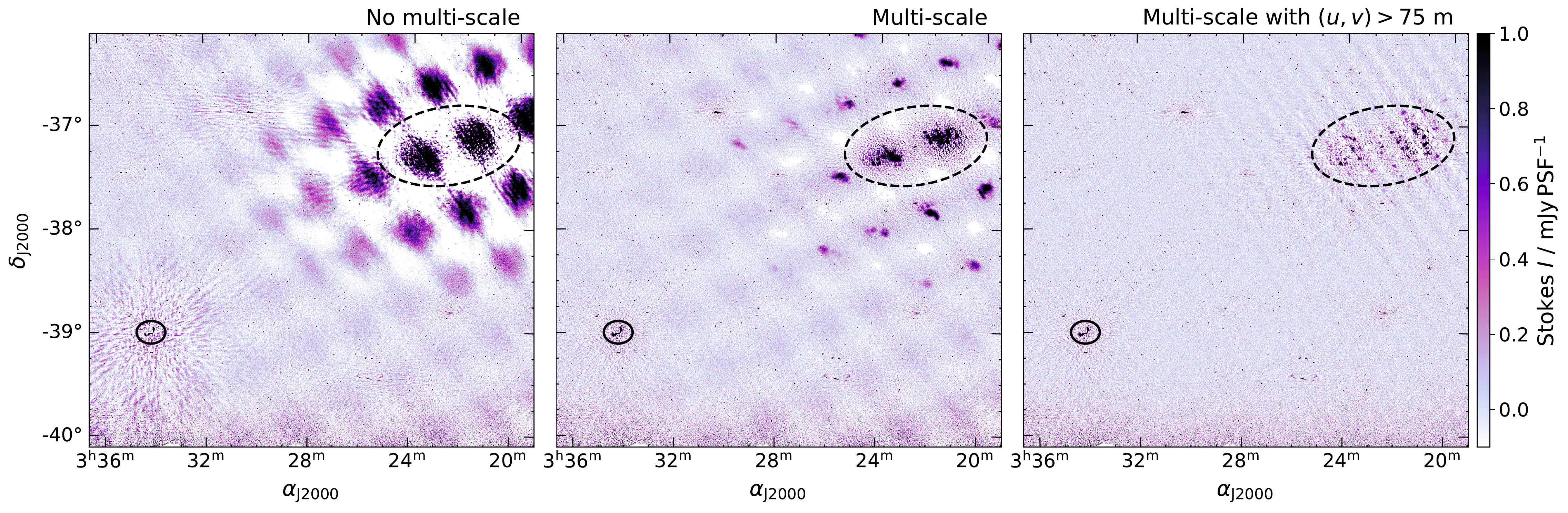}
\caption{\label{fig:multiscale} \corrs{Comparison of the field containing Fornax~A (\texttt{RACS\_0329-37}, enclosed with the black, dashed ellipse) without multi-scale deconvolution (\emph{left}), with multi-scale deconvolution (\emph{centre}), and with multi-scale deconvolution after application of the $(u,v)<75$\,m cut to the data (\emph{right}). Another miscellaneous extended source is highlighted in the black, solid circle.}}
\end{figure*}

\corrs{At the central observing frequency of {1295.5}\,MHz, RACS-mid has a smaller field-of-view (FoV) than RACS-low. The frequency-averaged full-width at half maximum (FWHM) for a single primary beam of the PAF is $\sim$1.3\,deg. To achieve more uniform sensitivity across the footprint, we use the \texttt{closepack36} footprint with a 0.9\,deg separation between beams. Figure~\ref{fig:footprint} shows the \texttt{closepack36} footprint layout at 50\% and 12\% beam attenuation, plotted on top of the RACS-low \texttt{square\_6x6} footprint for comparison. Figure~\ref{fig:fwhm} shows the FWHM of the PAF beams as a function of frequency alongside the beam-averaged FWHM. The grey, shaded region in Figure~\ref{fig:fwhm} is flagged (see Section~\ref{sec:spectralcoverage}).}

Figure~\ref{fig:tiling} shows the celestial sphere tiled with the RACS-mid footprints. The number of tiles required has increased to 1493 (cf.~903 for RACS-low). Aside from this increase in pointings, the tiling is similar to RACS-low and features the same quasi-rectangular grid over the South Celestial Pole (SCP). With \corrs{only a minimal increase in noise in the} higher-declination strips of RACS-low~\footnote{RACS-low was observed up to $\delta_\text{J2000} = +37^\circ$, with image sensitivity up to $+41^\circ$, while the resulting catalogue is restricted to $\delta_\text{J2000} \leq +30^\circ$.}, we opted to observe further north for subsequent RACS passes. For RACS-mid, the survey pointings continue to $\delta_\text{J2000} = +46^\circ$ to test the performance at the elevation limits of the telescope, with the image edges reaching $\delta_\text{J2000} \approx + 49^\circ$. These high-declination observations are performed at much lower elevation than the rest of the survey, and consequently have a \corrs{significantly extended point-spread function (PSF).} The angular resolution becomes similar to the \corrs{1.4-GHz NRAO \footnote{National Radio Astronomy Observatory.} VLA \footnote{Very Large Array.} Sky Survey \citep[NVSS;][]{ccg+98}} for this northern-most set of observations. Details of the resulting PSF over the full survey are described in Section~\ref{sec:psf}. RACS-mid observations are additionally scheduled with a limit to the maximum hour angle, $\text{HA} \pm 1$\,h, to help ensure a well-behaved PSF and to help with overall consistency in data quality.

RACS-mid included a shift from semi-automated observing carried out for RACS-low towards autonomous scheduling and observing via the SAURON scheduler (Moss et al. in prep), incorporating improvements in scheduling based on RACS-low and early ASKAP survey observations. Consequently, while observations for RACS-low collected multiple target fields into single scheduling blocks with unique identifiers (hereinafter SBIDs), for RACS-mid---and other ASKAP observations---individual fields are now generally observed under a unique SBID. This change was mainly driven by the need for more reliable control over timing of field observations and for better resilience to interrupted observations. It also simplifies processing and makes it easier for users to find information in the database regarding particular fields. Each field for RACS-mid is named \texttt{RACS\_HHMM$\pm$DD}, and survey data products include both SBID and field name as unique identifiers.

Over the course of the survey, 35 SBIDs were found to be affected by instrumental errors. For most, this was due to the lagged updating of delays which generally affected the first SBID after a delay calibration scan, and has since been corrected in the system. For one day of observing, delays were inconsistent between the fields and the bandpass observation, rendering the 21 SBIDs observed that day unusable. The 35 fields observed under these SBIDs were all re-observed and given updated SBIDs. Images are not provided for the original 35 SBIDs, only for the the re-observations. In the database they are marked as \verb|OBSERVED| rather than \verb|IMAGED|. For a small subset of fields, a significant fraction of the data were flagged due to an unwrap of the antennas over the course of the observation. Ten fields with $<720$\,s of observing time were also re-observed. An additional subset of fields, particularly at high declination, were also re-observed to try to reduce the size of the PSF which may be affected by missing antennas or other flagged data. Images for the short observations and other miscellaneous re-observed fields are provided with this data release as they still have scientific use.

\subsection{Spectral coverage and effective frequency}\label{sec:spectralcoverage}
RACS-mid is observed at {1295.5\,MHz} using the full instantaneous bandwidth available to ASKAP (i.e.~288\,MHz) similar to RACS-low. Due to significant and persistent broadband radio frequency interference (RFI) in the lower half of the band the RACS-mid data are restricted to a bandwidth of {144\,MHz}. This is illustrated in figure~1 from \citetalias{racs1}. The resulting central frequency for the survey products (namely images and resulting spectral measurements) is shifted to 1367.5\,MHz as a result of this flagging. \corrs{The subset of the band that is flagged is shaded in Figure~\ref{fig:fwhm}.} 

\subsection{Data-processing}\label{sec:processing}

Calibration and imaging follows almost identically to RACS-low. This process is done at the Pawsey Supercomputing Research Centre~\footnote{\url{https://pawsey.org.au/}.} located in Perth, using the \texttt{galaxy} supercomputer. Processing, including calibration, imaging, and mosaicking is performed through \texttt{ASKAPSoft} \citep{Guzman2019}~\footnote{\url{https://www.atnf.csiro.au/computing/software/askapsoft/sdp/docs/current/index.html}.}, which is built specifically as a collection of software to process ASKAP data on Pawsey systems with an associated pipeline for ease of use. For most of the survey imaging and calibration, \texttt{ASKAPSoft} version 1.5.0 was used except for a small subset of observations taken beginning May 2022 which used version 1.6.0. For mosaicking and post-imaging work, only version 1.6.0 is used. Changes implemented for the 1.6.0 version of the pipeline largely included a difference in how small supercomputer jobs were arranged to make better use of available compute nodes and does not affect the data quality.

\subsubsection{General calibration and imaging}\label{sec:calibration:imaging}

Bandpass, flux density scale, on-axis leakage, and initial gain calibration are determined using the flux standard of PKS~B1934$-$638, which is normally observed once each day or for each observing configuration.  \CORRS{For on-axis leakage correction, it is assumed PKS~B1934$-$638 is completely unpolarized. \citet{Rayner2000} report $\sim +0.03$\% fractional circular polarization for PKS~B1934$-$638, though as noted by \citet{Sullivan2013} the variable nature of circularly polarized emission means this may have changed in the time since that measurement was made.} Once bandpass and initial gain calibration solutions are applied to each science observation, data from each beam are self-calibrated over three loops, independently. Each loop decreases the CLEAN threshold during the imaging step to generate a deeper field model. Self-calibration normalises gain amplitudes, creating a phase-only--equivalent calibration self-solution.

Imaging with deep deconvolution then follows for each beam independently. The \texttt{ASKAPSoft} imager makes use of a $w$-projection gridding algorithm, with multi-scale CLEAN and multi-frequency synthesis (MFS) deconvolution. An equivalent of `Briggs' \citep{bri95} robust 0.0 weighting is achieved through preconditioning \citep{Rau2010} of the data. As part of the MFS deconvolver, the sky brightness distribution is expanded into two Taylor terms to account for \corrs{the normal power law spectral dependence of sources in total intensity and the spectral-dependence of some instrumental features, e.g.~the PSF and primary beam. As the effective fractional bandwidth is small ($\sim 10$\%), the most significant spectral effect is the primary beam FWHM variation across the band (see Figure~\ref{fig:fwhm}) which is accounted for with the second Taylor term.} At this stage, both Stokes I and V continuum products are produced for each beam. \corrs{The same number of Taylor terms are used for Stokes V imaging, though circularly polarized emission mechanisms have more variation.} 

\CORRS{For Stokes I, the final CLEAN run uses up to ten major iterations with a minor cycle threshold of 45\% and minor cycle gain of 30\%. The major cycle stopping threshold is 0.75\,mJy\,PSF$^{-1}$, with a final minor cycle threshold of 0.5\,mJy\,PSF$^{-1}$. With fewer sources, for Stokes V deconvolution we use a maximum of only three major iterations with a 30\% and 20\% minor cycle threshold and gain, respectively. The \texttt{ASKAPsoft} imager does not restrict where CLEAN components can be found except in the final major cycle, where CLEAN components below the major cycle threshold can be found if they lie within pixels of the model generated during the previous major cycles. For both Stokes I and V images, peak positive residuals after CLEANing are $\sim 0.6$--0.8\,mJy\,PSF$^{-1}$, though vary depending on beam and field. The individual restored Stokes I and V beam images are convolved to a common resolution prior to mosaicking.} 

\corrs{The \texttt{ASKAPsoft} imaging of Stokes V results in the sign of the circularly polarized emission to be consistent with the IAU standard (right-hand circularly polarized light is positive, and left-hand circularly polarized light is negative). This is opposite to the convention generally adopted in pulsar astronomy \citep[see e.g.][]{vanStraten2010}.}

\subsubsection{Extended sources, the Galactic Plane, and large-scale ripples}\label{sec:gp}

Figure~\ref{fig:uvcover}\label{figuse:uvcover:1} shows the $(u,v)$ coverage for a nominal RACS-mid observation compared to a similar observation from RACS-low, highlighting the minimal $(u,v)$ coverage \corrs{as angular scales increase beyond $3$\,arcmin. Both example observations represent the highest-elevation pointings in RACS-mid and RACS-low.} The Galactic Plane poses a significant challenge in imaging even with good $(u,v)$ sampling for most modern instruments (e.g. the MWA, \citealt{HurleyWalker2019a,HurleyWalker2022,Tremblay2022}; MeerKAT, \citealt{Heywood2022}; and deep ASKAP observations, \citealt{Umana2021}), more so for the snapshot observations described here. 

Issues arise from incomplete sampling of the inner $(u,v)$ plane: flux density on large angular scales is not well measured, and artefacts around bright, extended sources can dominate images. \corrs{Multi-scale deconvolution is used to help in modelling extended sources, though this can result in `ghost' sources appearing due to the CLEAN algorithm misinterpreting artefacts and source sidelobes as real sources. In Figure~\ref{fig:multiscale} we show that despite this, multi-scale deconvolution is still an appropriate choice to ensure extended sources are modelled well during deconvolution, however, the residual ghost sources are not local to the extended source, and can appear throughout the full image. The center panel of Figure~\ref{fig:multiscale} highlights an example of the ghost sources, in this case originating from nearby radio galaxy Fornax~A. In this instance, a single ghost source has an absolute integrated flux density of up to $\sim 0.5$\,Jy, compared to the $\sim 6$\,Jy integrated flux density of a single lobe of Fornax~A from the same image.}

\corrs{Despite their prominence during visual inspection of the image, ghost sources are typically not detected and characterised by the \texttt{selavy} \citep{selavy} or \texttt{aegean} \citep{hmg+12,hth18}~\footnote{\url{https://github.com/PaulHancock/Aegean}.}  source-finders used in this work, as they use a position-dependent noise and $5\sigma_\text{rms}$-thresholding and are not optimized for detection and modelling of faint extended sources.  Other source-finders such as \texttt{PyBDSF} \citep[][as used in \citetalias{racs2}]{pybdsf} may detect them, depending on user-settings. In the Fornax~A example shown in Figure~\ref{fig:multiscale}, no ghost sources are detected by \texttt{selavy} or \texttt{aegean}.}

\corrs{To help reduce the number of ghost sources we set a minimum $(u,v)$ cut corresponding to 75\,m baselines during imaging for a selection of affected observations. This removes large-scale ripples and other sidelobe features, reducing the number of ghost sources. An example of the effect of the $(u,v)$ cut is shown in the right panel of Figure~\ref{fig:multiscale}. While this $(u,v)$ cut does not improve imaging quality directly at the location of the extended source, it helps the reduce artefacts in the sky within a few degrees of the source, reducing the mean root-mean-square (rms) noise over the image (from 191 to 183~\textmu Jy\,PSF$^{-1}$ across the Fornax~A image) and local rms noise at the affected locations (e.g.~from 267 to 170~~\textmu Jy\,PSF$^{-1}$ around a `ghost' source in the Fornax~A image). Most selected observations are within or near the Galactic Plane, though we also select a small number of extra-Galactic fields like the field containing Fornax~A. Some additional extra-Galactic fields are also affected by solar interference, which results in a similar problem along with large-scale ripples and a $(u,v)$ cut is used for those observations as well.}

The $(u,v)$ cut (in metres) used for each SBID is provided in the RACS database and as an additional item in the FITS header under the \texttt{MINUV} keyword. \corrs{These SBIDs still feature the largest number of residual ghost sources and general artefacts as they are typically in the Galactic Plane, and we recommend users exercise caution when inspecting the images if interested in real extended sources.}

\subsubsection{Peeling}\label{sec:peeling}

\begin{figure*}[t!]
    \centering
    \includegraphics[width=1\linewidth]{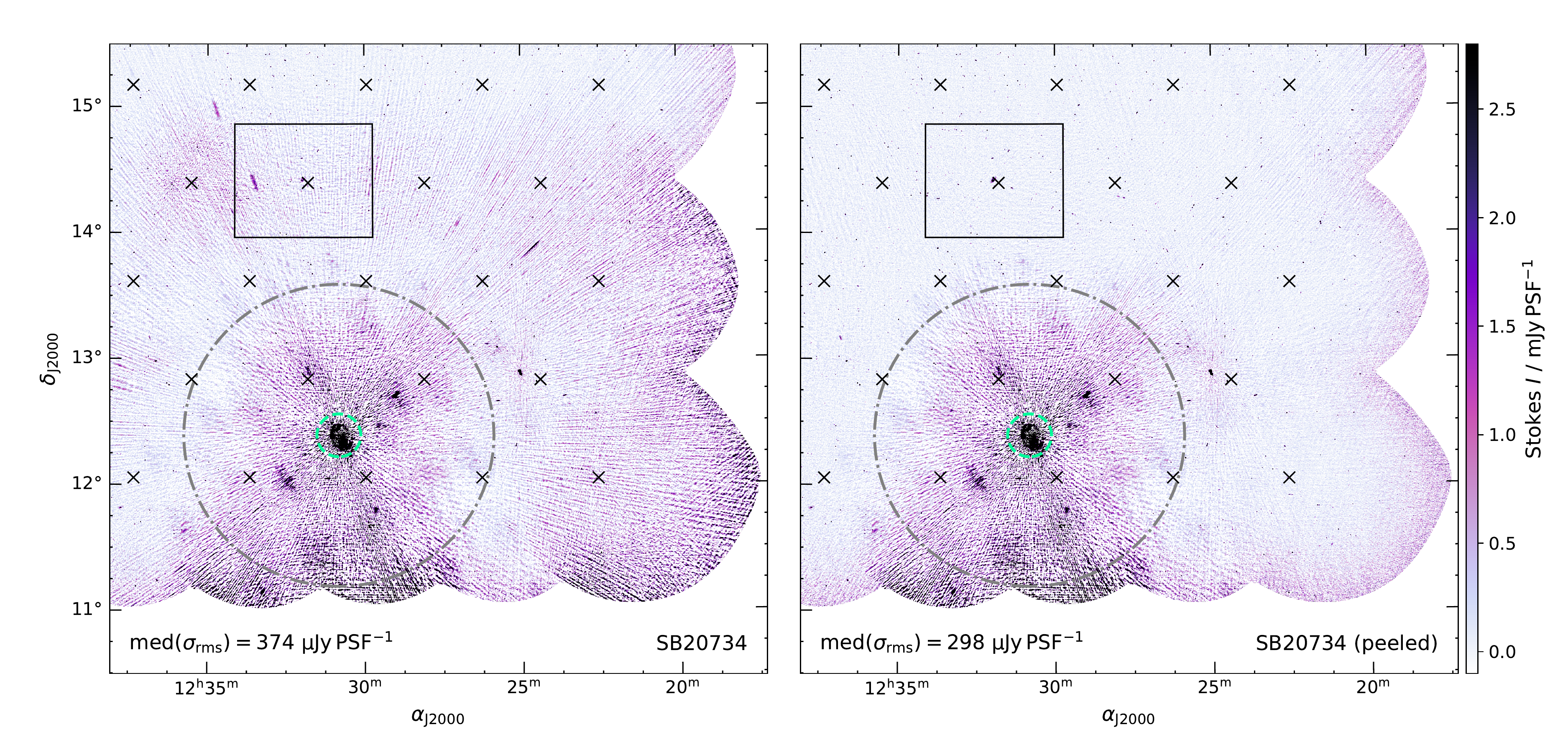}\\
    \includegraphics[width=1\linewidth]{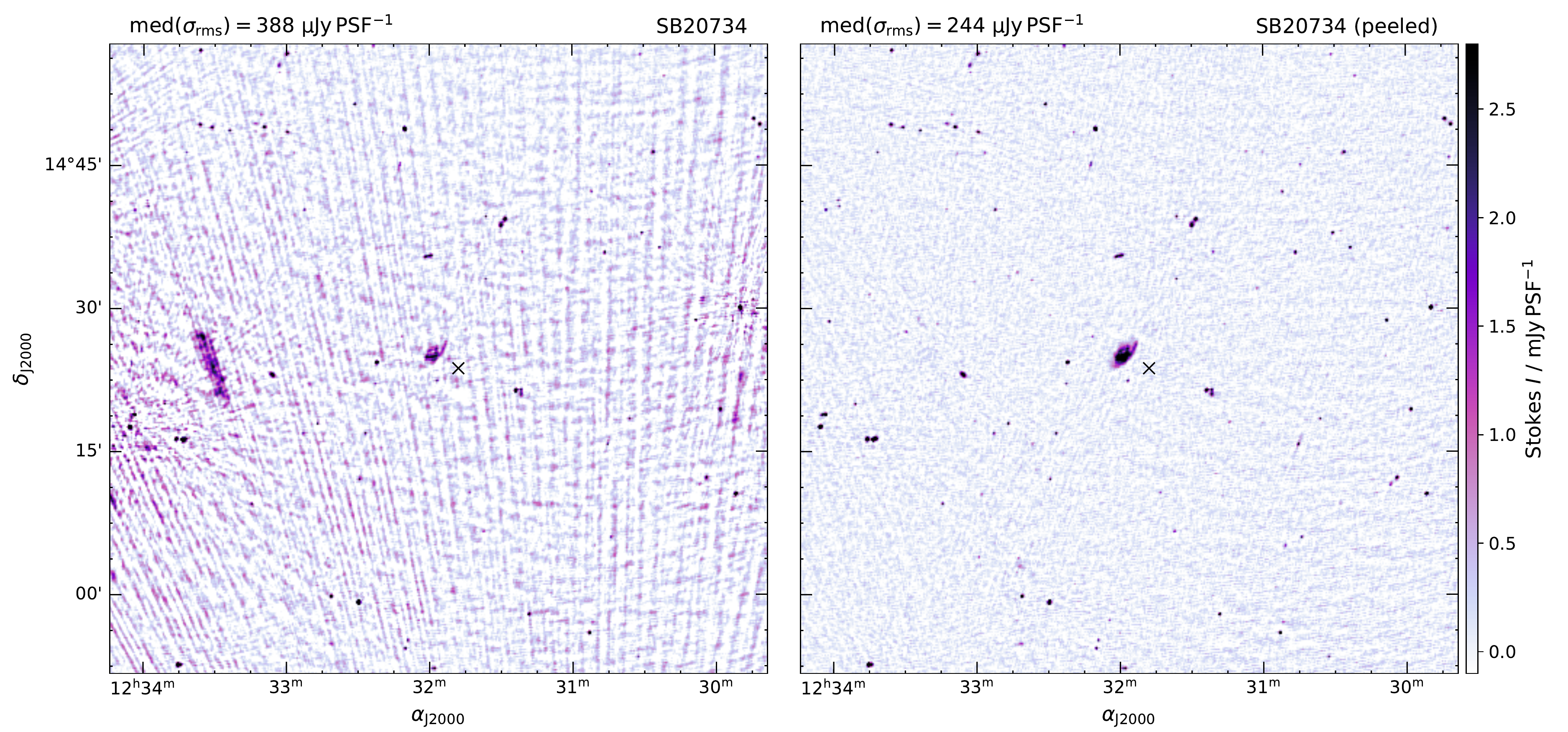}
    \caption{\label{fig:peel} SB20734 containing Virgo~A before peeling (\textit{left}) and after peeling (\textit{right}). A zoom-in of the region above Virgo~A before and after peeling is shown in the bottom row, with a solid, black box indicating its location in the top panels. The dot-dash, grey circle has a 1.2\,deg radius: beams with centres within this radius are excluded from peeling. The dashed, green circle indicates the radius within which Virgo~A is modelled. Black crosses indicate the beam centres. The median rms noise, $\sigma_\text{rms}$, in the top panels is quoted for the full tile excluding the 1.2\,deg circle containing Virgo~A (which is unchanged after peeling). In the bottom panels $\sigma_\text{rms}$ is quoted for the zoomed-in region only.}
\end{figure*}

\begin{table}
    \centering
    \begin{threeparttable}
    \caption{\label{tab:peeling} List of peeled sources, aperture within which they are eventually subtracted (see main text) and SBIDs they are subtracted out of for beams where they are $\geq 1.2$\,deg from the beam centre.}
    \begin{tabular}{r c c l}\toprule
         Source & Coordinates & Radius & SBIDs  \\
         & (hh:mm:ss dd:mm:ss) & ($^\prime$) & \\\midrule
         Centaurus~A & 13:25:28 -43:00:40 & 6.60 \tnote{a} & \makecell[l]{21760,21761,\\ 21856,21857} \\
         Taurus~A & 05:34:31 +22:00:59 & 4.20 & \makecell[l]{20712,20776,\\21399,21400} \\
         Cygnus~A & 19:59:28 +40:44:02 & 1.80 & \makecell[l]{20490,20611,\\20612,20613,\\20614,22818,\\22819} \\
         Hercules~A & 16:51:08 +04:59:33 & 2.34 & 21929 \\
         Hydra~A & 09:18:05 -12:05:42 & 1.20 & 21746,21842 \\
         Orion~A \tnote{b} & 05:35:18 -05:23:11 & 7.11 &  \\
         Orion~B \tnote{b} & 05:41:41 -01:53:58 & 4.95 & \multirow{-2}{*}{\makecell[l]{21057,21058,\\21653,21654,\\21655}} \\
         Pictor~A & 05:19:48 -45:45:53 & 6.30 & 21898,22029 \\
         Virgo~A & 12:30:49 +12:23:28 & 10.2 & 20734,20799 \\
         PKS~B0407$-$658 & 04:08:20 -65:45:09 & 0.36 & \makecell[l]{21946,22023,\\22445,33405} \\
         PKS~B0409$-$752 & 04:08:48 -75:07:19 & 0.78 & 22024,22025 \\
         PKS~B0723$-$008 & 07:25:50 -00:54:55 & 0.36 & 21064 \\
         PKS~B1932$-$464 & 19:35:56 -46:20:39 & 0.63 & 20276,20363 \\         
         PKS~B1934$-$638 & 19:39:25 -63:42:45 & 0.78 & \makecell[l]{20147,20272,\\20511} \\
         PKS~B2152$-$699 & 21:57:07 -69:41:15 & 1.38 & 21937,21940 \\
         PKS~B2356$-$611 & 23:59:02 -60:54:52 & 4.50 & 21710 \\
         3C~48 & 01:37:41 +33:09:34 & 0.60 & 20336,20338 \\
         3C~84 & 03:19:48 +41:30:42 & 1.44 & 20259,20260 \\
         3C~111 & 04:18:21 +38:01:52 & 2.82 & 20298 \\
         3C~119 & 04:32:36 +41:38:26 & 1.02 & 20262,20263 \\
         3C~123 & 04:37:04 +29:40:15 & 0.78 & \makecell[l]{20386,20387,\\20523} \\
         3C~147 & 19:59:28 +40:44:02 & 0.90 & 25458 \\
         3C~196 & 08:13:36 +48:13:00 & 1.38 & 25465 \\
         3C~273 & 04:37:04 +29:40:15 & 0.78 & 20800,21079 \\
         3C~298 & 12:29:06 +02:03:08 & 0.66 & 21978,21979 \\
         3C~345 & 16:42:58 +39:48:32 & 1.02 & 20176 \\
         3C~380 & 18:29:31 +48:44:46 & 1.02 & 22883 \\
         3C~409 & 20:14:27 +23:34:53 & 0.62 & 20824 \\
         3C~433 & 21:23:44 +25:04:14 & 1.32 & 20752,20753\\
         3C~446 & 22:25:47 -04:57:01 & 0.36 & 20456,20457 \\
         3C~454.3 & 22:53:57 +16:08:53 & 0.90 & \makecell[l]{20758,20832,\\20833} \\\bottomrule

    \end{tabular}
    \begin{tablenotes}[flushleft]
    {\footnotesize
    \item[] \item[a] Only the inner lobes and core are included as the large-scale outer lobes are largely undetected in the RACS-mid data (see Section~\ref{sec:gp} for some discussion of the angular scale sensitivity). \item[b] Both Orion A and B are peeled/subtracted in the same tiles in that order, as they have an angular separation of $\sim 3.8$\,deg.
    }
    \end{tablenotes}
    \end{threeparttable}
\end{table}

A significant source of artefacts in the initial RACS-mid imaging is contribution from bright off-axis sources. Both large, extended radio sources as well as compact sources radiate sidelobes and additional direction-dependent artefacts through the imaged field of view when they fall within one of the sidelobes of the primary beam. The first primary beam sidelobe is $\sim2$\,deg from the beam centre at 1367.5\,MHz. In extreme cases, sidelobes or artefacts from such sources can cause the deconvolution to diverge rendering a single beam image unusable. To mitigate this issue, we opt for a `peeling' and subtraction approach for particularly problematic sources. This \corrs{by-eye selection} typically includes sources $\gtrsim 10$\,Jy at 1367.5\,MHz. This `peeling' largely follows the definition and process described by \citet[][see also \citealt{Smirnov2011b}]{Noordam2004}, though we use a combination of (1) direct visibility subtraction, (2) true directional peeling, and (3) temporary mainlobe subtraction prior to peeling, depending on signal-to-noise ratio (SNR) and complexity of the offending source. The modes are generally used together---directional subtraction follows a round of peeling to remove residual emission due to, e.g., a difference in the directional gain amplitudes with the original data.

\emph{Direct visibility subtraction.} For direct subtraction, the visibilities are phase-rotated to the direction of the bright source to be removed, and the source is imaged using the widefield imager \texttt{WSClean} \citep{wsclean1,wsclean2}~\footnote{The choice of \texttt{WSClean} over the \texttt{ASKAPSoft} imager is simply due to the need to retain, modify, and otherwise replace the various columns of data in the MeasurementSet (i.e. \texttt{DATA}, \texttt{CORRECTED\_DATA}, and \texttt{MODEL\_DATA}) as well as temporarily write these to disk (or store in memory, depending on size). \texttt{ASKAPSoft} software is generally designed around minimal disk-writing and minimal column editing so it is currently not compatible with this process.}. A mask is created to exclude the sky outside a circular aperture enclosing the source, and a CLEAN component model is derived from that masked image, and from it corresponding model visibilities $M$. Using the Jones matrix formalism of the radio interferometer measurement equation \citep{Hamaker1996,Smirnov2011a}, the modified visibilities are then computed as
\begin{equation}\label{eq:subtraction}
    V_{pq}^\prime = V_{pq} - M_{pq} \, ,
\end{equation}
for correlated visibilities formed by antennas $p$ and $q$. The resultant visibility data, $V^\prime$, is then phase-rotated back to the original direction. This subtraction procedure is always run if the source is outside of the specified FoV for a given SBID. There is no benefit in subtracting within the main lobe FoV as this is functionally similar to deconvolving the source during normal imaging and would result in a missing source in the image.

\emph{Directional peeling.} With a sufficient SNR for the off-axis source, a round of gain calibration on the derived CLEAN model can be reliably performed. The CLEAN model, $M$, is then subtracted after applying the inverse of the derived gains, $G$. Thus the source-subtracted visibilities, $V^\prime$, are \begin{equation}
    V_{pq}^\prime = V_{pq} - G_{p}^{-1} M_{pq} (G_{q}^{H})^{-1} \, ,
\end{equation}
where the superscript $H$ is the Hermitian transpose. This is a form of direction-dependent calibration, but is only applied to the source model and not the data itself. As this involves solving for gain solutions, a sufficiently high SNR is required to determine reliable gains for all antennas whether or not the mainlobe model has been subtracted. The choice of cut-off SNR is dependent on source structure, and can typically be lower for point sources. As the default mode of gain calibration here is to solve for both phases and amplitudes (which tend to produce the best results), an additional directional subtraction round is always run afterwards.

\emph{Mainlobe subtraction.} If the SNR of the source is not sufficient for good solutions during directional peeling, but direct subtraction by itself is not sufficient to remove all unwanted artefacts, then subtracting the field model within the mainlobe of the primary beam prior to peeling can help. In this process, we start with a temporary directional subtraction of the bright off-axis source followed by shallow imaging of the mainlobe. The field model within the mainlobe is subtracted, and the bright off-axis source is returned to the data for gain calibration. The field model is then also returned to the data once gain solutions are derived, and the bright off-axis source is peeled as per usual.

Table~\ref{tab:peeling} lists all SBIDs within which sources have been peeled and/or subtracted. Generally, only sources off the Galactic Plane are included as the Galactic Plane poses additional imaging challenges (see Section~\ref{sec:gp}). Sources are chosen after a full round of imaging to visually identify which sources cause problems. For each SBID with a problematic source, all 36 beams are independently re-processed and the source is only peeled or subtracted if it is greater than 1.2\,deg away from the beam centre of a given beam. Because we do not peel the source from all beams, artefacts persist for some beams; though they are now localised to within $\sim 1.2$\,deg of the source. Figure~\ref{fig:peel} shows the peeling result for SB20734 that contains Virgo~A. \corrs{The dot-dash, grey circle indicates the 1.2~deg radius outside of which peeling is performed, and the dashed, green circle indicates the aperture within which Virgo~A is modelled. In the figure we also show a zoom-in of a region directly to the north of Virgo~A, highlighting the improvements.} A \texttt{python}-based pipeline~\footnote{\texttt{PotatoPeel}: \url{https://gitlab.com/Sunmish/potato}.} for peeling is used as a manual intermediate step in the \texttt{ASKAPSoft} pipeline prior to self-calibration once a problematic source has been identified, requiring its location and approximate size. 

\subsection{Tile mosaics and primary beam measurements}\label{sec:holography}

\begin{table*}[t]
    \centering
    \begin{threeparttable}
    \caption{\label{tab:beamformer} Periods of common beam weights (BWT).}
    \begin{tabular}{cccccc}\toprule
        Name  & Date range & RACS-mid SBID range \tnote{a} &  $N_\text{obs}$ \tnote{b}  & BWT SBID \tnote{c} & Holography SBID \\\midrule
        BWT-1 \tnote{d} & 2020-12-20 to 2021-02-24 & 20147--23538 & 1425 & 19933 & - \\
         BWT-2 \tnote{d} & 2021-02-27 to 2021-03-15 & 23708--24691 & 29 & 23605 & - \\
         BWT-3 \tnote{d} & 2021-03-31 to 2021-04-07 & 25440--25872 & 46 & 24965 & - \\
         BWT-4 & 2021-07-29 to 2021-08-01 & 29328--29584 & 29 & 28469 & 28507 \\
         BWT-5 \tnote{d} & 2021-10-29 to 2021-11-10 & 33088--33423 & 19 & 33082 & - \\
         BWT-6 & 2021-11-16 to 2022-02-27 & 33588--37725 & 6  & 33468 & 37202 \\
         BWT-7 & 2022-03-04 to 2022-03-05 & 37889--37898 & 3  & 37864 & 37997 \\
         BWT-8 & 2022-05-25 to 2022-06-04 & 40925--41287 & 18 & 38292 & 39161 \\
         BWT-9 & 2022-06-09 to 2022-06-11 & 41469--41554 & 7 & 41410 & 41458 \\
         \bottomrule
    \end{tabular}
    \begin{tablenotes}[flushleft]
    {\footnotesize
    \item[a] Inclusive, but sparsely sampled in this range. \item[b] Number of observations including \mbox{re-observations}, but excluding the \corrs{35} SBIDs not imaged. \item[c] SBID for beam-forming observation. \item[d] No appropriate holography for these beam weights.
    }
    \end{tablenotes}
    \end{threeparttable}
\end{table*}

\begin{figure*}[t]
\centering
\includegraphics[width=0.8\linewidth]{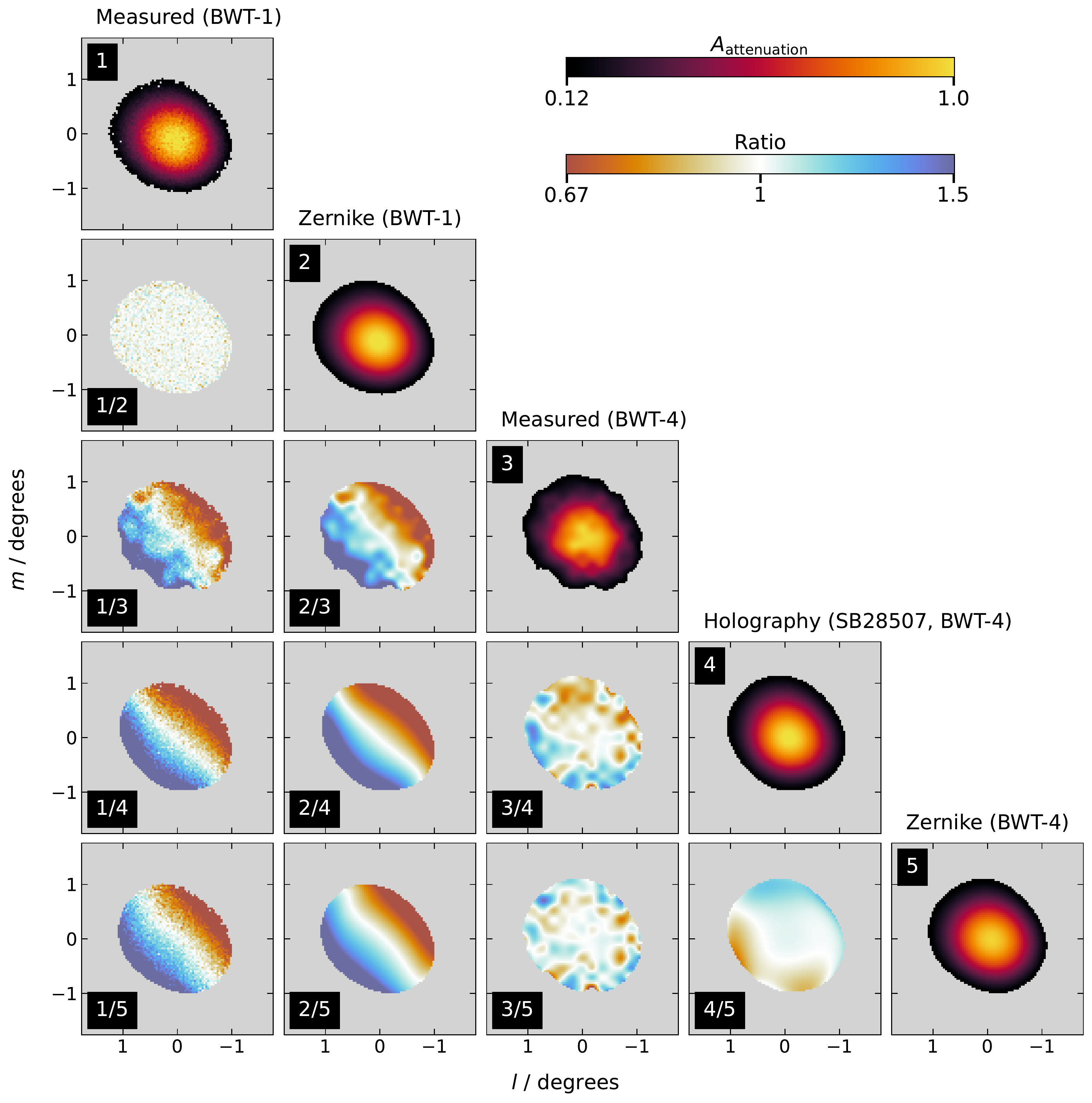}
\caption{\label{fig:icomparison} \corrs{Measured and model $A_\text{attenuation}$ of beam 35 for BWT-1 and BWT-4. Diagonal panels show, from top to bottom, (1) the binned, measured attenuation from BWT-1, (2) the best-fit Zernike polynomial model for BWT-1, (3) the binned, measured attenuation from BWT-4 (after regridding and interpolation), (4) the holographic measurements from SB28507 used for BWT-4, and (5) the best-fits Zernike polynomial model for BWT-4. Plots underneath the diagonals are the ratio difference between the various patterns (as \textit{top model} / \textit{bottom model}). All patterns are clipped at 0.12 to reflect the cutoff used during mosaicking.}}
\end{figure*}

\begin{figure*}
    \centering
        \begin{subfigure}[b]{0.5\linewidth}
    \includegraphics[width=1\linewidth]{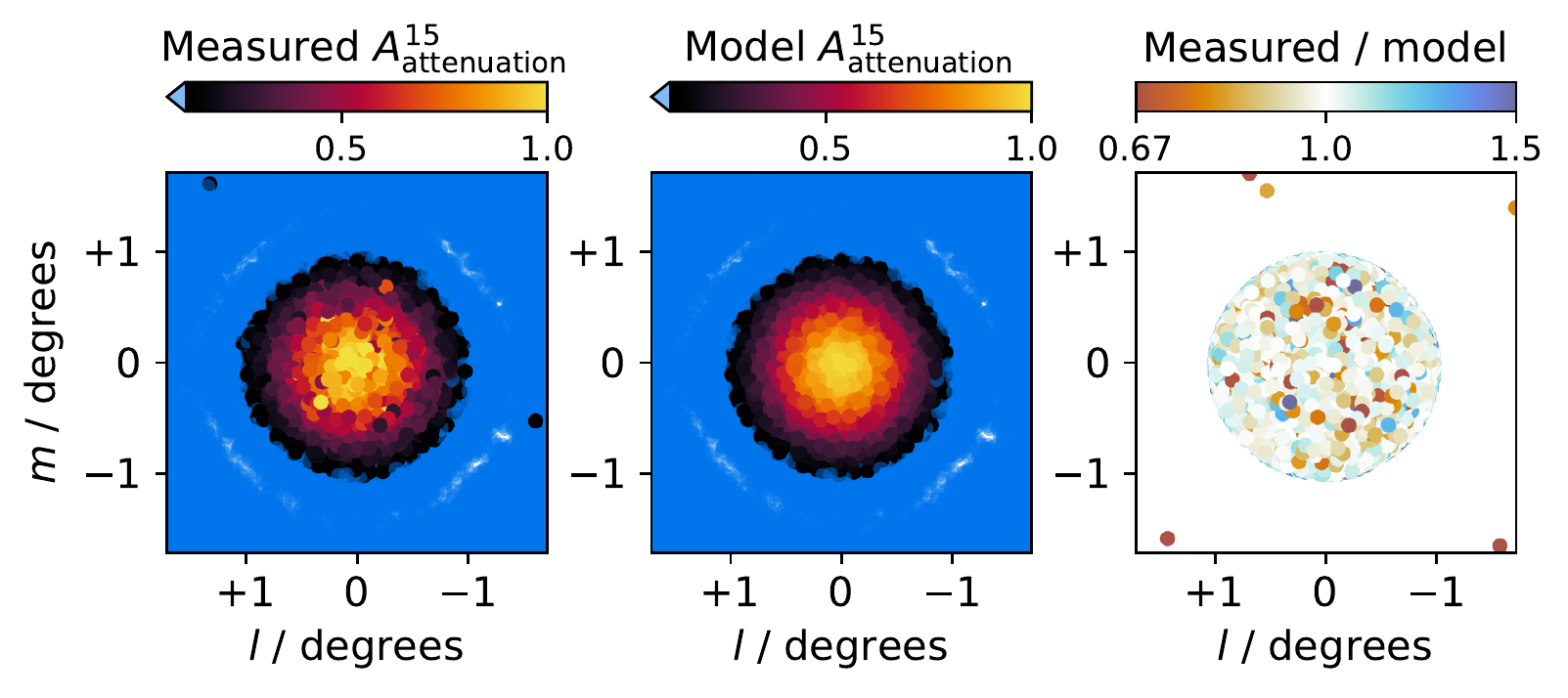}
    \caption{\label{fig:zernike:single:beam15:run1} BWT-1 beam 15.}
    \end{subfigure}%
    \begin{subfigure}[b]{0.5\linewidth}
    \includegraphics[width=1\linewidth]{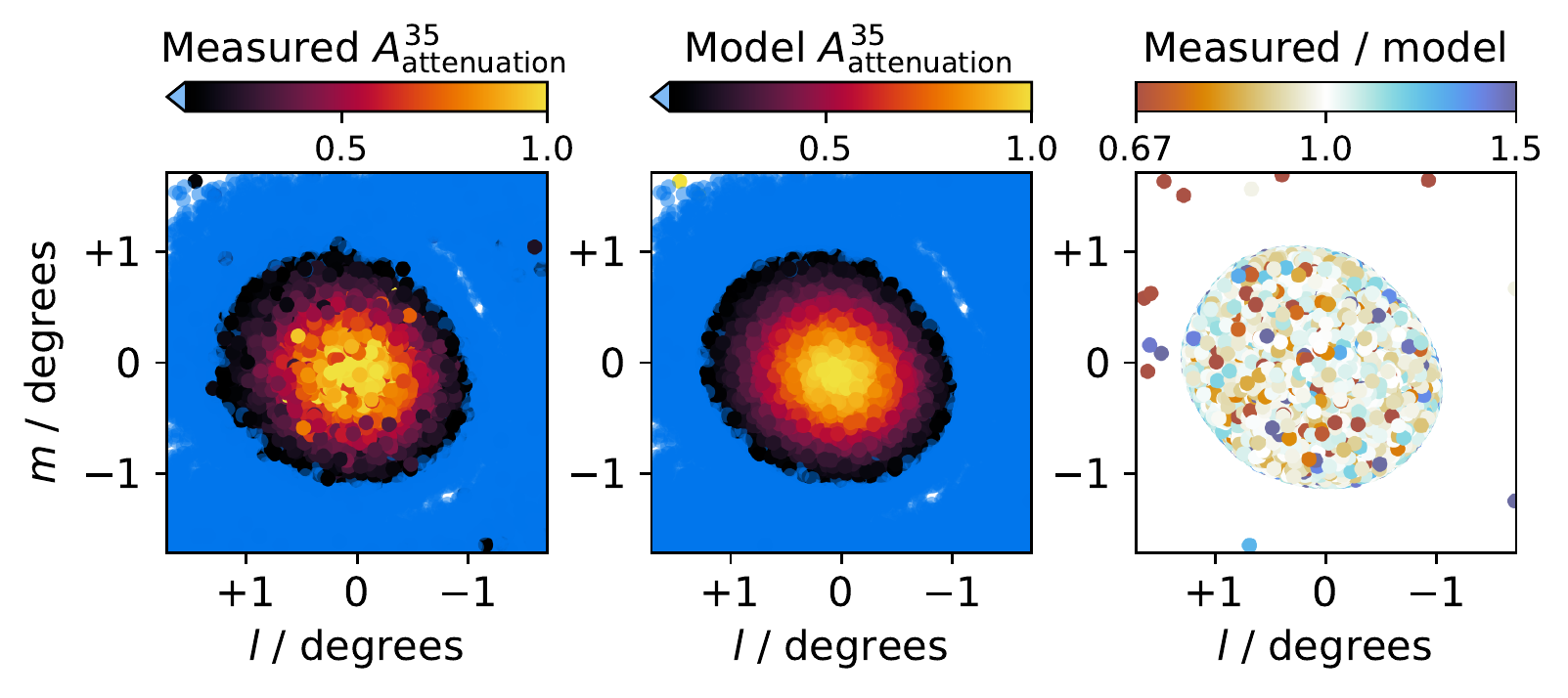}
    \caption{\label{fig:zernike:single:beam35:run1} BWT-1 beam 35.}
    \end{subfigure}\\%

    \begin{subfigure}[b]{0.5\linewidth}
    \includegraphics[width=1\linewidth]{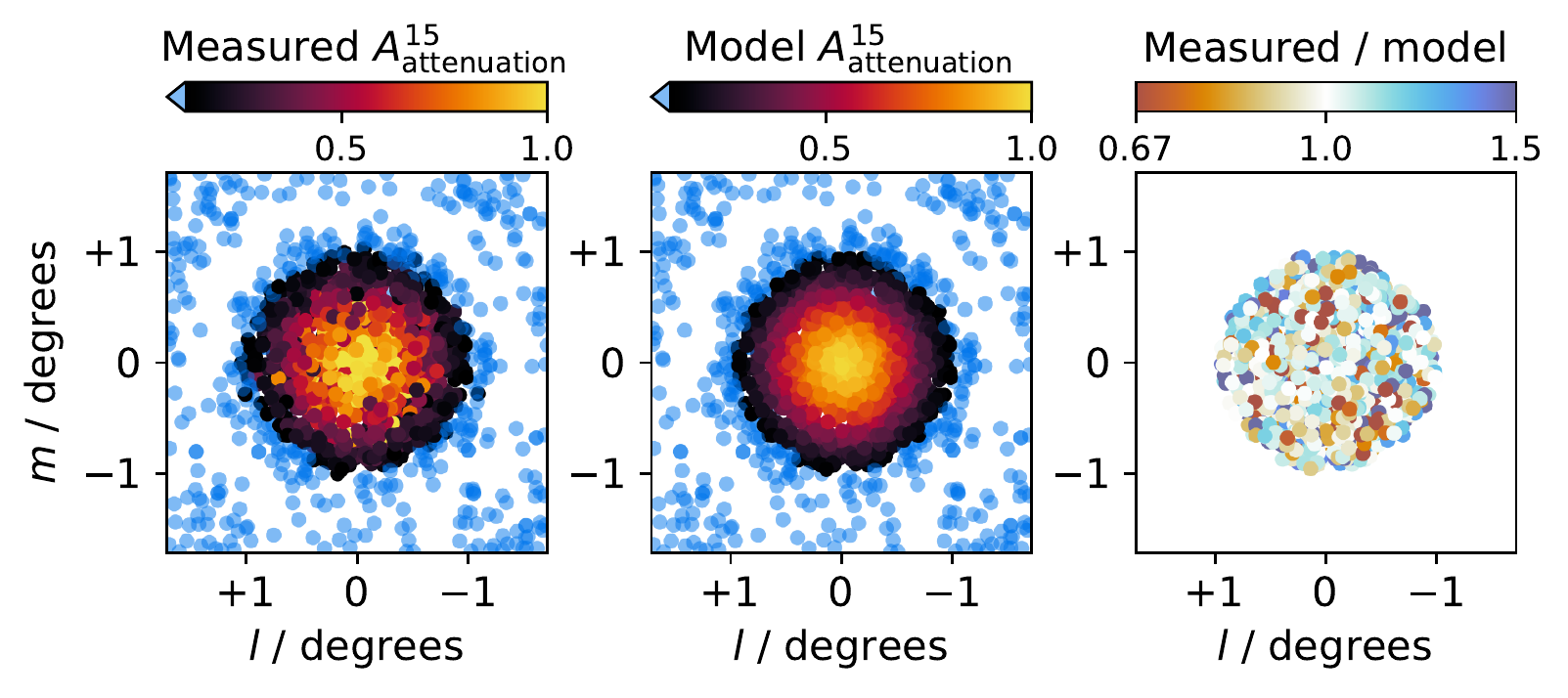}
    \caption{\label{fig:zernike:single:beam15:run2} BWT-2 beam 15.}
    \end{subfigure}%
    \begin{subfigure}[b]{0.5\linewidth}
    \includegraphics[width=1\linewidth]{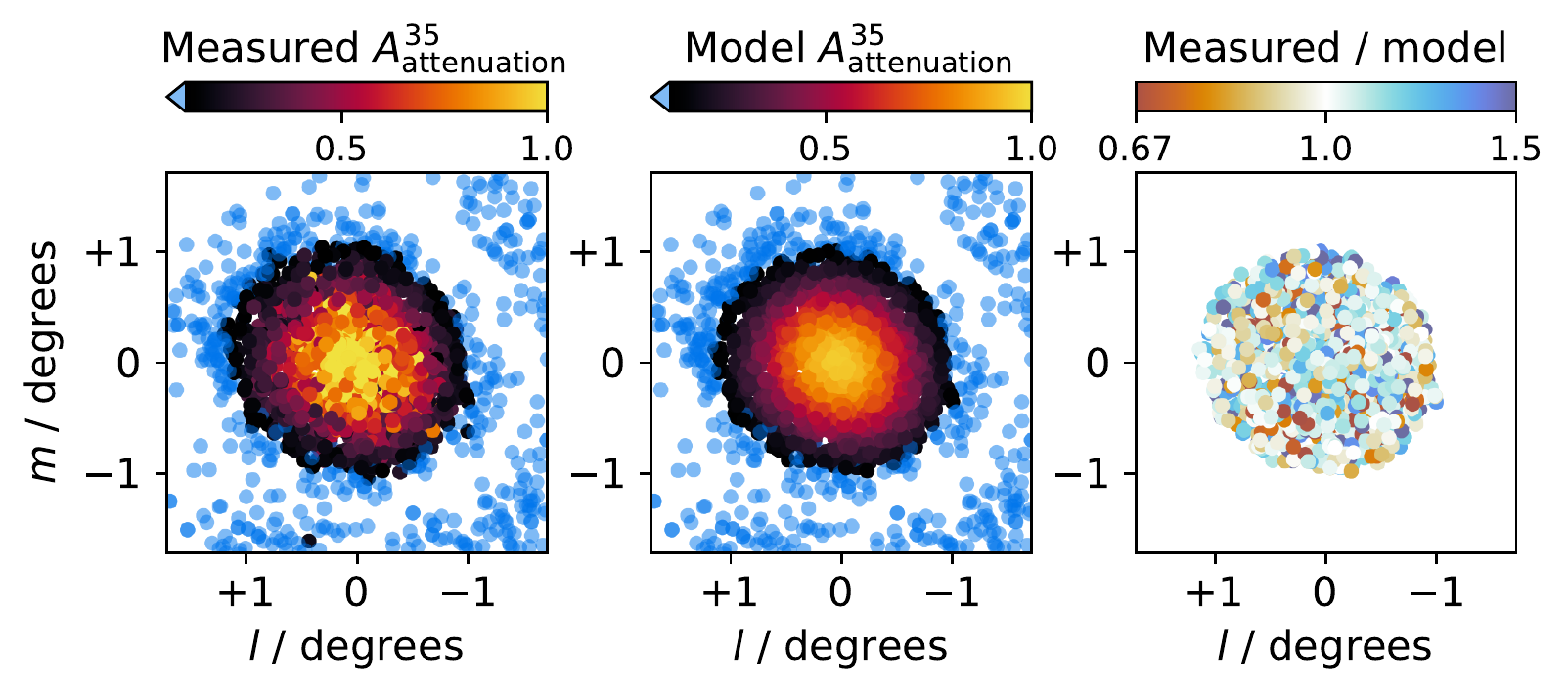}
    \caption{\label{fig:zernike:single:beam35:run2} BWT-2 beam 35.}
    \end{subfigure}\\%

        \begin{subfigure}[b]{0.5\linewidth}
    \includegraphics[width=1\linewidth]{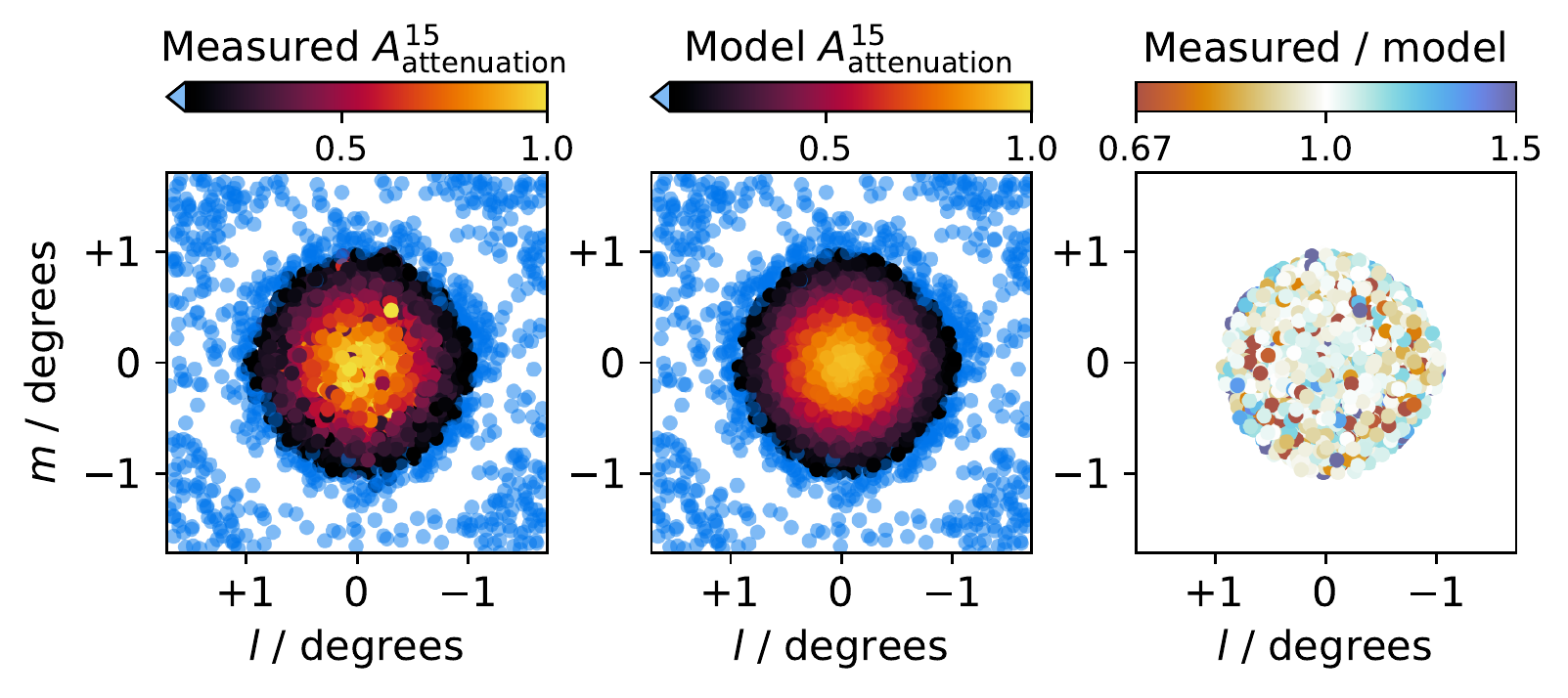}
    \caption{\label{fig:zernike:single:beam15:run3} BWT-3 beam 15.}
    \end{subfigure}%
    \begin{subfigure}[b]{0.5\linewidth}
    \includegraphics[width=1\linewidth]{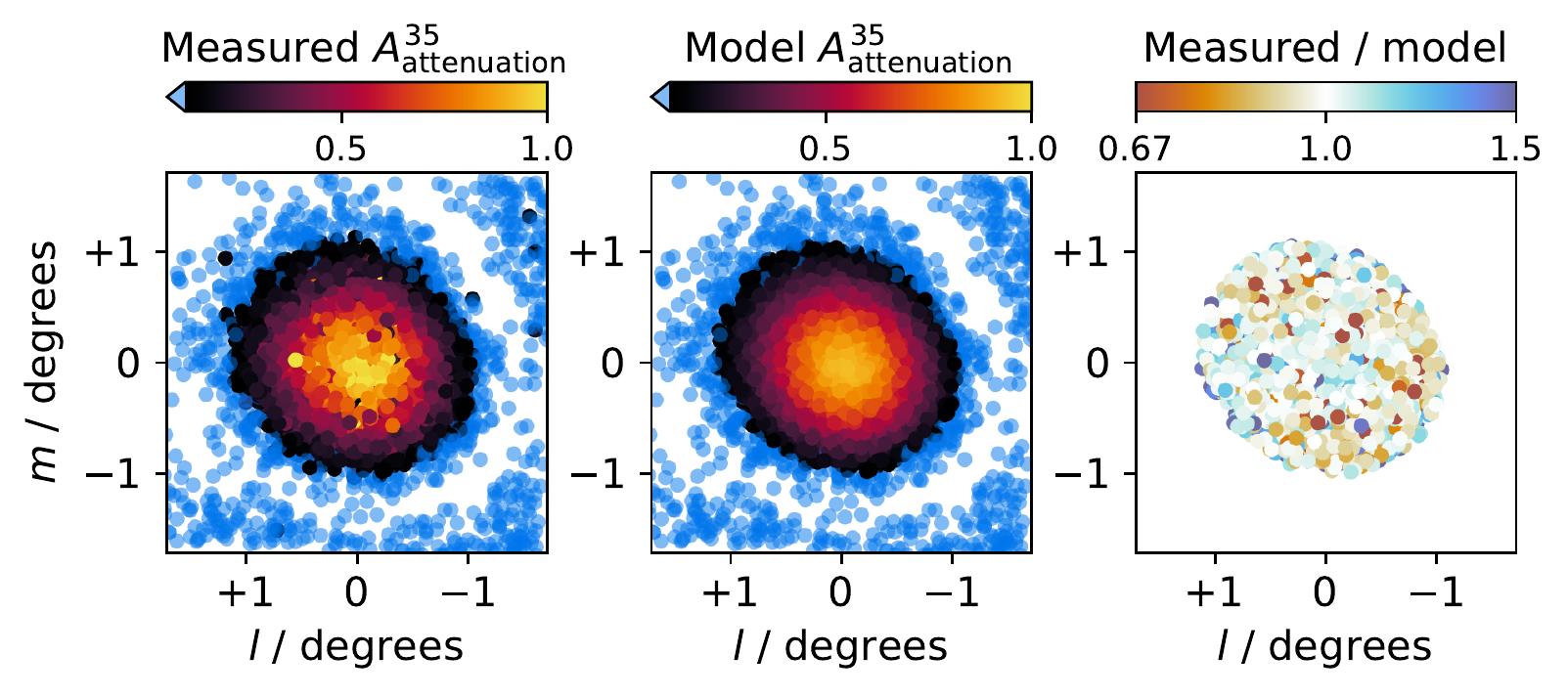}
    \caption{\label{fig:zernike:single:beam35:run3} BWT-3 beam 35.}
    \end{subfigure}\\%
        \begin{subfigure}[b]{0.5\linewidth}
    \includegraphics[width=1\linewidth]{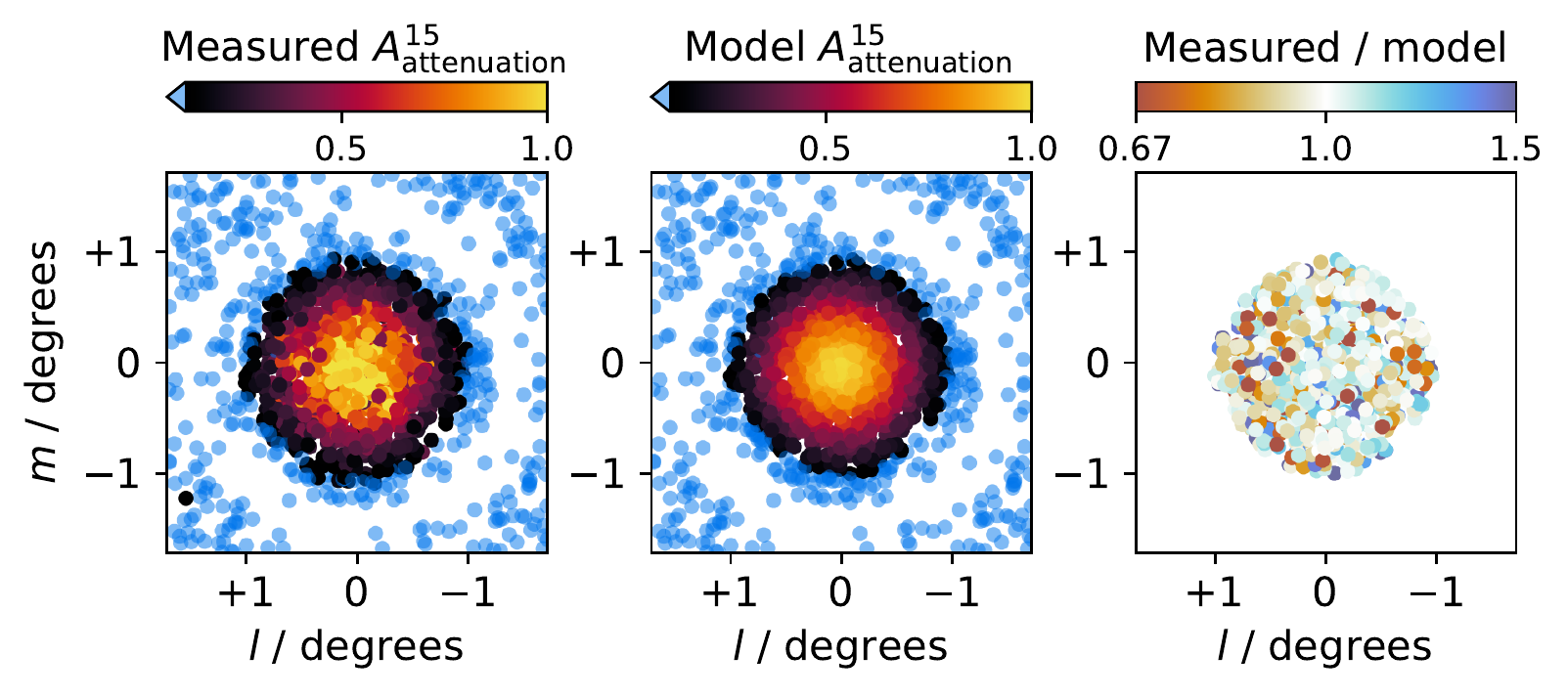}
    \caption{\label{fig:zernike:single:beam15:run5} BWT-5 beam 15.}
    \end{subfigure}%
    \begin{subfigure}[b]{0.5\linewidth}
    \includegraphics[width=1\linewidth]{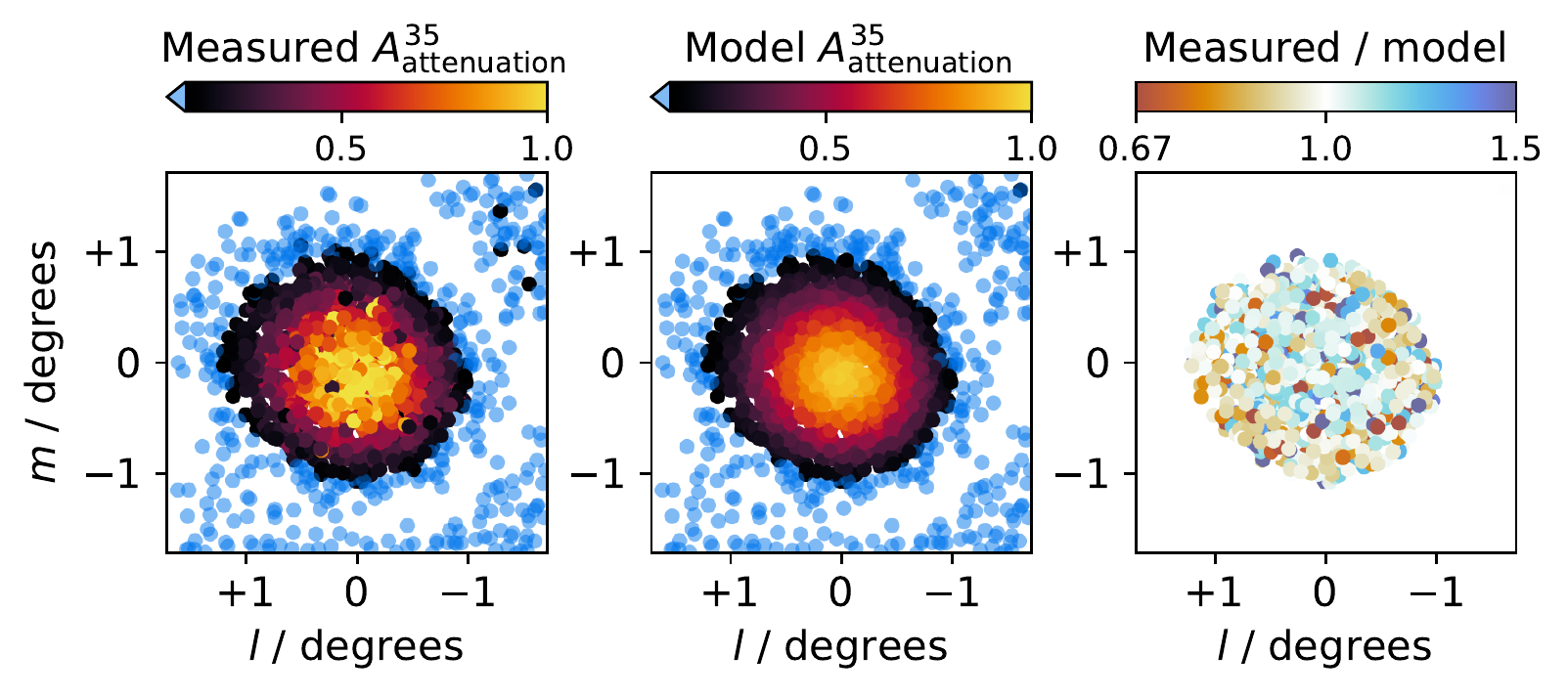}
    \caption{\label{fig:zernike:single:beam35:run5} BWT-5 beam 35.}
    \end{subfigure}\\%
    \caption{\label{fig:zernike:single} Central beam 15 [\subref{fig:zernike:single:beam15:run1}, \subref{fig:zernike:single:beam15:run2}, \subref{fig:zernike:single:beam15:run3}, \subref{fig:zernike:single:beam15:run5}] and corner beam 35 [\subref{fig:zernike:single:beam35:run1}, \subref{fig:zernike:single:beam35:run2}, \subref{fig:zernike:single:beam35:run3}, \subref{fig:zernike:single:beam35:run5}] Stokes I $A^b_\text{attenuation}$ modelling for BWT-1 (\textit{top row}), BWT-2 (\textit{second row}), BWT-3 (\textit{third row}), and BWT-5 (\textit{bottom row}). \textit{Left panels.} Measured attenuation pattern showing individual sources. \emph{Centre.} The fitted Zernike model at the location of the individual sources. \emph{Right.} Ratio of the measured and model attenuation patterns representing residuals. The colourmap for $A^b_\text{attenuation}$ is clipped at 0.12, corresponding to the blue sources.}
\end{figure*}

\begin{figure*}
    \centering
    \includegraphics[width=\linewidth]{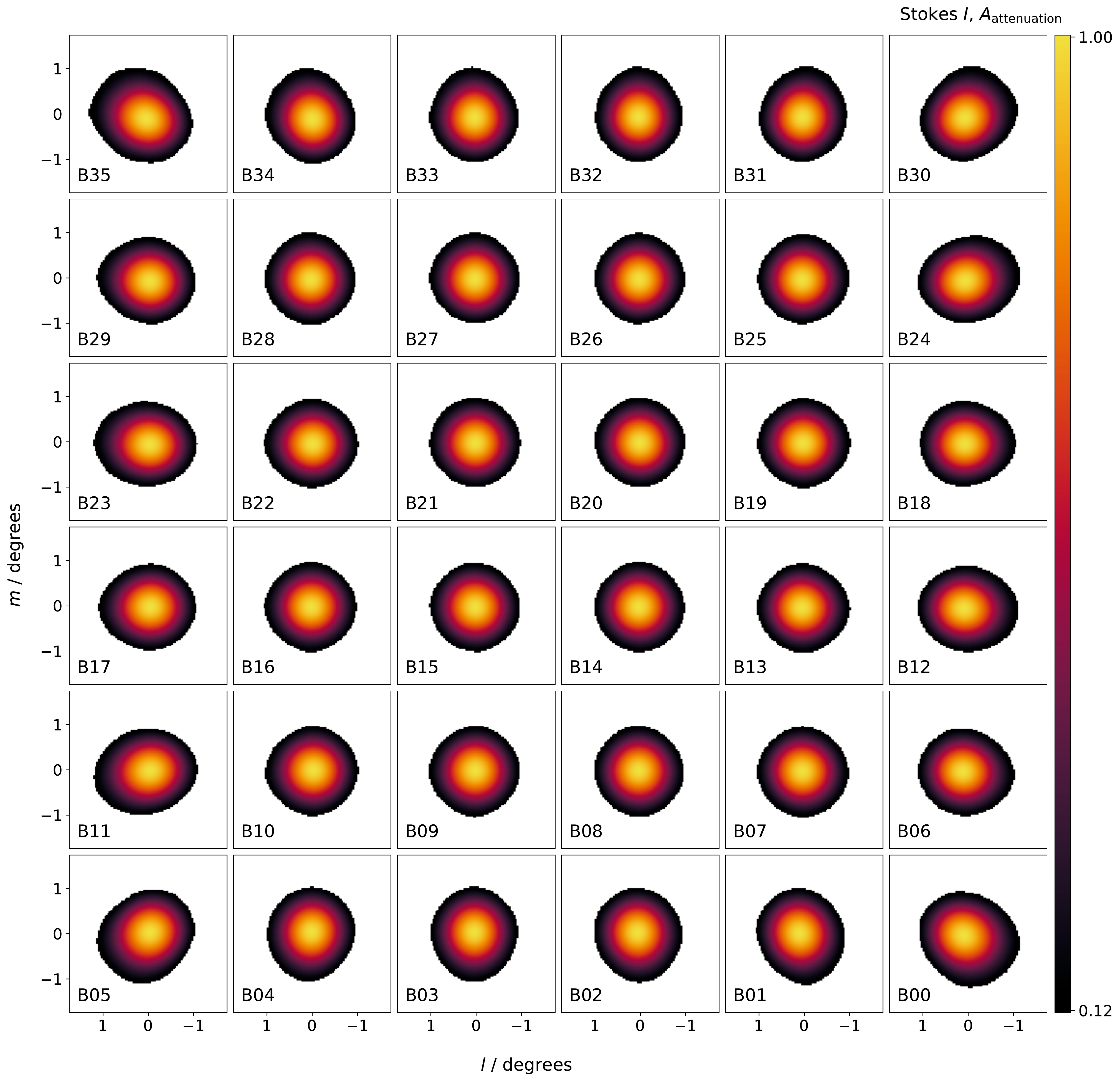}
    \caption{\label{fig:fullstokesi} \corrs{Stokes I model beams for BWT-1 for all beams in the footprint. Beams are clipped at 12\% attenuation and are arranged to match the footprint (Figure~\ref{fig:footprint}).}}
\end{figure*}

\begin{figure*}[t]
    \begin{subfigure}[b]{0.5\linewidth}
    \includegraphics[width=1\linewidth]{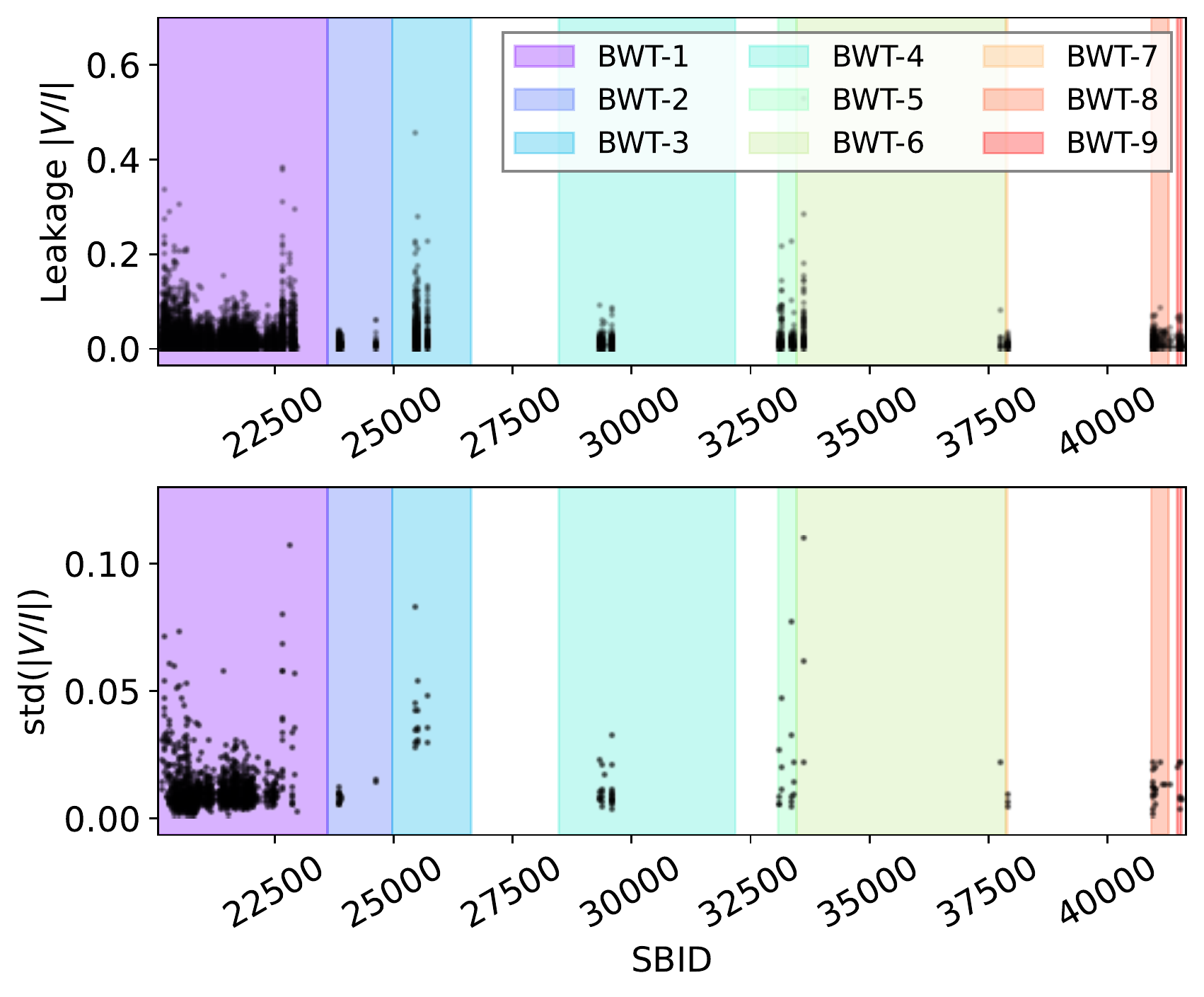}
    \caption{\label{fig:v_leakage_sbid:pre} Pre-correction, beam 0.}
    \end{subfigure}%
    \begin{subfigure}[b]{0.5\linewidth}
    \includegraphics[width=1\linewidth]{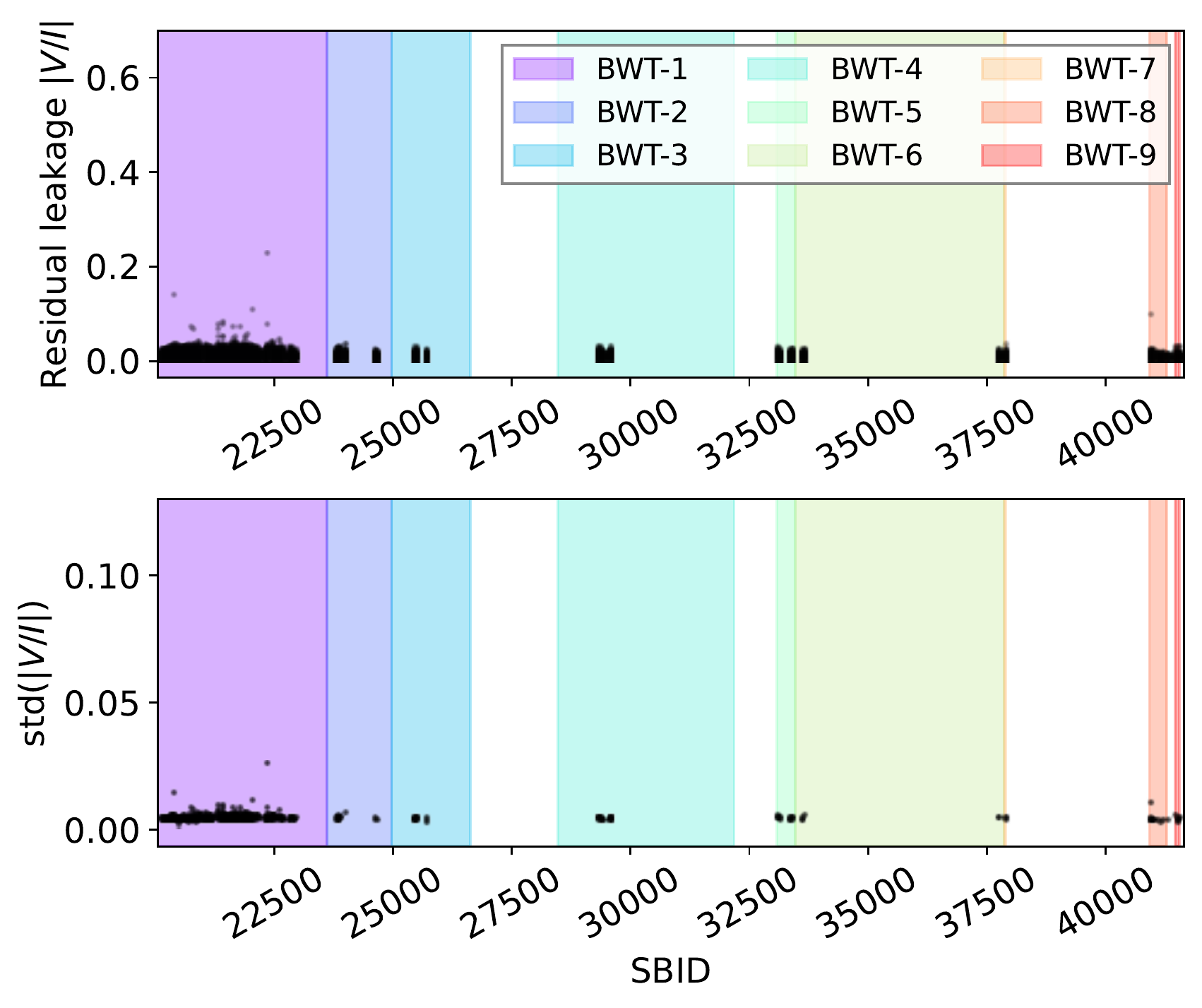}
    \caption{\label{fig:v_leakage_sbid:post} Post-correction, full tile.}
    \end{subfigure}%
    \caption{\label{fig:v_leakage_sbid} $V$/$I$ across all SBIDs for beam 0 prior to mosaicking and leakage correction \subref{fig:v_leakage_sbid:pre} and for the full tiles after mosaicking and applying leakage correction \subref{fig:v_leakage_sbid:post}. \textit{Top panels.} $|V/I|$ for all sources with $S_I > 100\sigma_{\text{rms},I}$. \textit{Bottom panels.} The standard deviation of $V/I$ for each observation. Each BWT (advancing from left to right; see Table~\ref{tab:beamformer}) is shaded a different colour.}
\end{figure*}

\begin{figure*}[t]
\includegraphics[width=0.8\linewidth]{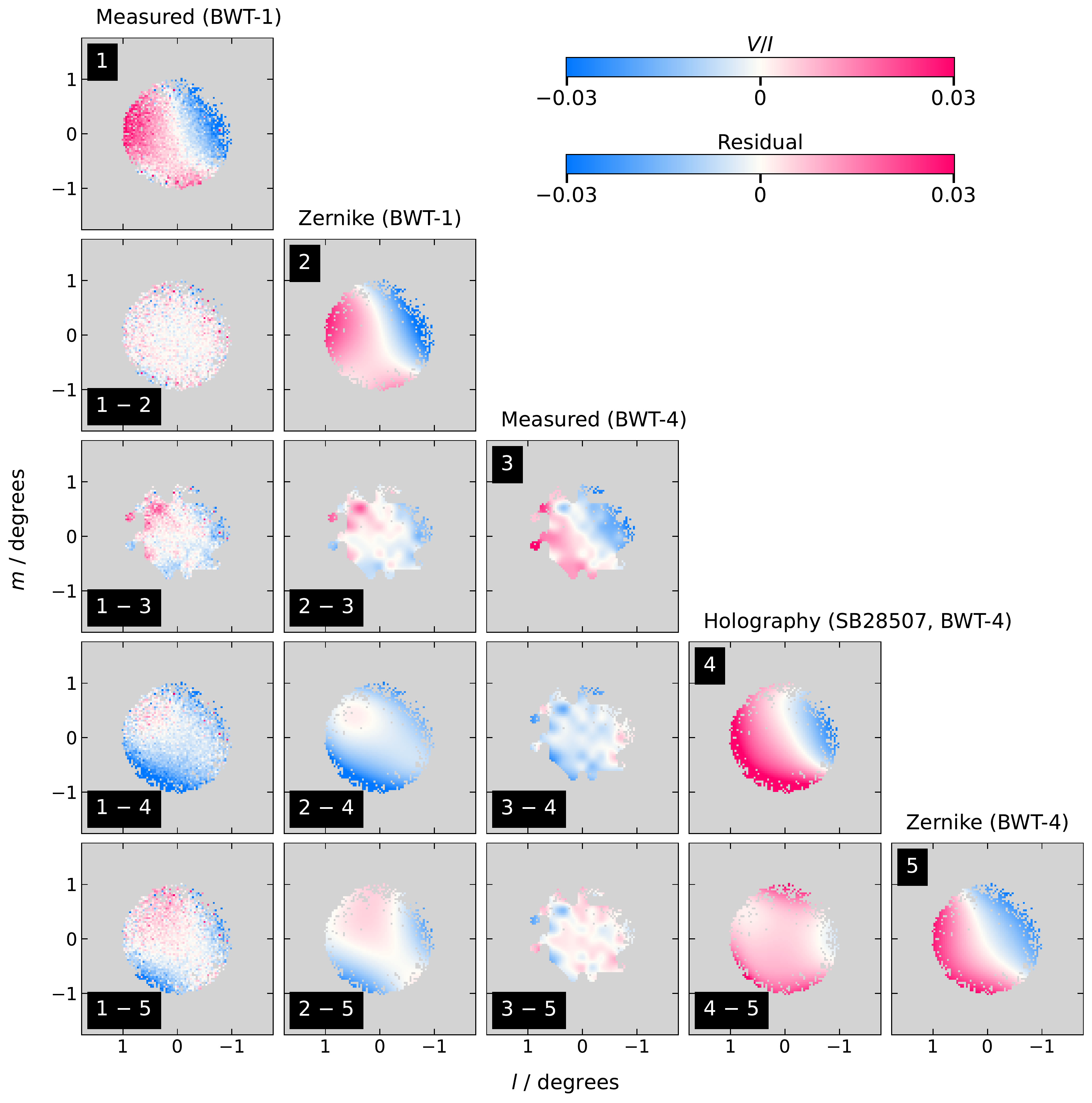}
\caption{\label{fig:vcomparison}\corrs{A comparison of the measured and modelled leakage of Stokes I into V ($V/I$) for beam 35 in BWT-1 and BWT-4. Diagonal panels show, from top to bottom, (1) the binned, measured leakage from BWT-1, (2) the fitted Zernike polynomial for BWT-1, (3) the binned, measured leakage from BWT-4, (4) the leakage model derived from holographic measurements, and (5) the fitted Zernike polynomial for BWT-4. Plots underneath the diagonals are the residual differences (as \emph{top} $-$ \emph{bottom}) and all patterns are clipped with reference to (1).}}
\end{figure*}

\begin{figure*}
    \centering
    \begin{subfigure}[b]{0.5\linewidth}
    \includegraphics[width=1\linewidth]{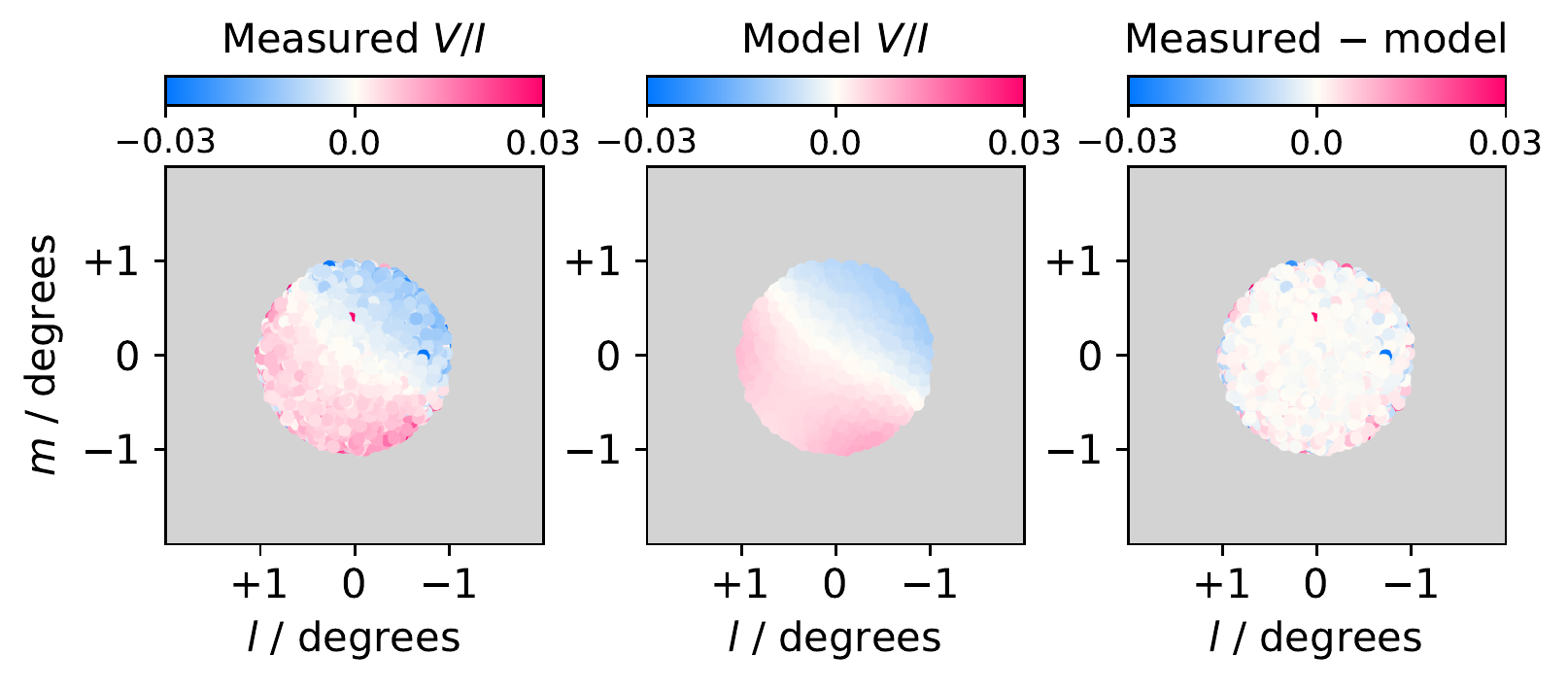}
    \caption{\label{fig:leakage:run1:beam15} Beam 15 Stokes \textit{V} leakage, BWT-1.}
    \end{subfigure}%
    \begin{subfigure}[b]{0.5\linewidth}
    \includegraphics[width=1\linewidth]{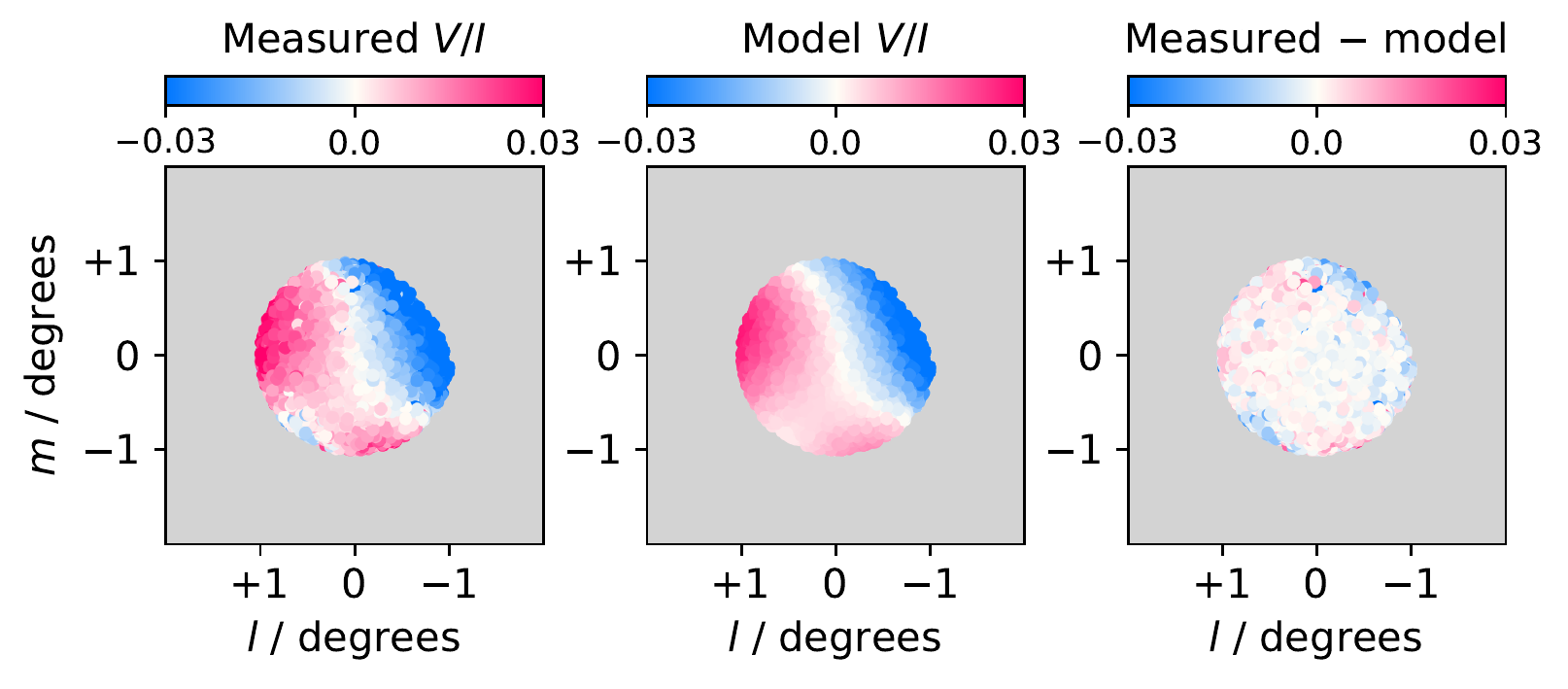}
    \caption{\label{fig:leakage:run1:beam35} Beam 35 Stokes \textit{V} leakage, BWT-1.}
    \end{subfigure}\\%
        \begin{subfigure}[b]{0.5\linewidth}
    \includegraphics[width=1\linewidth]{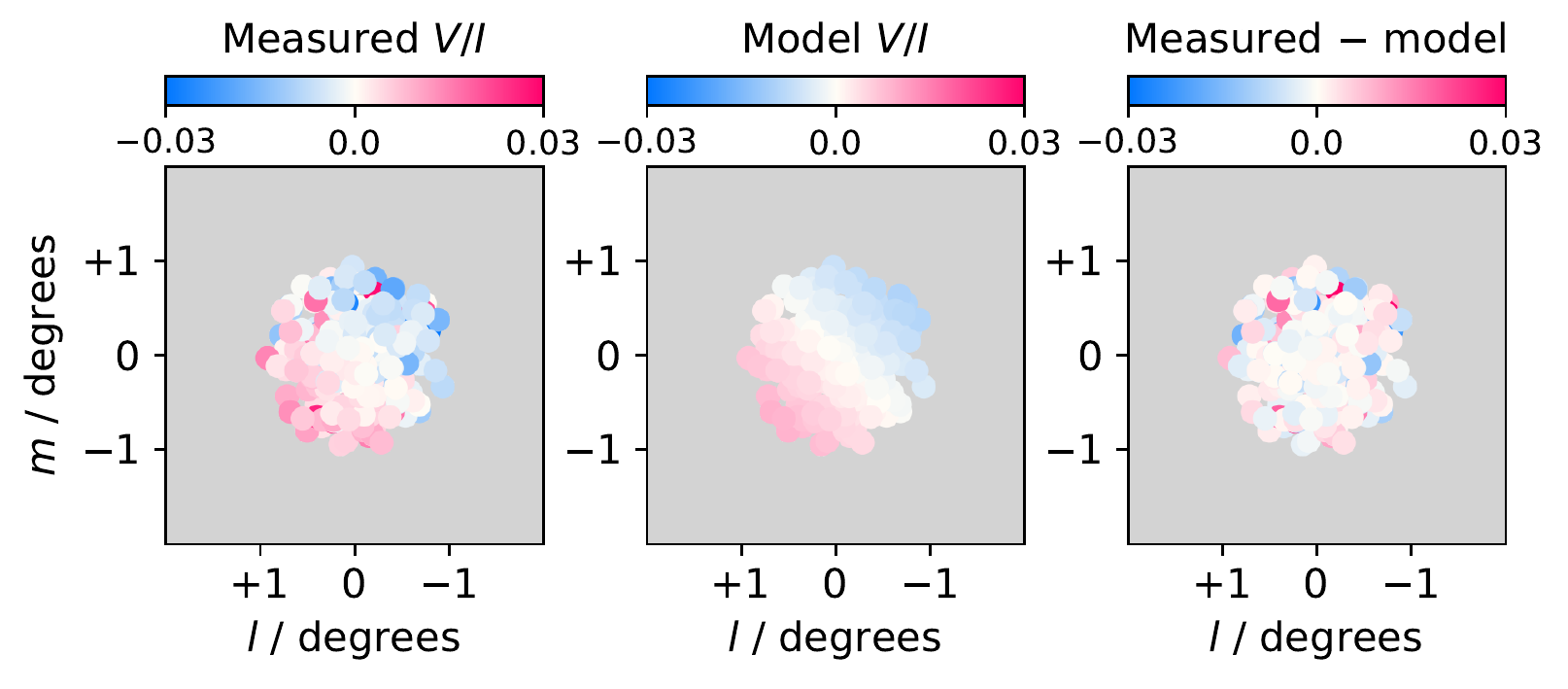}
    \caption{\label{fig:leakage:run2:beam15} Beam 15 Stokes \textit{V} leakage, BWT-2.}
    \end{subfigure}%
    \begin{subfigure}[b]{0.5\linewidth}
    \includegraphics[width=1\linewidth]{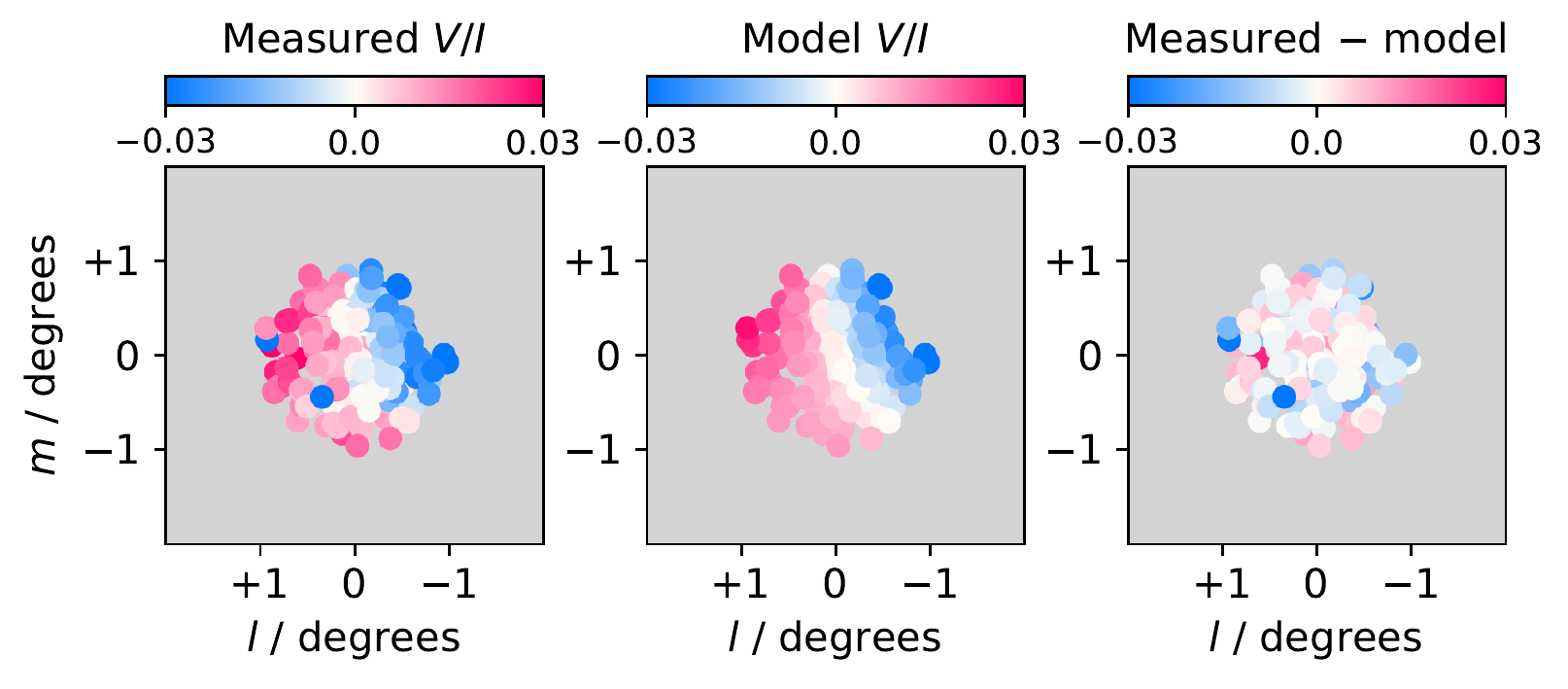}
    \caption{\label{fig:leakage:run2:beam35} Beam 35 Stokes \textit{V} leakage, BWT-2.}
    \end{subfigure}\\%
        \begin{subfigure}[b]{0.5\linewidth}
    \includegraphics[width=1\linewidth]{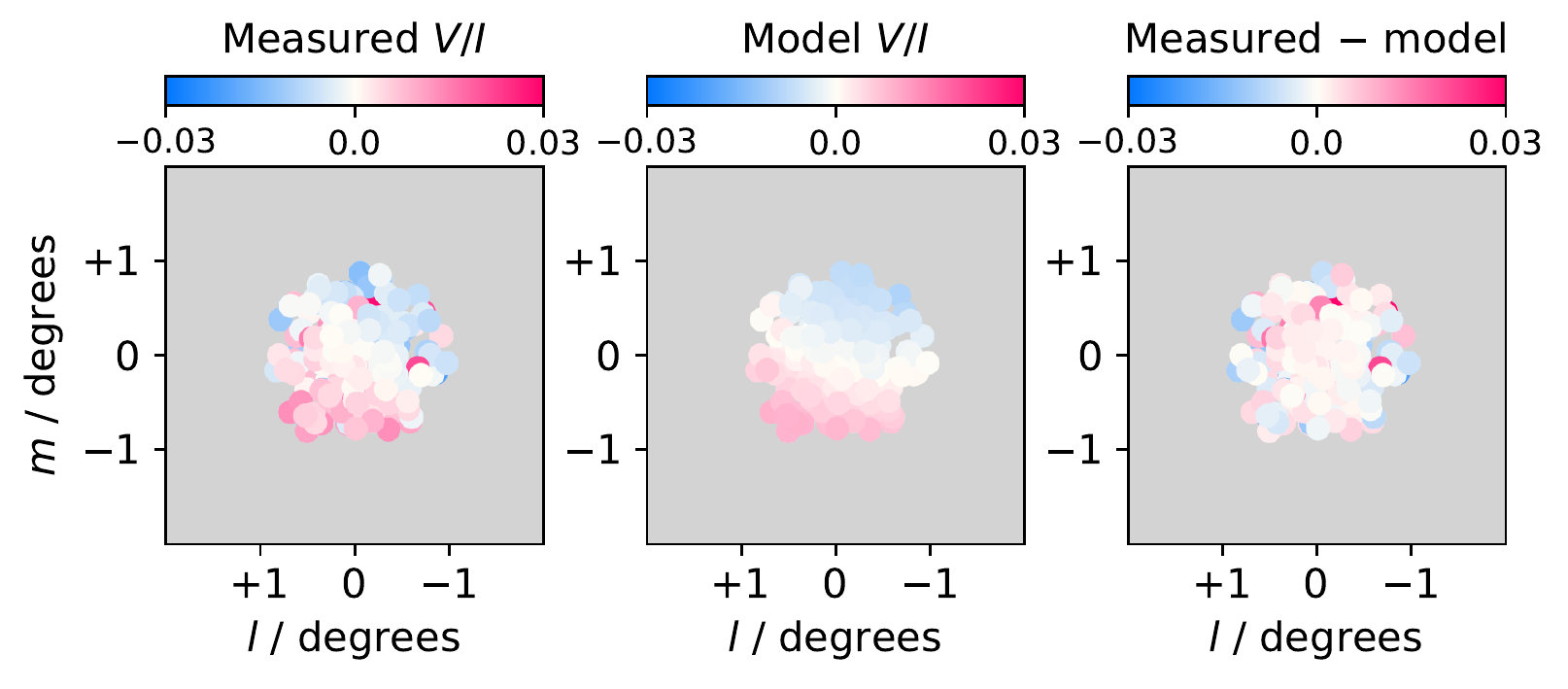}
    \caption{\label{fig:leakage:run3:beam15} Beam 15 Stokes \textit{V} leakage, BWT-3.}
    \end{subfigure}%
    \begin{subfigure}[b]{0.5\linewidth}
    \includegraphics[width=1\linewidth]{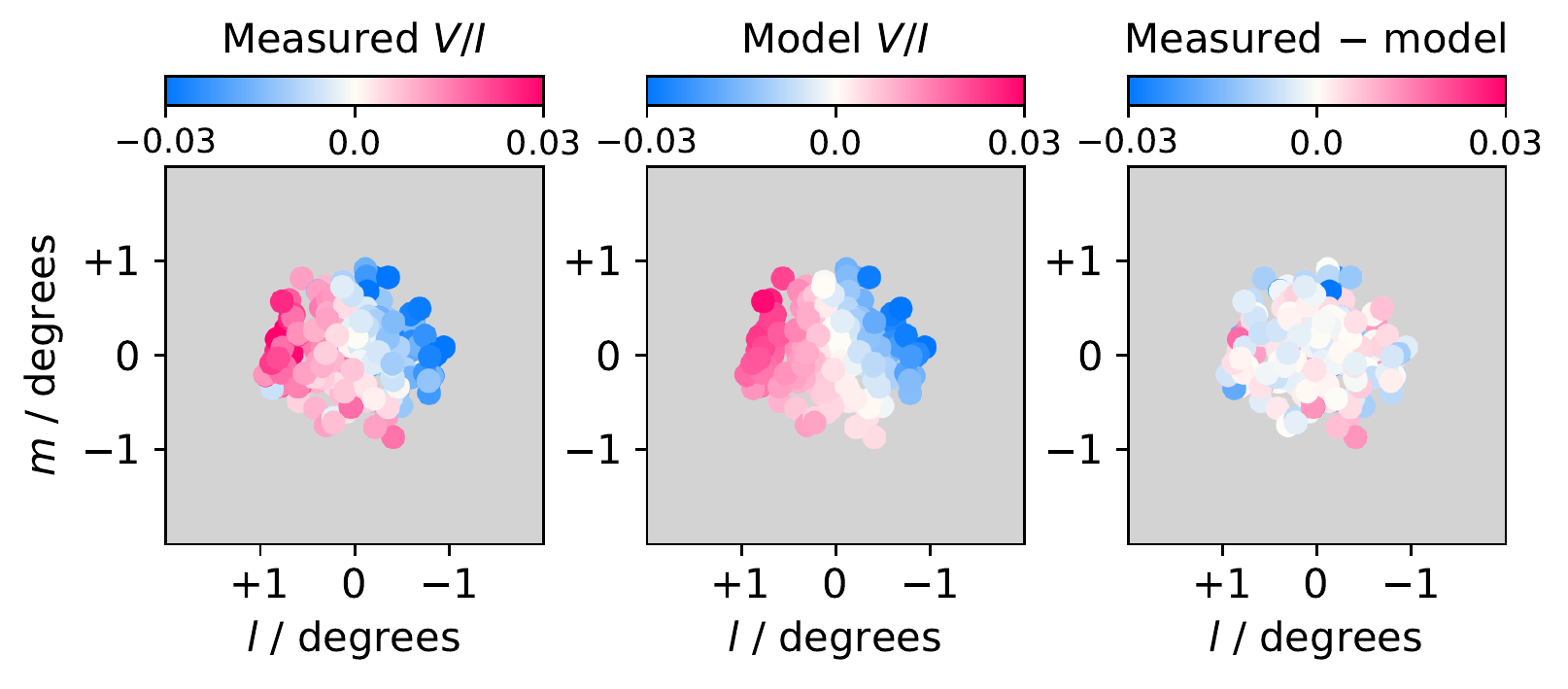}
    \caption{\label{fig:leakage:run3:beam35} Beam 35 Stokes \textit{V} leakage, BWT-3.}
    \end{subfigure}\\%
    \begin{subfigure}[b]{0.5\linewidth}
    \includegraphics[width=1\linewidth]{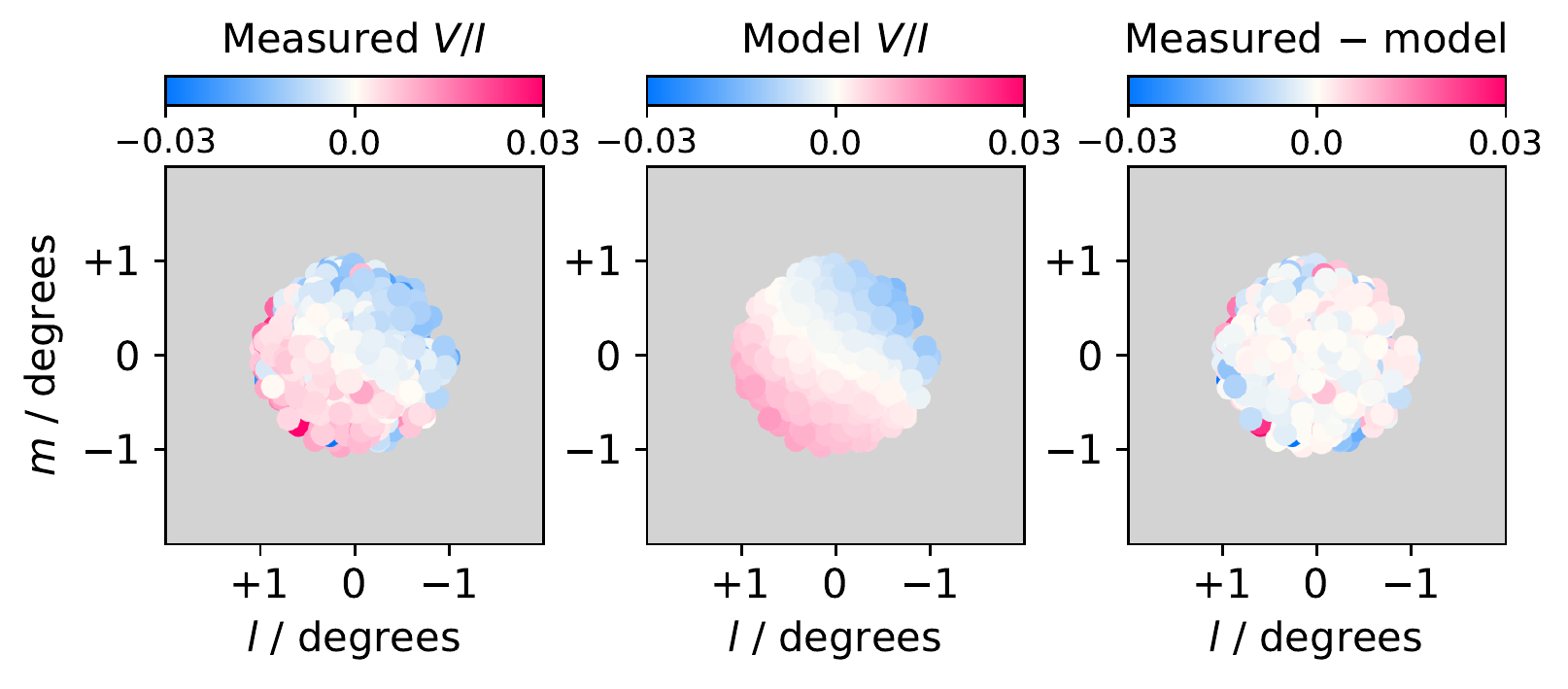}
    \caption{\label{fig:leakage:run5:beam15} Beam 15 Stokes \textit{V} leakage, SB21610--SB21710.}
    \end{subfigure}%
    \begin{subfigure}[b]{0.5\linewidth}
    \includegraphics[width=1\linewidth]{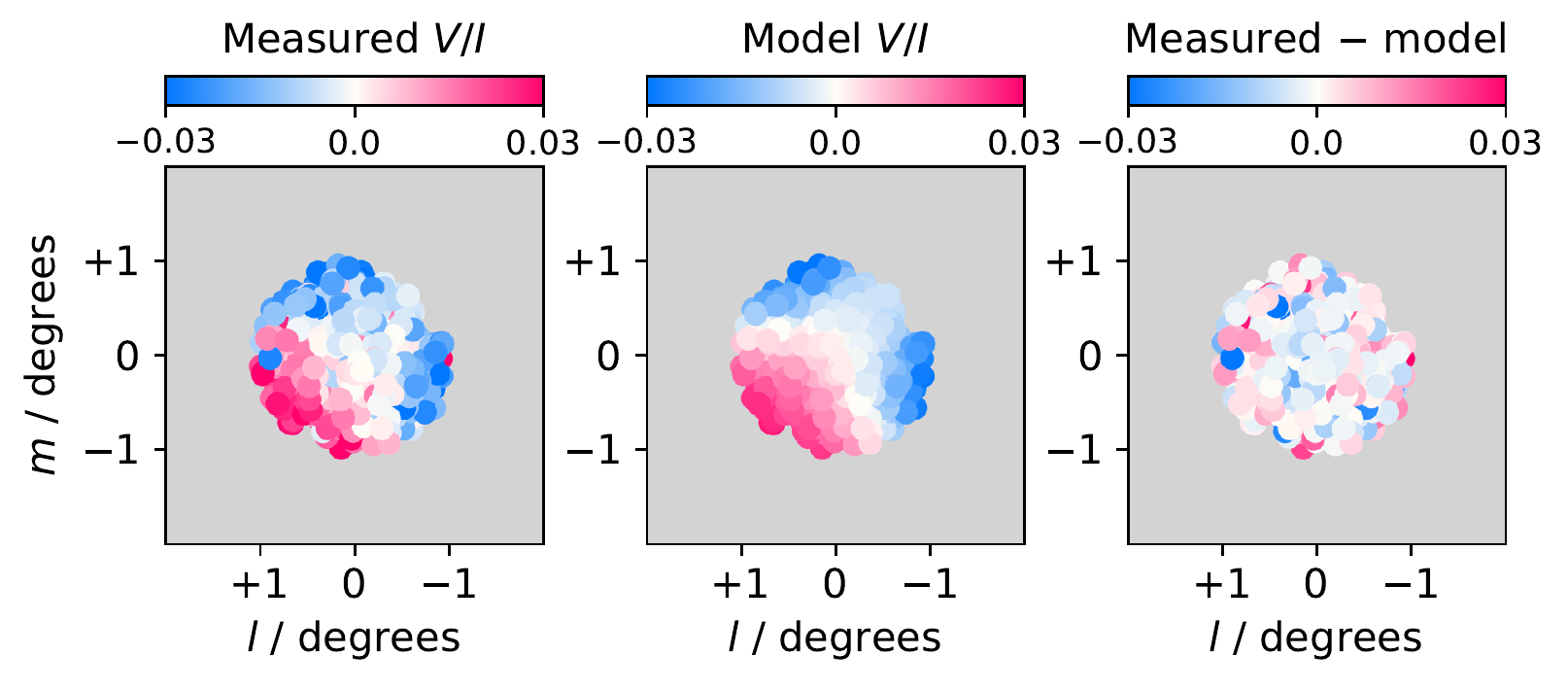}
    \caption{\label{fig:leakage:run5:beam35} Beam 35 Stokes \textit{V} leakage, SB21610--SB21710.}
    \end{subfigure}\\%
    \caption{\label{fig:leakage:run1} Central beam 15 [\subref{fig:leakage:run1:beam15}, \subref{fig:leakage:run2:beam15}, \subref{fig:leakage:run3:beam15}, \subref{fig:leakage:run5:beam15}] and corner beam 35 [\subref{fig:leakage:run1:beam35}, \subref{fig:leakage:run2:beam35}, \subref{fig:leakage:run3:beam35}, \subref{fig:leakage:run5:beam35}] Stokes V leakage modelling results for BWT-1 (\textit{top row}), BWT-2 (\textit{second row}), BWT-3 (\textit{third row}), and the subset SB21616--SB21710 from BWT-1 (\textit{bottom row}). \textit{Left panels.} Measured $V/I$ leakage pattern with individual sources. \emph{Centre.} The fitted Zernike model at the location of the individual sources. \emph{Right.} Residual leakage patterns. Only sources within 1\,deg of the beam centre are included. \corrs{Note \subref{fig:leakage:run1:beam35} is similar to the top three panels of Figure~\ref{fig:vcomparison}.}}
\end{figure*}

\begin{figure*}
    \centering
    \includegraphics[width=\linewidth]{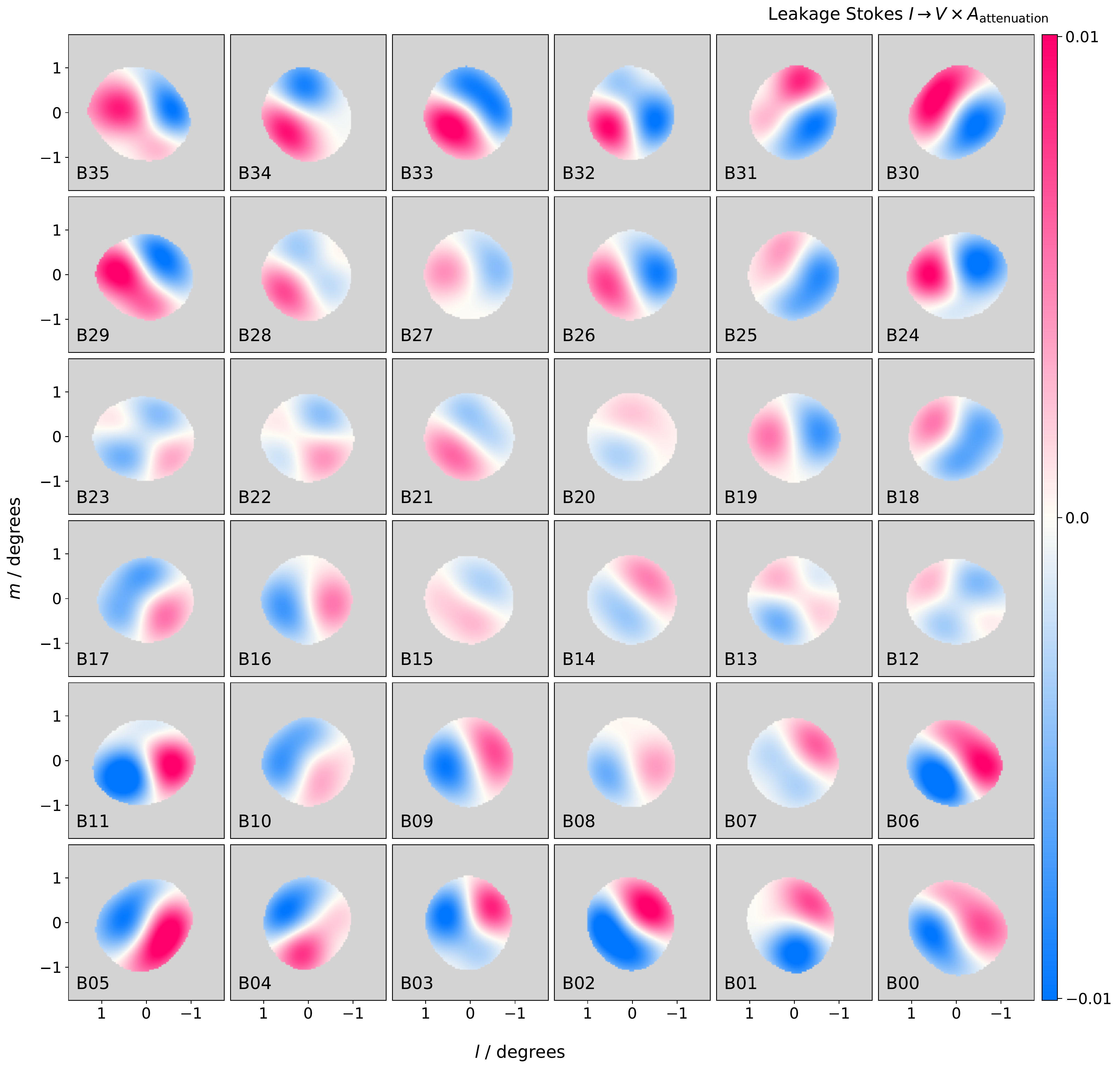}
    \caption{\label{fig:fullstokesv} \corrs{Stokes V model beams ($V/I \times A_\text{attenuation}$) for BWT-1 for all beams in the footprint. Beams are clipped at 12\% Stokes I attenuation and are arranged to match the footprint (Figure~\ref{fig:footprint}).}}
\end{figure*}

\begin{figure*}
\includegraphics[width=1\linewidth]{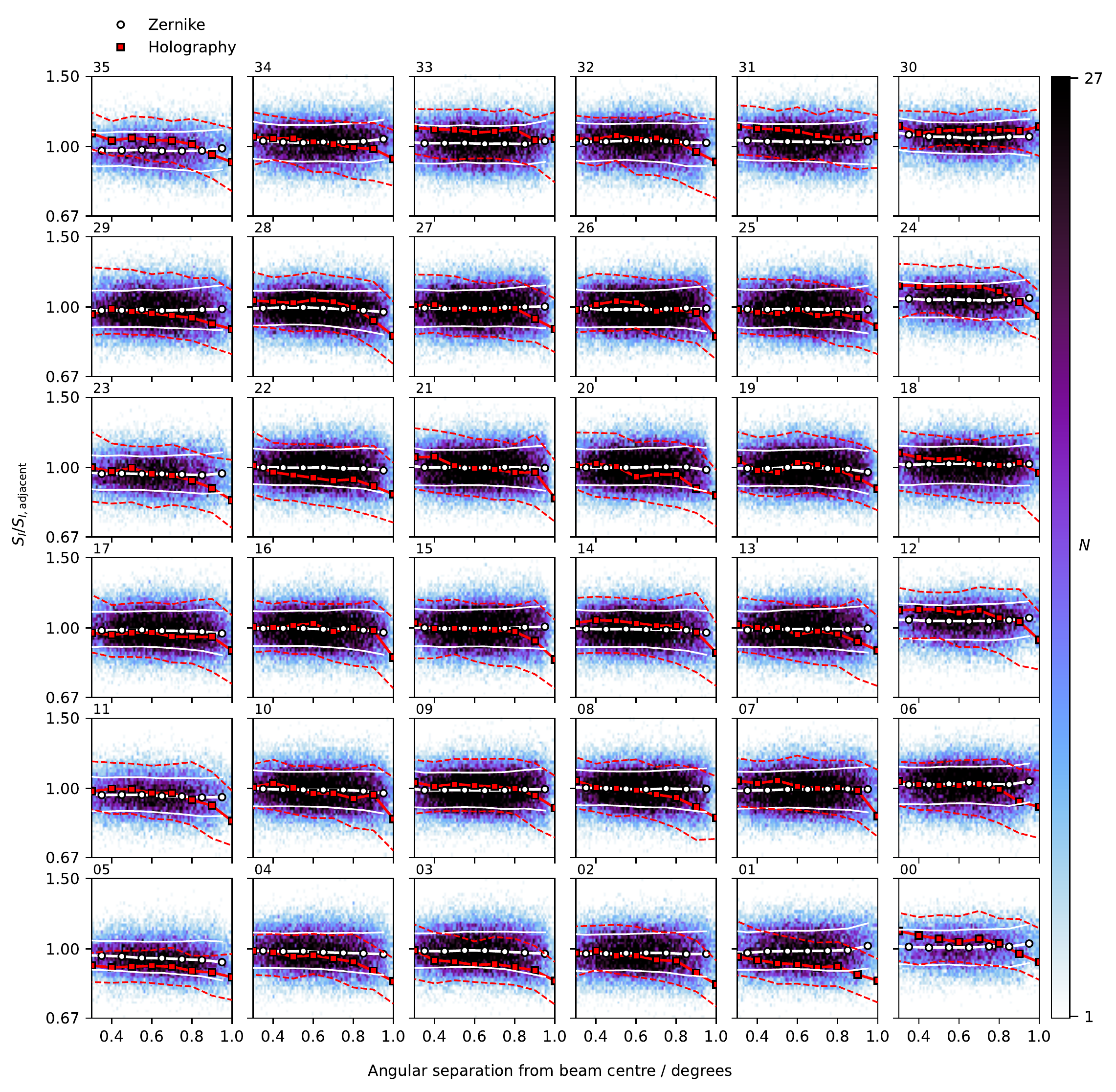}
\caption{\label{fig:ibeamoverlap}\corrs{2-D histogram of Stokes I flux density ratios of sources detected across adjacent beams as a function distance from the reference beam centre. Binned median flux density ratios for the Zernike models (white circles) and holography models (red squares) are shown, along with $16^\text{th}$ and $84^\text{th}$ percentiles for the corresponding bins. The per-beam plots are arranged to match the footprint (Figure~\ref{fig:footprint}).} \CORRS{The colour scale is linear in the reported range.}}
\end{figure*}

\begin{figure*}
\includegraphics[width=1\linewidth]{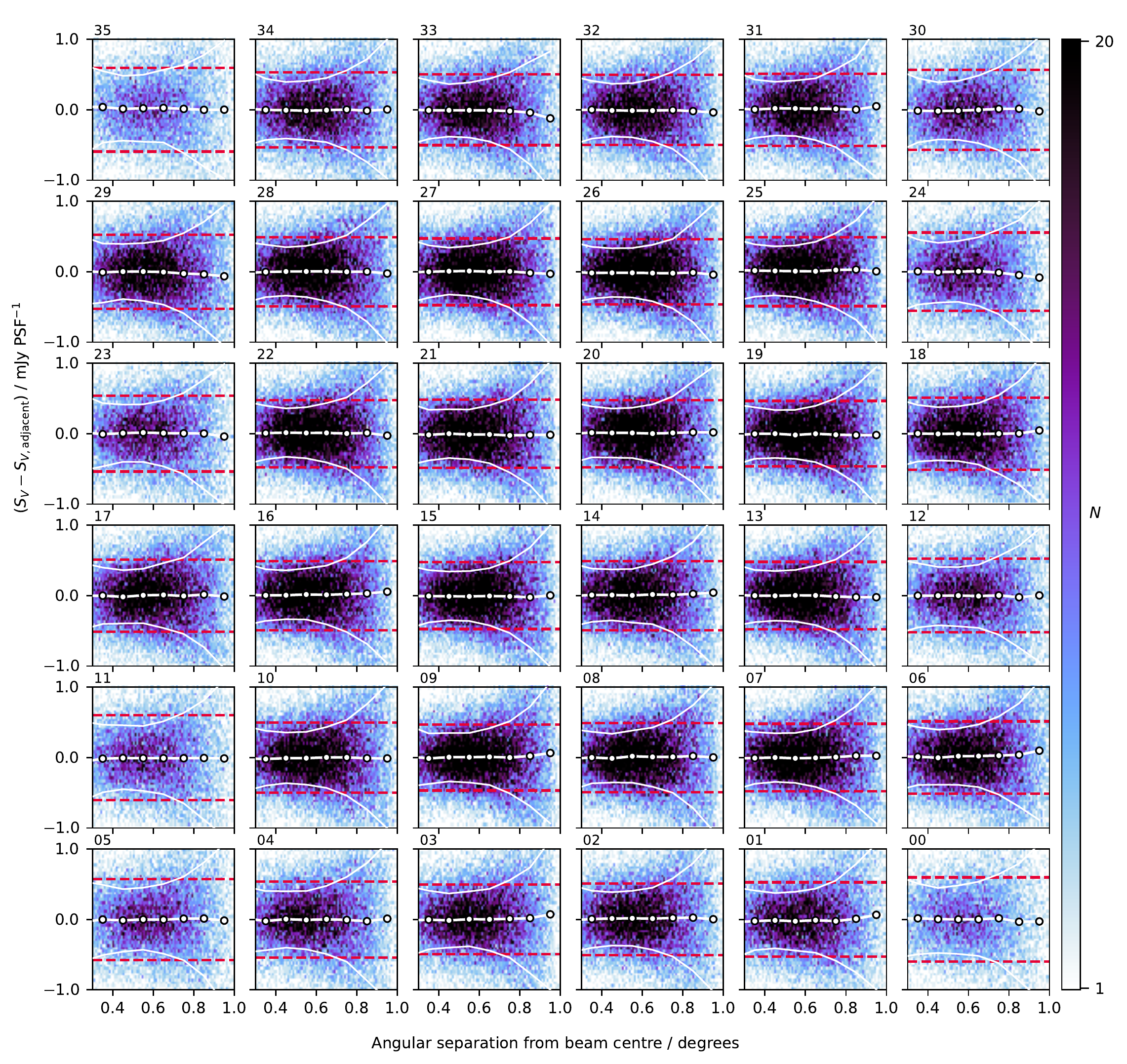}
\caption{\label{fig:vbeamoverlap1} \corrs{Difference in Stokes V measurements between adjacent for unpolarized ($S_V < 3\sigma_{\text{rms},V}$ in the reference beam) Stokes I sources. The white circles are medians in bins, and the white, solid lines are the corresponding $16^\text{th}$ and $84^\text{th}$ percentiles. The dashed, red lines indicate the median $3\sigma_{\text{rms},V}$. The per-beam plots are arranged to match the footprint (Figure~\ref{fig:footprint}). } \CORRS{The colour scale is linear in the reported range.}}
\end{figure*}

\begin{figure}[t]
\includegraphics[width=1\linewidth]{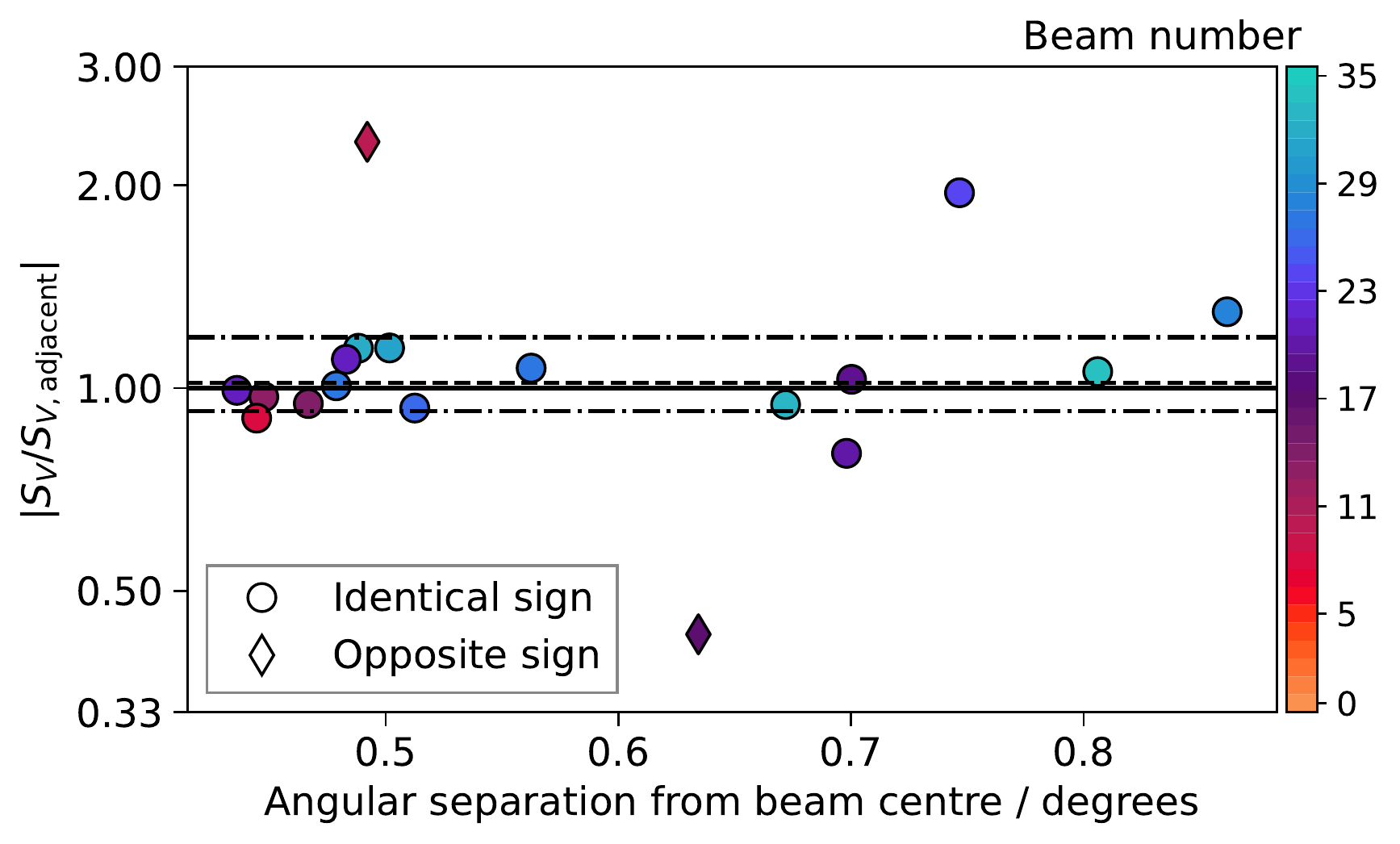}
\caption{\label{fig:vsingleoverlap}\corrs{Absolute Stokes V flux density ratios for sources detected in adjacent beams, as a function of angular separation from the reference beam centre. The sources are coloured by reference beam number. The black, solid line indicates a ratio of 1, the black, dashed line is the median ratio, and the black, dot-dash lines indicate the 16$^{\text{th}}$ and 84$^{\text{th}}$ percentiles.}}
\end{figure}

After imaging and peeling, an image of the full tile for each SBID is created via linear mosaic of the 36 individual beam images. The beam images are weighted by a combination of image sensitivity and primary beam attenuation. \CORRRS{A beam- and position-dependent model of the leakage of Stokes I into V is also applied during linear mosaicking of the Stokes V images to remove widefield leakage.}

Prior to RACS-low, primary beam attenuation (and corrections) were assumed to approximately follow a 2-D circular Gaussian model. This was found to be inadequate to represent the low-band primary beam response \citepalias{racs1}. A post-imaging correction was made by \citetalias{racs1} derived from holographic measurements of the shape of primary beams after completion of the survey, and these patterns were applied to the final tile mosaics. Further comparison to other surveys showed general agreement in overall brightness scale \citepalias{racs1,racs2}. For RACS-mid, the intention was to use the observatory-derived holography to provide primary beam corrections \CORRRS{and widefield leakage corrections} from the start, which would also allowing appropriate weighting when mosaicking the individual beams to form the tile mosaics. A description of the holographic measurement process is provided by \citet{askap:holography}.

\subsubsection{The effect of PAF beam-forming}\label{sec:holography:beamform}

Over the course of processing data for RACS-mid, we found that the primary beam response measured from holography was not constant in time. Significant changes appear to occur after the digital beam-former weights are re-measured. This process of measuring the parent beam-former weights is performed with a cadence of a few weeks to a few months. \citet[][see also \citealt{Hotan2014,McConnell2016}]{Hotan2021} describes the digital beam-forming procedure and how the beam-former weights are also updated using the on-dish calibration (ODC) system to prevent degradation of the beams between parent weight measurements \citep[see also][for a description of the beam-forming for the PAFs of Apertif on the Westerbork Synthesis Radio Telescope]{vanCappellen2022,Denes2022}. The derivation of beam-forming weights uses a method to maximise SNR when pointing at the Sun. Changes to solar features may cause the resulting digital beam response to shift and/or change shape if the centroid of the solar emission is not constant between the beam-forming observations. RACS-mid was observed over 9 disjoint sets of dates, each with a different set of parent beam weights (hereinafter we refer to these periods as BWT-1 to BWT-9), and for a majority of the observations no appropriate matching holographic measurement of the primary beam is available. Application of mismatched primary beam responses can result in brightness scale errors up to a factor of 2 at the beam edges for all beams. While the digital beam-former weights are updated more frequently with the ODC system, we do not see significant brightness scale discrepancies between these ODC updates. 

For BWTs with applicable holographic measurements, the holographic Stokes I primary beam measurements are used (five periods), but these constitute only $\sim 4\%$ of the total survey. Most of the observations were taken during BWT-1, with $\sim 90\%$ of the total observations. Table~\ref{tab:beamformer} summarises the BWTs and indicates the range of SBIDs (and associated observation dates) applicable and whether matching holographic primary beam measurements are available.

\subsubsection{Measurement of the Stokes I primary beam response}\label{sec:holography:zernike}

In lieu of post-mosaicking corrections \citepalias[cf.][]{racs1}, we opt to measure the primary beam response per beam for BWT-1--3 and BWT-5 which do not have appropriate holographic measurements. A similar, non-holographic approach has been adopted by \citet{Kutkin2022} for Apertif post-imaging primary beam corrections. We use in-field sources extracted from apparent brightness images of the $\sim 40\,000$ individual beams and compare these to NVSS measurements where available. Beams from tiles that lie outside of the NVSS coverage are excluded. We also exclude beams from SBIDs in the Galactic Plane and those with peeled sources due to the increase in artefacts in those fields. We use the \texttt{aegean} source-finder with a detection threshold of $10\sigma$ to create per-beam source lists---this yields of order 50--200 sources per beam per SBID prior to cross-matching. We cross-match these individual single-beam source-lists to the NVSS~\footnote{Using \texttt{match\_catalogues} packaged with \texttt{flux\_warp} \citep{Duchesne2020a}: \url{https://gitlab.com/Sunmish/flux_warp}.}, including only compact~\footnote{Using a simple cut to the ratio of integrated flux density, $S_\text{int}$ to peak surface brightness $S_\text{peak}$, $S_\text{int} / S_\text{peak} < 1.2$.} and isolated~\footnote{No neighbours within 25\,arcsec.} sources. The final \corrs{per-beam} cross-matched \corrs{source-lists} contain $\sim 50$--100 sources per SBID. We assume a beam attenuation of the form,
\begin{equation}
    A^{b}_\text{attenuation} = \frac{S_{b}\left(1400/1367\right)^\alpha}{S_{\text{NVSS}}} \,
\end{equation}
for sources with apparent flux density $S_{b}$ in RACS-mid beam, $b$. We also assume a nominal $\alpha = -0.7$ to scale NVSS measurements to 1367.5\,MHz, though the final attenuation patterns are normalised after modelling.

The measured $A^{b}_\text{attenuation}$ is median-binned in tile $(l,m)$ coordinates, stacking all SBIDs for a given BWT. Bin sizes range to $2.1 \times 2.1$\,arcmin$^2$ for BWT-1 to $6.9 \times 6.9$\,arcmin$^2$ for BWT-2, BWT-3, and BWT-5. The larger bin size used for the later BWTs is to account for more sparsely sampled beams.
We fit 2-D models to the binned measurements of $A^{b}_\text{attenuation}$ as a function of $(l,m)$ using standard least-squares methods. While generic 2-D polynomial and elliptical Gaussian models are tested we find these do not represent the attenuation patterns for all beams. Instead we find Zernike polynomial models \citep{Zernike1934}~\footnote{We use the Zernike models implemented in \texttt{galsim}, \citet{galsim}: \url{https://github.com/GalSim-developers/GalSim}.} fit the primary beam main lobe patterns well. \corrs{Zernike polynomials have been used for modelling holographic primary beam measurements from the VLA \citep[e.g.][]{Iheanetu2019,Sekhar2022} and MeerKAT \citep[e.g.][]{Asad2021,Sekhar2022}. A brief comparison of some alternate beam attenuation models are shown in Appendix~\ref{app:beams}.}

Each binned $A^{b}_\text{attenuation}$ dataset is fit with a Zernike polynomial of reasonably high Noll index \citep{Noll1976}. We use the Akaike information criterion \citep[AIC;][]{Akaike1974} to select the appropriate Noll index for each beam. The choice of \corrs{Noll} index varies per beam and per BWT, \corrs{increasing slightly for BWT-1 with larger bins, and range from 38--99 for BWT-1, 32--40 for BWT-2, 38--58 for BWT-3, and 22--41 for BWT-5. The different BWTs and beams have significant differences in the density of sources in the sidelobes, accounting for some of the variation seen in the selected Noll indices. Consequently, the sidelobes of the primary beam are generally poorly modelled and are clipped in the final beam models. For comparison, \citet{Sekhar2022} use a Noll index of 66 to model both VLA and MeerKAT beams, though in that case the sidelobes are well-modelled with their holographic measurements.} Attenuation patterns for each beam are additionally clipped below 12\% which reflects the clip used during mosaicking. The individual beam images are incorporated into the FITS file format used by the observatory to store the holographic primary beam measurement and are used by the \texttt{ASKAPSoft} mosaicking software. 

\corrs{Figure~\ref{fig:icomparison} shows the binned, measured and Zernike model Stokes I response for beam 35 from BWT-1, as well as the binned, (and regridded) measured, Zernike model, and holographic model response for beam 35 from BWT-4. The BWT-1 beam 35 measured data use a $2.1 \times 2.1$\,arcmin$^{2}$ bin size, and the BWT-4 beam 35 measured data use a $10.4 \times 10.4$\,arcmin$^{2}$ bin size. For display purposes the BWT-4 data are regridded to the same bin size as the BWT-1 data, including interpolation. The ratios between the measured and model responses are also shown to highlight the offset in brightness scale that would be introduced when using the holographic model from a different BWT. The main difference we see between BWTs is a shift in peak position of the beam, but there is also a small deviation in the shape that is more difficult to account for in simply shifting the holographic model beam positions. A small additional offset is observed between the BWT-5 holographic model and the BWT-5 Zernike model which results in a 10--20\% variation in brightness scaling towards the beam edges.} \corrs{Further examples of beams 15 and 35 for BWT1--3\&5 are shown in Figure~\ref{fig:zernike:single}, highlighting the measured attenuation pattern per source, the resulting model at each source's location, and residuals after application of the model to the measured sources. Figure~\ref{fig:fullstokesi} shows the Stokes I beam models for all beams of BWT-1 to highlight the variation in the beam shapes across the full footprint.}

We do not have the SNR to accurately measure the spectral-dependence of the beams. \corrs{As shown in Figure~\ref{fig:fwhm}, the beam-averaged FWHM as measured by holography varies from 1.30 to 1.18~degrees at the low- and high-frequency ends of the (unflagged) band. As the Zernike beam models are determined from the MFS apparent brightness images, any residual frequency dependence is implicitly captured by the Zernike models and the 0$^\text{th}$-order Taylor maps after primary beam correction will not have residual frequency-dependent spectral effects.} Mosaicking for SBIDs that use these in-field measured beams does not provide a frequency-dependent correction for the 1$^\text{st}$-order Taylor maps that are created during imaging. These 1$^\text{st}$-order Taylor maps will only be correct at the centre of each beam. Mosaic images are trimmed to remove additional primary beam sidelobe components present above 12\% for the SBIDs weighted using holographic measurements.

\subsubsection{Measurement of the Stokes V widefield leakage}\label{sec:stokesv}

To characterise the widefield leakage from Stokes I into V for all BWTs, we follow a similar method to our characterisation of the total intensity response described above. This method is also being used for leakage characterisation of Stokes I into Q and U by {Thomson et al. (submitted)} for SPICE-RACS \corrs{, and has been used for widefield leakage correction of the Westerbork Synthesis Radio Telescope \citep{Farnsworth2011}, the MWA \citep{Lenc2017}, and other ASKAP data (e.g. POSSUM; West et al., in prep; Gaensler et al., in prep)}. We begin using the same total intensity source catalogues used in Section~\ref{sec:holography:zernike}, and extract the corresponding uncorrected Stokes V flux density at the peak total intensity pixel location from each beam image. \CORRS{The number of detectable circularly polarized sources is expected to be low \citep[e.g.][]{Lenc2018,Callingham2022} and we assume the selected sources are unpolarized with no preferred handedness \citep[as suggested by previous Stokes V surveys, e.g.][]{Rayner2000,Lenc2018,Callingham2022}.}  As above, we model the 2-D widefield leakage surface using a Zernike polynomial using least-squares for each BWT. In contrast to our total intensity modelling, we cut out components that are \corrs{$<100\sigma_{\text{rms},I}$} or are separated from the beam centre by more than $1^\circ$. In addition to this sample cut, model-fitting here is also performed on individual sources rather than binned data \corrs{in contrast to the approach taken with the Stokes I beam modelling}. \CORRRS{As we do not normalise the leakage surface, any residual position-independent leakage introduced or left-over from on-axis corrections using PKS~B1934$-$638 are also included in these widefield models.}

Initially inspecting the distribution of $V$/$I$, we find some points show spuriously high fractional circular polarization, despite our cuts. In Figure~\ref{fig:v_leakage_sbid} we show the standard deviation of $V$/$I$ as a function of SBID. We see that there are some observations with high variance, including some entire BWT. For our purposes, we initially excluded SBIDs with a $V$/$I$ standard deviation greater than the $84^\text{th}$ percentile of the entire set of observations, however, due to small number of remaining SBIDs for most BWTs we opt to relax this to the $99.7^\text{th}$ percentile for BWT-3--9. After excluding outlying sources, we fit Zernike polynomials up to a maximum Noll index of 10, and select the best model according to the AIC. Finally, we regrid and interpolate our fitted models to exactly match the corresponding holography images produced for the Stokes I beams. 

\corrs{Figure~\ref{fig:vcomparison} shows the leakage surface for beam 35 for BWT-1 and BWT-4, and compares to the BWT-4 holographic measurement of the same beam as in Figure~\ref{fig:icomparison} for the Stokes I response. In Figure~\ref{fig:vcomparison} the data are binned as in the Stokes I case, though we do not bin the data for fitting. The residual differences between the measured and model leakage surfaces are shown. There is an offset between the holographic model and Zernike models, though we do not use the holographic model for widefield leakage correction.} We show \corrs{additional} examples of our fitted surfaces for beam 15 and 35 for BWT-1--3 in Figure~\ref{fig:leakage:run1}, \corrs{again showing the residual difference between the measured $V/I$ and the fitted leakage surfaces.} \corrs{Finally, model Stokes V beams for BWT-1 are shown in Figure~\ref{fig:fullstokesv} for all beams in the footprint.} \corrs{The partition between negative and positive leakage in the Zernike polynomial surfaces generally resemble those found for other instruments \citep[e.g. the VLA and MeerKAT;][]{Sekhar2022}, taking into account all beams being offset from the optical axis due to their arrangement in the \texttt{closepack36} PAF footprint. Some beams (e.g.~12/23 and 13/22, see Figure~\ref{fig:fullstokesv}) have non-standard shapes, though their symmetry in the PAF footprint suggests this is an accurate representation of the leakage for these beams. While PAF-based Stokes V beam modelling is not available in the literature for comparison, the leakage patterns vary between the PAF beams which is seen in leakage maps for Stokes Q and U for ASKAP (Thomson et al., submitted) and in Stokes Q for Apertif  \citep{Denes2022}.}

The SBID subset SB21616--SB21710 showed spuriously high residual leakage after correction compared to the remainder of BWT-1. This subset contains 91 SBIDs corresponding to a single calibrator, SB21637. We create models for this subset separate from BWT-1 for both the Stokes I response and the $V/I$ leakage. We find that some beams in this subset have notably different leakage patterns to the remainder of BWT-1. Figure~\ref{fig:leakage:run5:beam15} shows beam 15 for the SB21616--SB21710 and \ref{fig:leakage:run5:beam35} shows beam 35. While beam 15 in this subset resembles beam 15 from the full BWT-1 subset, beam 35 differs significantly. Other beams, including 12 and 23, were also found to have spuriously high ($|V/I| >0.03$) leakage sources when using the full BWT-1 leakage correction. For the SB21616--SB21710 subset, we use leakage models derived from those SBIDs only. We find no substantial difference in the Stokes I response in the SB21616--SB21710 subset and continue to use the full BWT-1 Stokes I model for these SBIDs. It is not clear what caused the change in leakage characteristics for select beams within this SBID subset.

The residual \corrs{per-SBID} $|V/I|$ leakage after mosaicking and application of the leakage surface is also shown in Figure~\ref{fig:v_leakage_sbid:post} \CORRS{to highlight the reduction of leakage}.

\subsection{{Validation of the individual primary beam models}}\label{sec:beamvalidation}

\begin{figure*}
    \centering
    \includegraphics[width=1\linewidth]{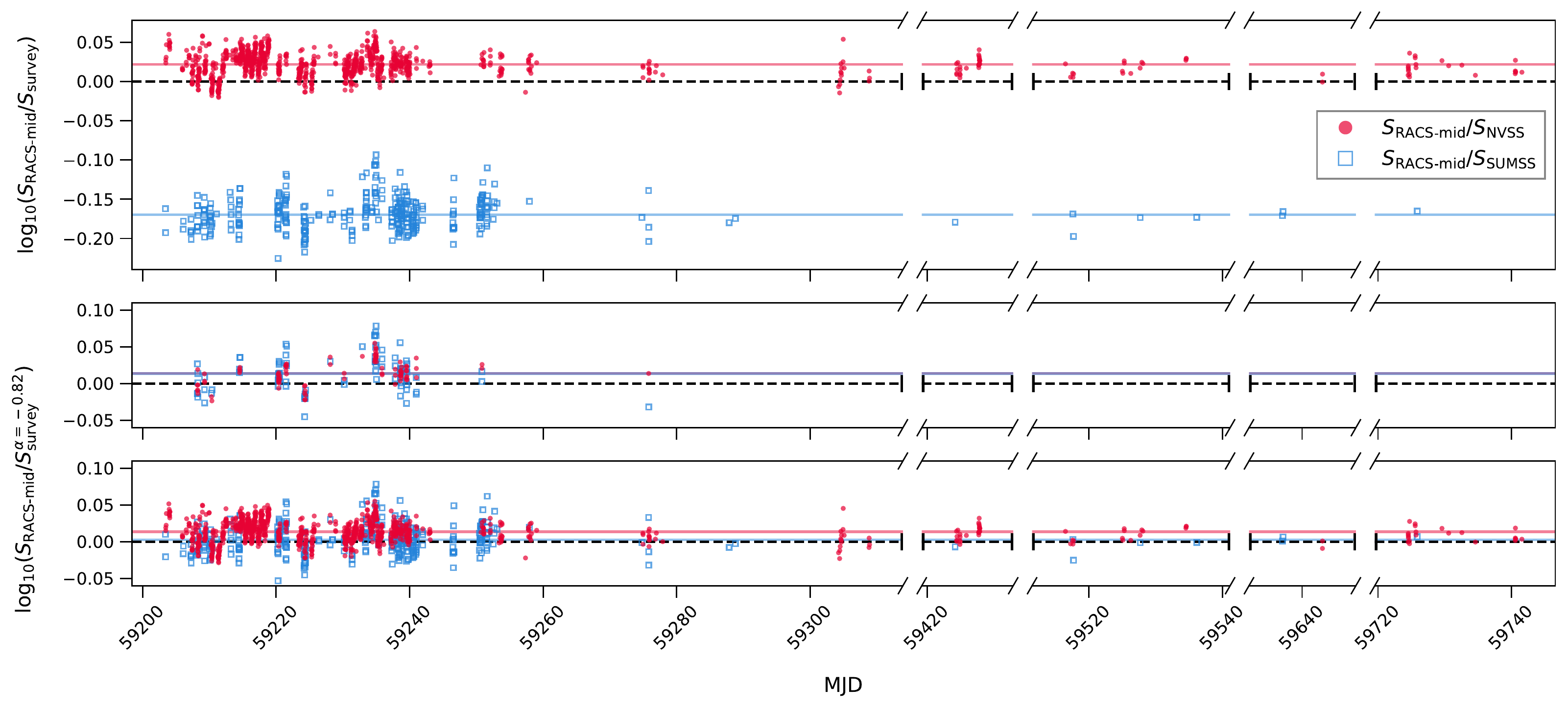}
   \caption{\corrs{\label{fig:timeflux} Time-dependence of the flux density scale across the survey with comparison to NVSS (red, filled circles) and SUMSS (blue, open squares). \emph{Top.} Logarithmic flux density ratios for comparisons with NVSS and SUMSS, with no frequency scaling. \emph{Middle.} SBIDs with sources in the central 2\,deg of the tile within the declination range $-40 \leq \delta_\text{J2000} \leq -30$ with both NVSS and SUMSS cross-matches. \emph{Bottom.} All SBIDs in the top panel but after scaling the source flux densities assuming a power law with index $\alpha = -0.82$. The black, dashed line in all panels corresponds to flux densities ratios of unity. The coloured lines correspond to medians for each survey comparison.}}
\end{figure*}

\begin{figure}[t]
    \centering
    \includegraphics[width=1\linewidth]{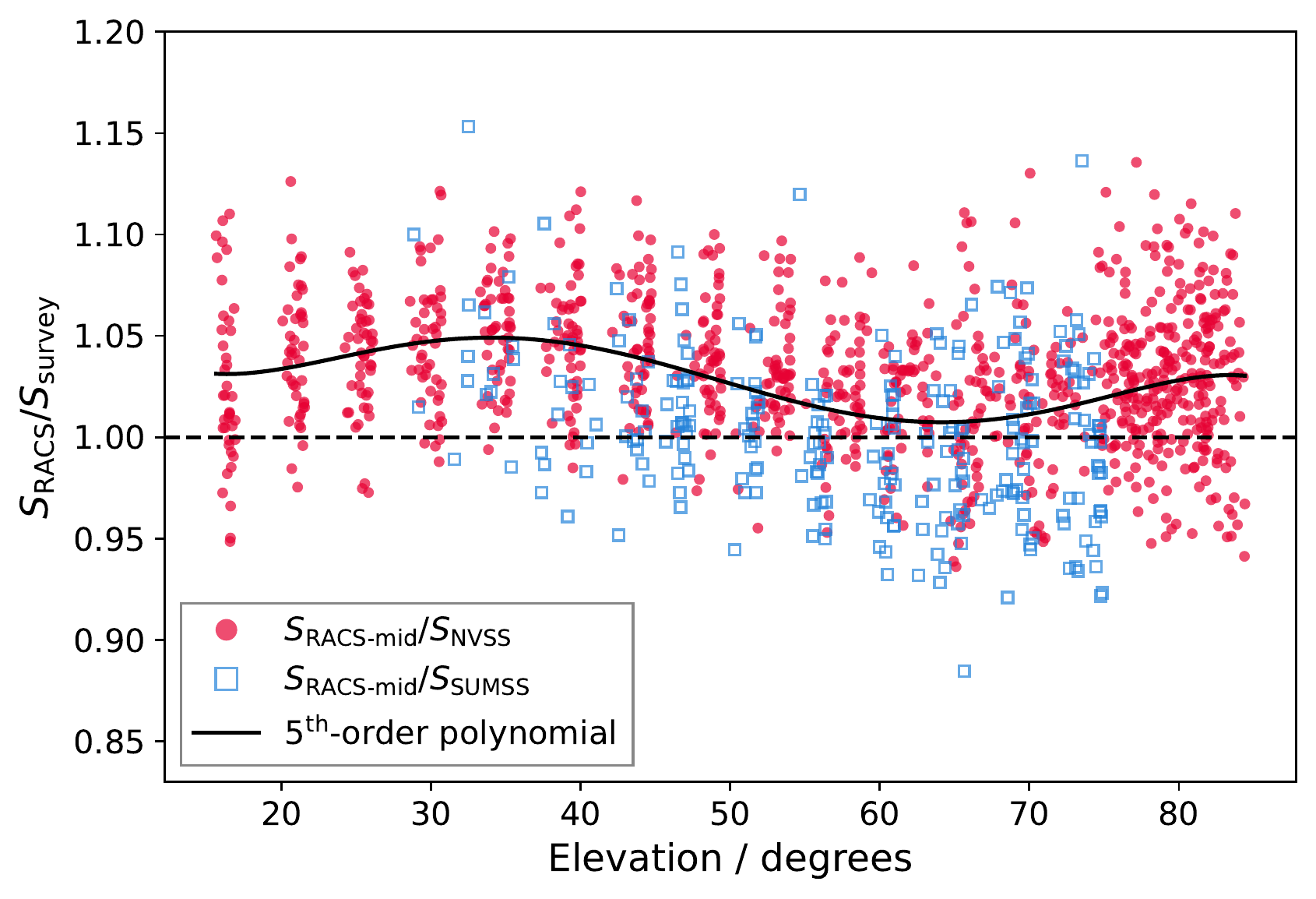}
    \caption{\label{fig:fluxelevation} Median tile flux density ratios between the NVSS (red, filled circles) and SUMSS (blue, open squares) after scaling in frequency assuming $\alpha = -0.82$ as a function of elevation. The solid black line shows the 5$^\text{th}$-order polynomial fit to the flux density ratios.
    }
\end{figure}

\corrs{To check the accuracy of the individual primary beam models, we inspect isolated (no neighbours within 45~arcsec) and compact ($0.8 < S_\text{int}/S_\text{peak} < 1.2$) sources detected across adjacent beams within single observations after application of the respective beam models. As we do not primary beam correct individual images as part of our processing, we instead apply the primary beam model to the per-observation, per-beam source-lists used during creation of the models with the compact and isolated source cuts described in Section~\ref{sec:holography:zernike}. As the Zernike models are defined to match the frequency of the MFS images, these are directly applied. For Stokes I, BWT-4\&6--9 make use of the frequency-dependent holography-based models and these are evaluated at 1367.5\,MHz before being applied. For Stokes V, leakage is removed using the Zernike models and the primary beam is then applied as per usual.}

\subsubsection{{Stokes I}}\label{sec:beamvalidation:i}

\corrs{The number of sources with Stokes I flux densities $>10\sigma_{\text{rms},I}$ per adjacent beam cross-match ranges from $\sim 25\,000$ to $\sim 53\,000$, with corner and edge beams featuring fewer sources due to fewer adjacent beams. Figure~\ref{fig:ibeamoverlap} shows the ratio of Stokes I flux density measurements of sources in adjacent beams as a function of distance from the beam centres. The plot shows the cross-match results for each beam separately, arranged to match the PAF footprint (see Figure~\ref{fig:footprint}). As each beam is cross-matched to each adjacent beam, the individual beam results are not independent. We calculate the median flux density ratio in bins separately for the Zernike-based models (white circles, with $\sim 24\,000$ to $\sim 50\,000$ sources per beam) and the holography-based models (red squares, with $\sim 1\,100$ to $\sim 2\,400$ sources per beam). The overall median ratio is $1.00_{-0.11}^{+0.12}$.  Generally there is good agreement in adjacent beams with two main exceptions: roll-off in the holography-derived models beyond $\sim 0.8$~deg from the beam centre of order $\sim 10$\% and offsets in beams 5 and 30. The corner beams 5 and 30 feature median ratios of $0.95_{-0.08}^{+0.10}$ and $1.07_{-0.10}^{+0.10}$ for the Zernike subset ($0.90_{-0.09}^{+0.09}$ and $1.12_{-0.12}^{+0.13}$, for the holography subset), respectively.}

\subsubsection{{Stokes V}}\label{sec:beamvalidation:v}

A similar process is repeated for Stokes V, though we first look at unpolarized sources \CORRS{in adjacent beams after application of the widefield leakage corrections and primary beam correction}. For this, we use sources detected at $>10\sigma_{\text{rms},I}$ in both beams, but restrict to sources with $|S_V| < 3\sigma_{\text{rms},V}$ in the reference beam. The adjacent beam Stokes V measurement is not restricted. This results in a similar number of sources per beam as in the Stokes I comparison in the previous section. Figure~\ref{fig:vbeamoverlap1} shows the unpolarized sources in the 36 beams with the difference in Stokes V measurements between adjacent beams. In Figure~\ref{fig:vbeamoverlap1} we also show the binned medians for the selected sources, along with the $16^\text{th}$ and $84^\text{th}$ percentiles.  The median $3\sigma_{\text{rms},V}$ for the relevant beam is shown in Figure~\ref{fig:vbeamoverlap1}, with the median $\sigma_{\text{rms},V}$ per beam ranging from 154--200~\textmu Jy\,PSF$^{-1}$. The overall median difference in Stokes V measurements from sources in adjacent beams is $0_{-503}^{+505}$~\textmu Jy\,PSF$^{-1}$, \CORRS{and shows an increase at small and large separations. This is consistent with elevated noise toward the edges of the adjacent and reference beams, respectively.}

\corrs{For circularly polarized sources, there are a total of 18 adjacent beam detections with $|S_V| > 10\sigma_{\text{rms},V}$ in each beam. Figure~\ref{fig:vsingleoverlap} shows the absolute ratio of Stokes V measurements from adjacent beams, coloured by beam number. Sources with the same sign in adjacent beams are indicated with circles, and sources with opposing signs are indicated with diamonds. The median ratio over all beams  is $1.02_{-0.09}^{+0.17}$. Two of the detections show opposing signs, which correspond to a single source (NVSS~J205111$+$081859) detected across beams 9 and 16 at $10.2\sigma_{\text{rms},V}$ and $28.3\sigma_{\text{rms},V}$, respectively, in SB21634. A single detection of PKS~B1340$-$373 in beam 23 from SB21680 shows $|S_V/S_{V,\text{adjacent}}| \approx 2$, corresponding to a residual leakage in the final mosaicked tile of $|V/I|\approx 0.015$. Both SB21634 and SB21680 are part of the subset noted in Section~\ref{sec:stokesv} that show different leakage patterns for some beams. This suggests the residual leakage in these SBIDs may differ from the remainder of the survey despite the separate leakage models used.}

\subsection{Additional brightness scaling of tiles}\label{sec:tilescale}

\begin{figure*}[t]
    \centering
    \includegraphics[width=1\linewidth]{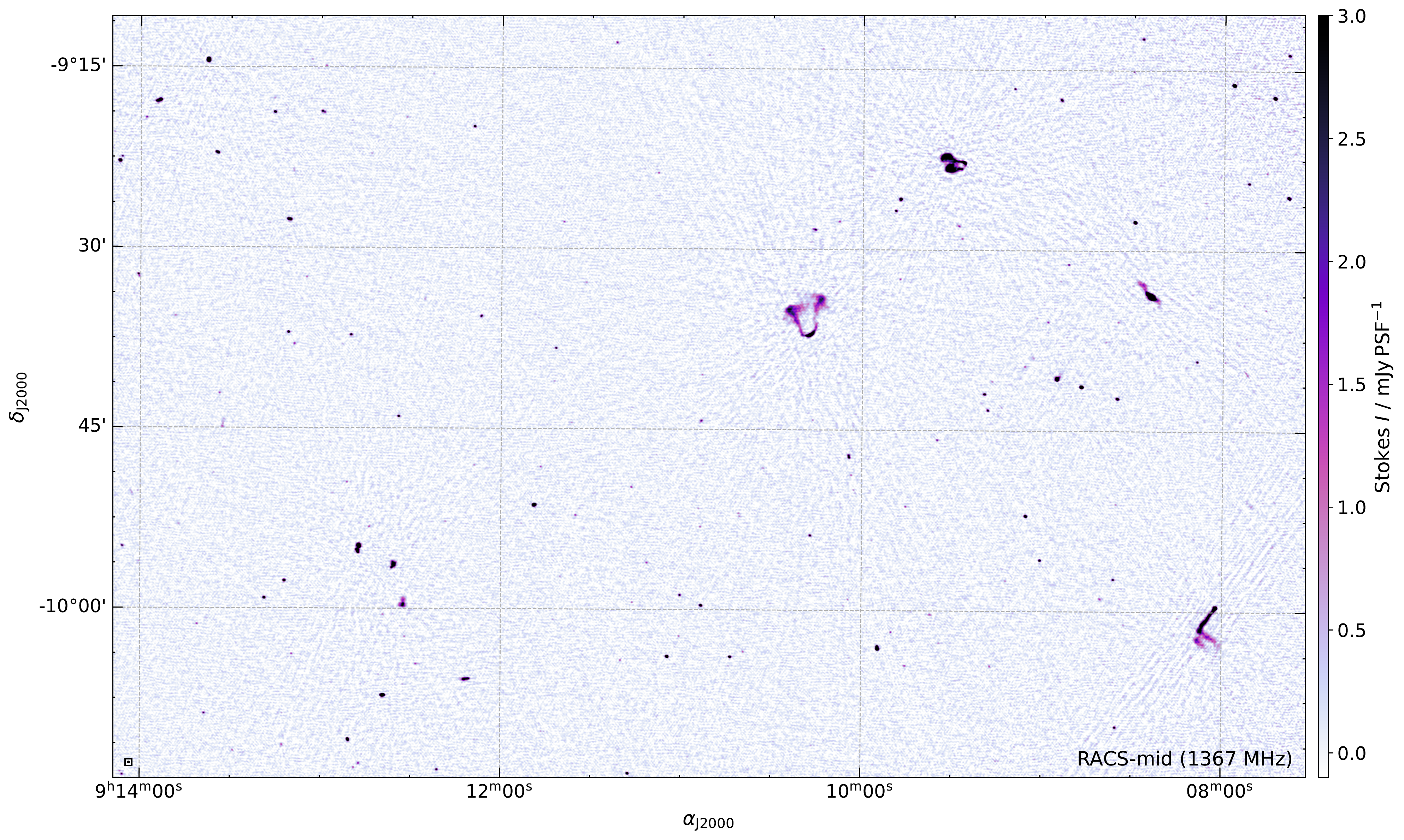}
    \caption{\label{fig:example:large}\corrs{Example RACS-mid image.}}
\end{figure*}

\begin{figure*}
    \centering
    \includegraphics[width=1\linewidth]{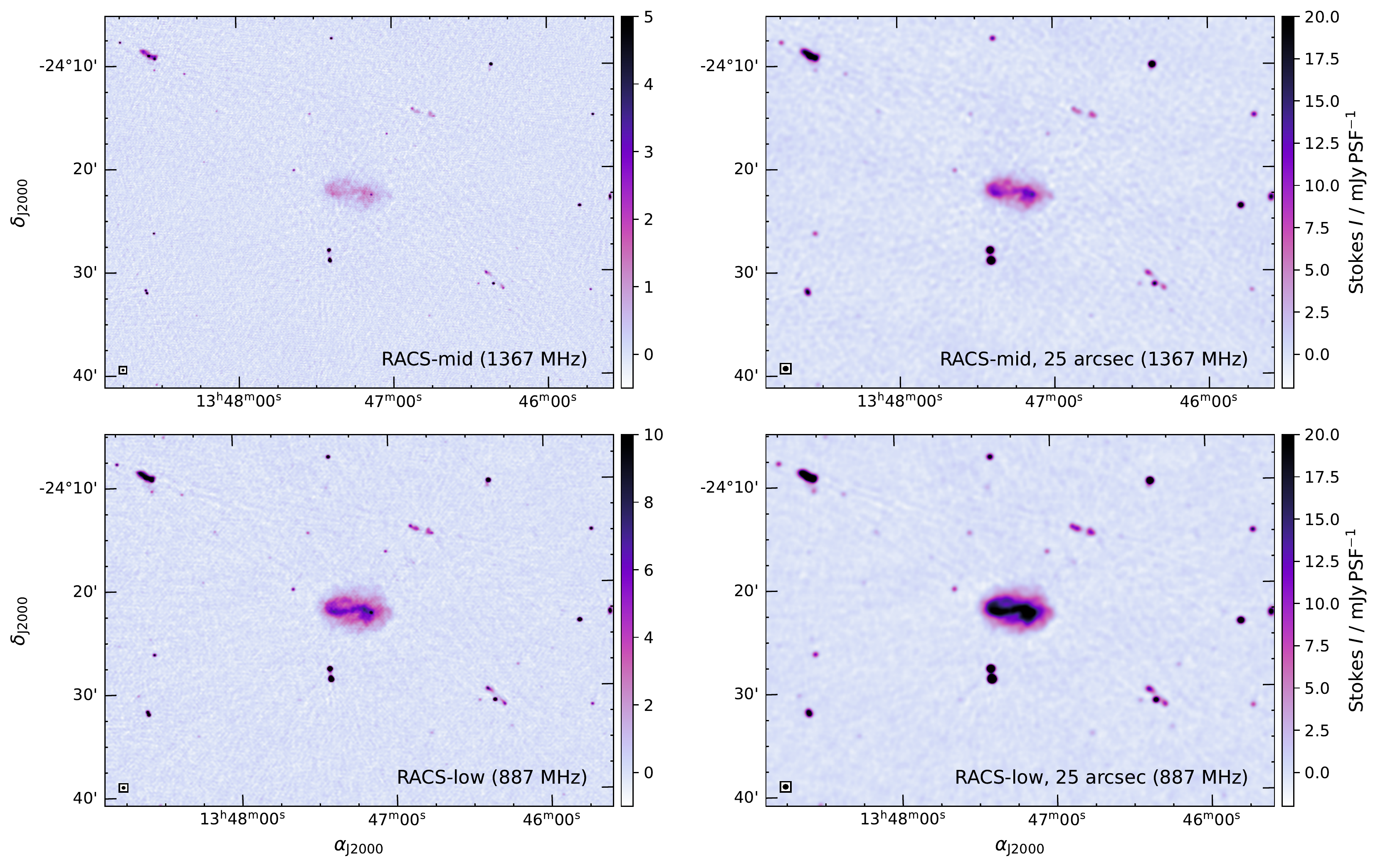}
    \caption{\label{fig:example:eso509} A comparison of RACS-mid (\textit{top left}) with RACS-low (\textit{bottom left}), and RACS-mid and RACS-low convolved to a 25~arcsec resolution (\textit{top right} and \textit{bottom right}, respectively), featuring a radio galaxy of $\sim 6$\,arcmin angular extent associated with ESO~509$-$G108. The size of the PSF for each image is shown in the bottom left as an ellipse. Note the colour scale varies between the images to reflect the change in angular resolution.}
\end{figure*}

\begin{figure*}
    \centering
    \includegraphics[width=1\linewidth]{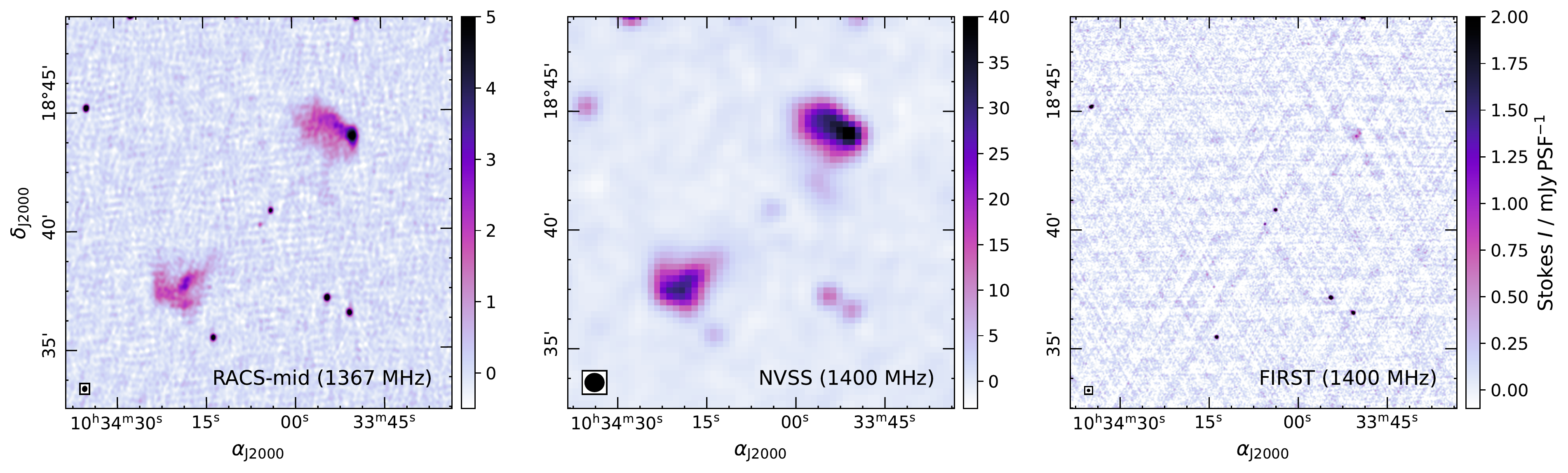}
    \caption{\label{fig:example:j1034}  \corrs{A radio galaxy associated with 2MASS~J10340385$+$1840490 as seen in the RACS-mid (\textit{right}), NVSS (\textit{centre}), and FIRST (\textit{right}) images.} Note the colour scale varies between the three images due to the difference in angular resolution.}
\end{figure*}

In \citetalias{racs1} it was shown that for RACS-low the flux density scale per SBID varies as a function of time  (their figure~7). We find a similar effect for RACS-mid. This is illustrated in Figure~\ref{fig:timeflux}, which shows the median flux density ratio between RACS-mid and \corrs{either the} NVSS or \corrs{the Sydney University Molonglo Sky Survey \citep[SUMSS;][]{bls99,mmb+03}} for individual SBIDs as a function of time\corrs{, observations in the Galactic Plane or those that underwent peeling.} We do not yet have a complete understanding of this variation, but we suspect the two main contributors to be the difference in elevation between PKS~B1934$-$638 and SBID target field and the temperature variation of the PAFs. Figure~\ref{fig:fluxelevation} highlights specifically the component of elevation-dependence in this brightness scale effect.  

As part of the processing pipeline, we use the source-finder \texttt{selavy} to produce a list of sources in the resulting Stokes I total intensity tile mosaics. To correct the observed time variation, we cross-match the per-SBID source-lists to the NVSS and SUMSS catalogues to determine the nominal median scale over each SBID independently. For this, we only use the central 2\,deg of the tile where edge effects left-over from the holography and primary beam correction do not have a significant impact. As with the measurement of the primary beam response, we do not use fields in the Galactic Plane and we also exclude the field containing Cygnus~A due to a lack of detected sources around Cygnus~A.

Because of the possible temperature and elevation dependence, we do not assume the tile median flux density ratios between RACS-mid and the surveys to be 1. To correct this variation accurately, particularly within the southern fields covered by SUMSS, we need to carefully choose the spectral index to scale the flux densities measured in either survey. To determine the nominal spectral index across the survey, we take tiles that contain sources within the overlap region between NVSS and SUMSS, namely where $-40^\circ \leq \delta_\text{J2000} \leq -30^\circ$, and determine the median spectral index by comparison of the two surveys with RACS-mid. As we choose tiles with both NVSS and SUMSS, we avoid contamination by the unknown temperature dependence. We find $\alpha \approx -0.82$ for this region between NVSS and SUMSS, consistent with the median spectral index found between the two catalogues by \citet{mmb+03}.

This spectral index is used for the remaining survey region to obtain median flux density ratios for the full survey. Figure~\ref{fig:timeflux} shows the flux density ratios as a function of time, for the full set of tiles, both scaled and non-scaled, as well for the overlap region used in determining the spectral index. For tiles in the Galactic Plane and those that underwent peeling, we only correct for an elevation dependence by fitting a $5^\text{th}$-order polynomial to $S_\text{RACS-mid}/S_\text{survey}$, where $S_\text{survey}$ is the combination of contributions from NVSS and SUMSS, using measurements from the remaining SBIDs. The result of this is shown in Figure~\ref{fig:fluxelevation}. In total, 63\% of tiles have a factor derived directly from NVSS cross-matches, 15\% from SUMSS, and 22\% from the elevation-dependent model.

Exact factors used for scaling for each SBID are recorded in the FITS header of the relevant images (both Stokes I and V) under the non-reserved keyword \texttt{FSCALE} (i.e. fixed scaled), and is applied as \begin{equation}
\texttt{IMAGE\_DATA = }\texttt{ORIGINAL\_DATA * }\texttt{FSCALE} \quad .
\end{equation} This is similar to the standard application of the \texttt{BSCALE} keyword except that in this case the image data have been explicitly scaled (i.e. the scale is now `fixed'), as opposed to the usage of \texttt{BSCALE} which is done on-the-fly when using the image data. The survey responsible for scaling is recorded in the FITS header under the \texttt{FMETHOD} keyword as either \texttt{NVSS}, \texttt{SUMSS}, or \texttt{ELEV}. While the flux density scale of a majority of SBIDs is dependent on NVSS due to its use in primary beam modelling, for work that requires independent measurements between RACS-mid and SUMSS the user may use the \texttt{FSCALE} parameter to revert to the original data. For this purpose we suggest including an additional uncertainty to flux density measurements corresponding to the $84^\text{th}$ percentile of the uncorrected medians: $\sim 5$\%. Scaling factors and their methods are also recorded in the RACS database under the same names.

\section{{Assessing the full survey images}}\label{sec:validation}

\begin{figure*}
    \centering
    \includegraphics[width=0.5\linewidth]{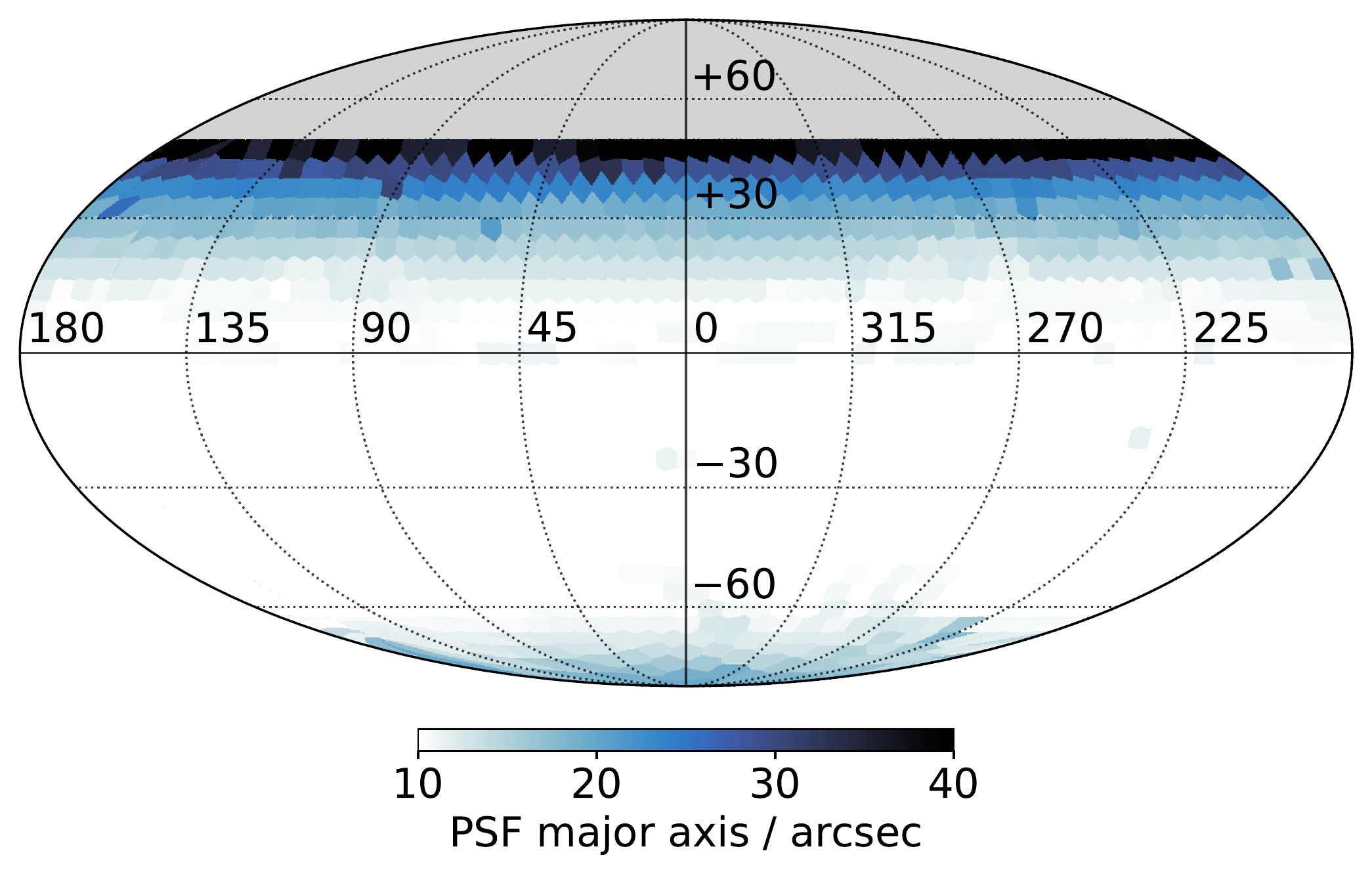}%
    \includegraphics[width=0.5\linewidth]{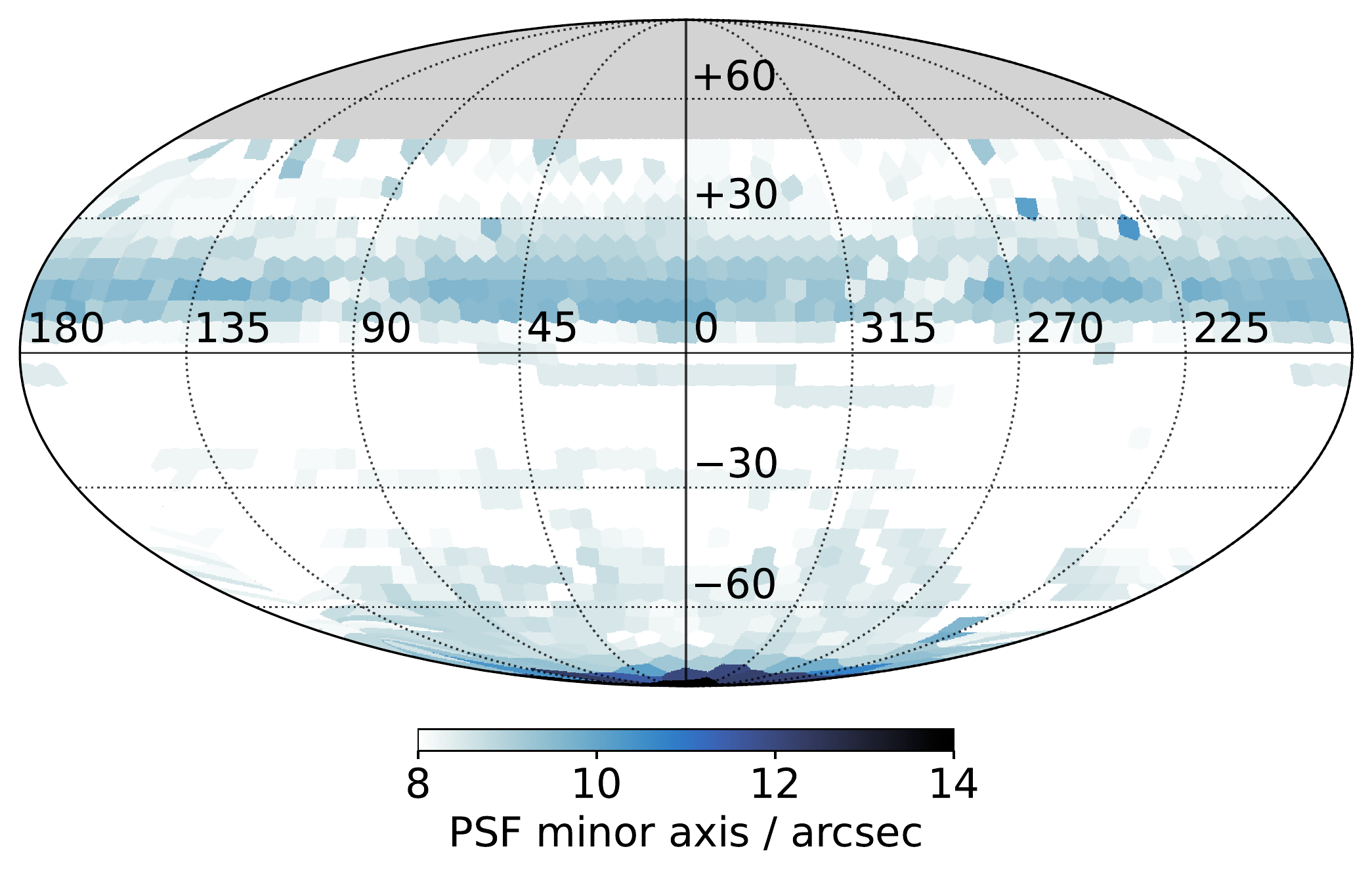}
    \caption{\label{fig:psf} PSF variation between tiles across the full survey area in equatorial coordinates. A single major axis (\textit{left panel}) and minor axis (\textit{right panel}) is associated with each tile. Note each panel has different colour scaling to highlight the variation in major and minor axes independently.}
\end{figure*}

\begin{figure}[t]
\centering
\includegraphics[width=1\linewidth]{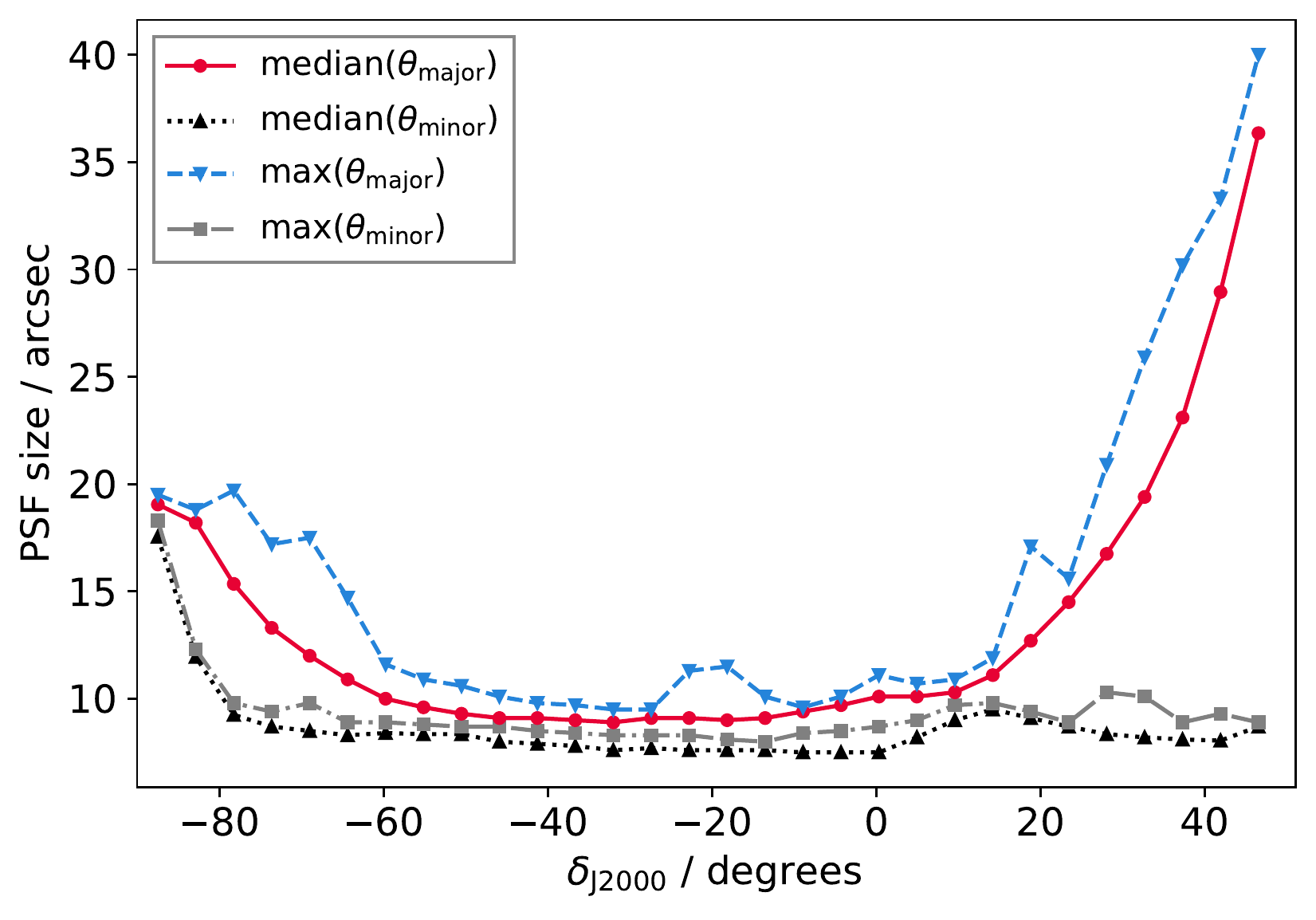}
\caption{\label{fig:psfcoords} The median and maximum PSF major and minor axes in declination, binned to approximately match tile separation.
}
\end{figure}

\begin{figure*}
    \centering
    \includegraphics[width=0.9\linewidth]{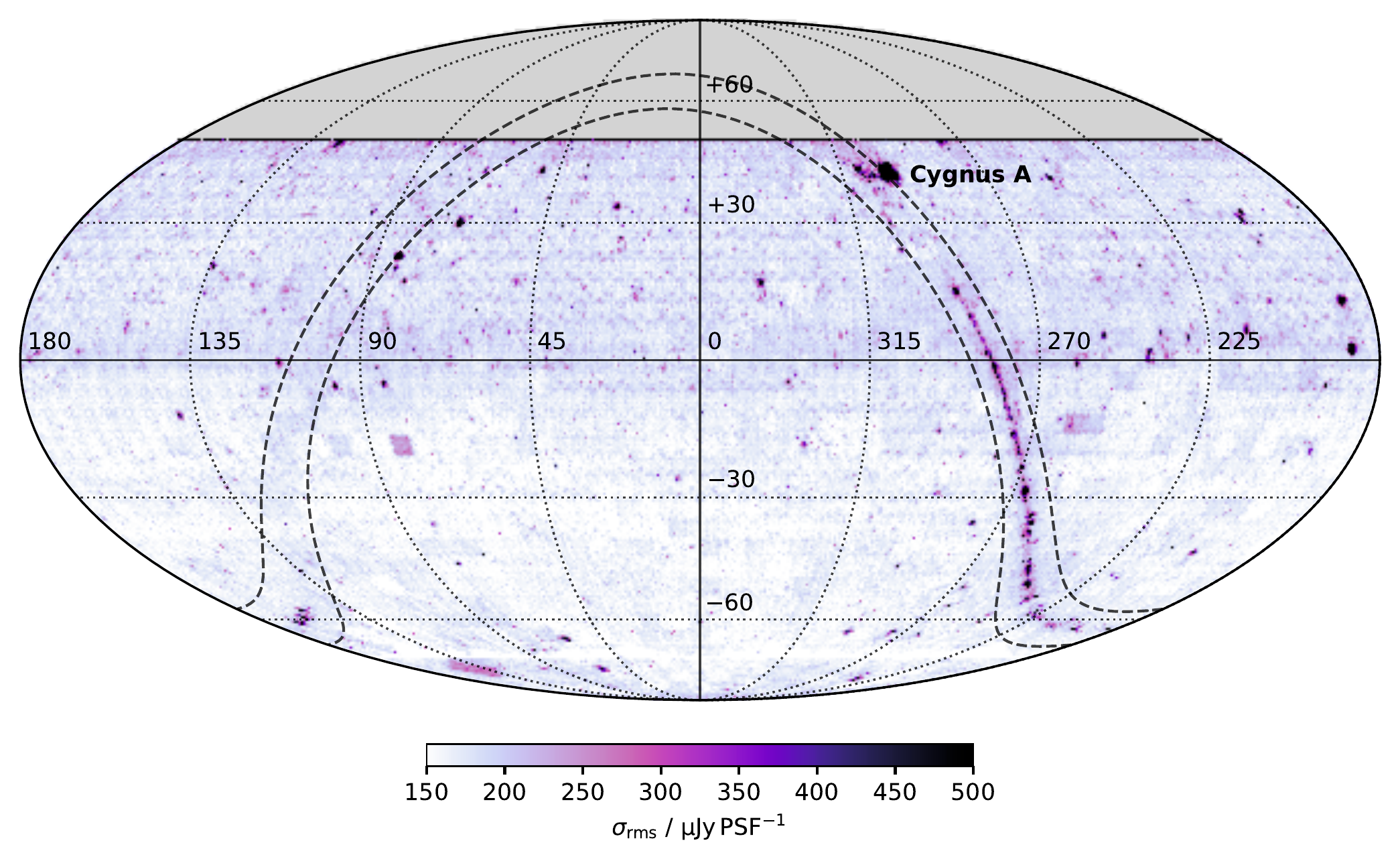}
    \caption{\label{fig:fullskyrms} Mosaic of individual \corrs{Stokes I} rms noise maps across the full survey in equatorial coordinates. The region of Galactic latitude covering $|b| < 5^\circ$ is bounded by the dashed, black lines.}
\end{figure*}

\begin{figure}
    \centering
    \includegraphics[width=1\linewidth]{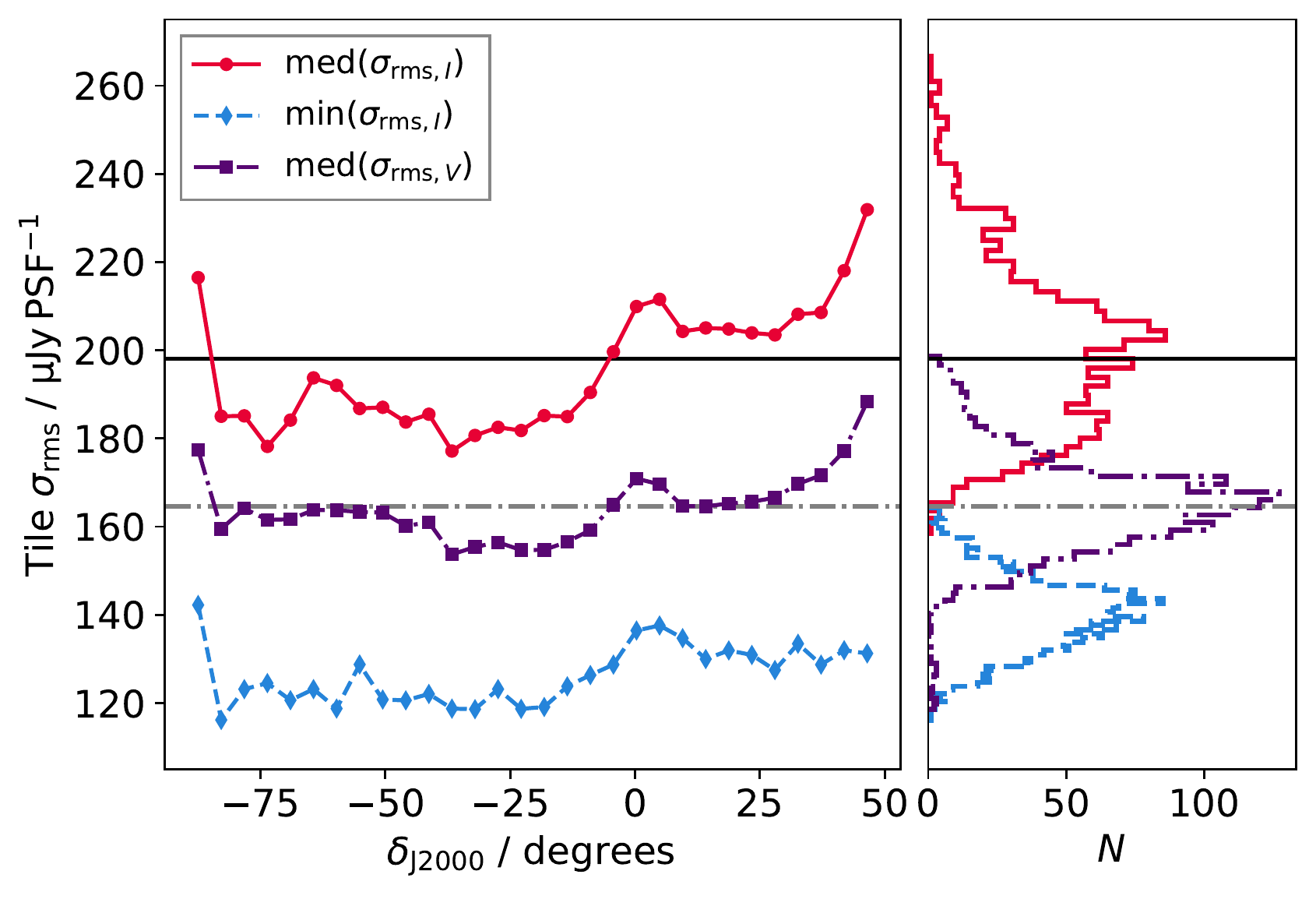}
    \caption{\label{fig:rmshist} Binned noise ($\sigma_\text{rms}$) across tiles as a function of declination for median and minimum values of $\sigma_\text{rms}$. \corrs{The solid, black and dot-dash, grey lines indicate median values over the full survey, for Stokes I and V, respectively.}}
\end{figure}

\begin{figure}
    \centering
    \includegraphics[width=1\linewidth]{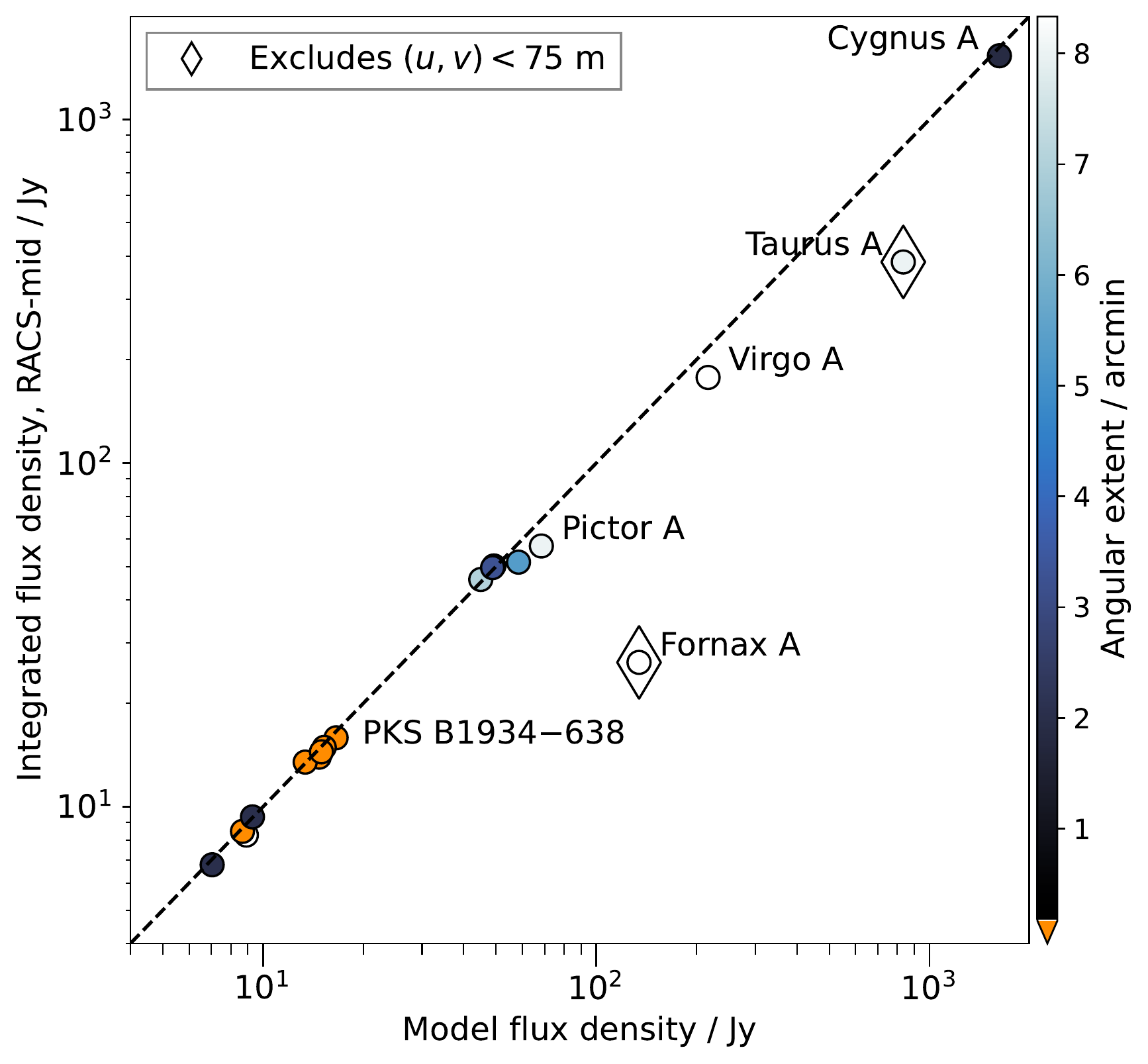}
    \caption{\label{fig:pbscale} Integrated flux density of bright calibrator sources modelled by \citet{Perley2017} with the inclusion of PKS~B1934$-$638 \citep{Reynolds1994} as a function of their model flux density. Sources are coloured by their angular extent, but clipped (orange) when smaller than the RACS-mid resolution. Sources in tiles with a $(u,v)<75$\,m visibility cut are enclosed by black diamonds. Measurement and brightness scale uncertainty are plotted but are smaller than the visible marker size.}
\end{figure}

\begin{figure*}
    \centering
    \begin{subfigure}[b]{0.5\linewidth}
    \includegraphics[width=1\linewidth]{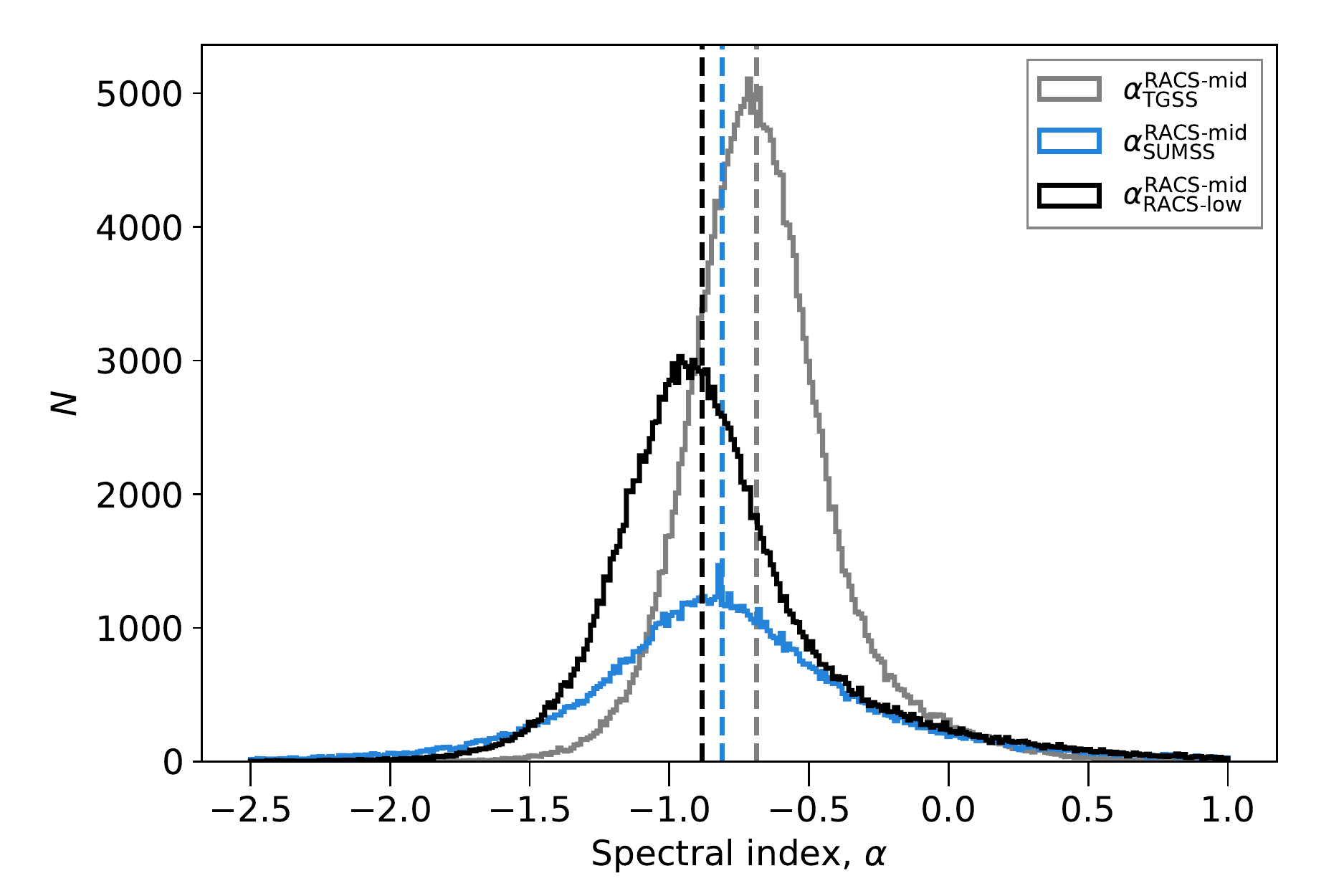}%
    \caption{\label{fig:tileflux:hist:alpha}}
    \end{subfigure}%
    \begin{subfigure}[b]{0.5\linewidth}
    \includegraphics[width=1\linewidth]{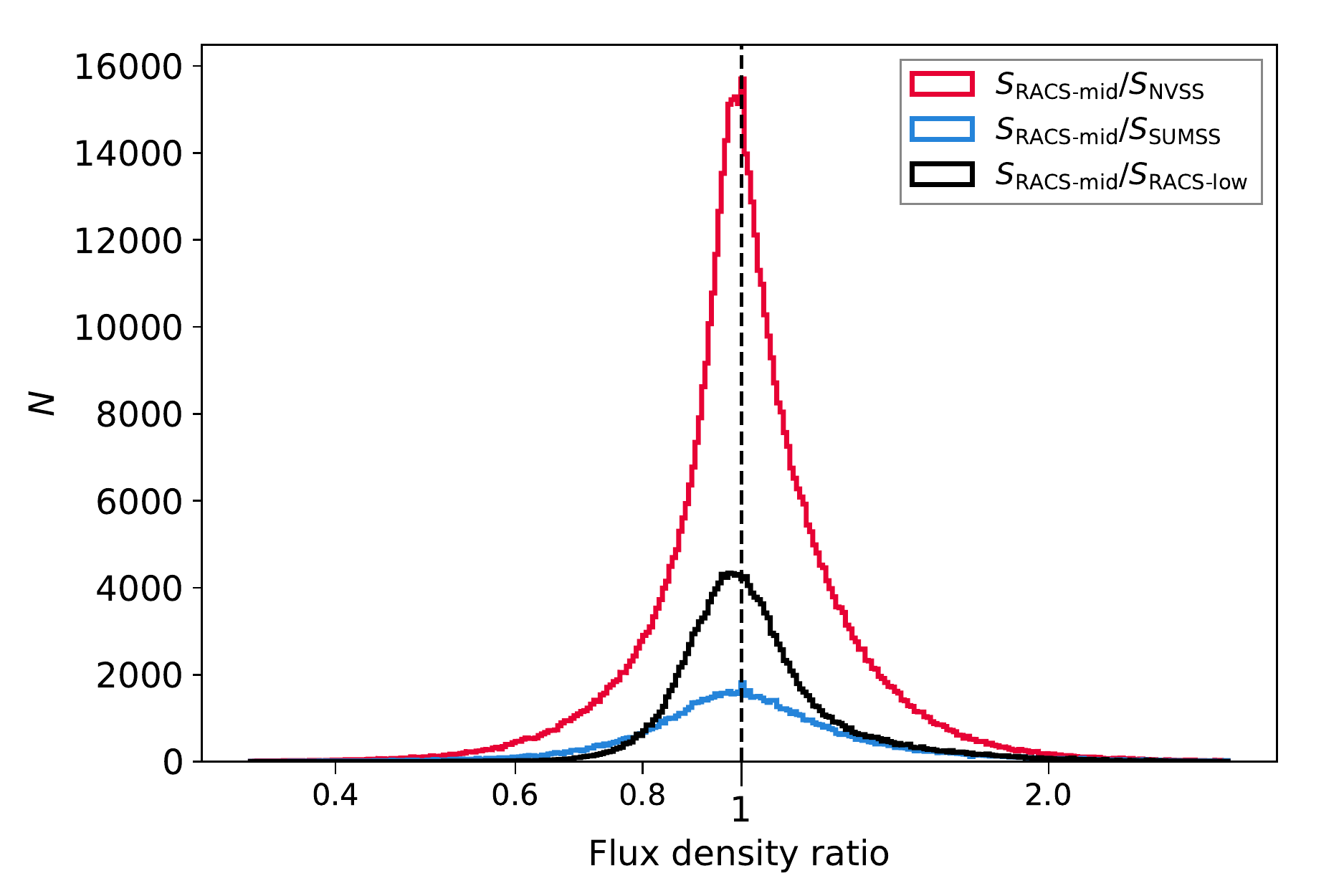}
    \caption{\label{fig:tileflux:hist:flux}}
    \end{subfigure}%
    \caption{\label{fig:tileflux:hist} \corrs{Histograms of spectral indices \subref{fig:tileflux:hist:alpha} and flux density ratios \subref{fig:tileflux:hist:flux} after correction of tile-specific medians (see Figure~\ref{fig:timeflux}) for tiles with NVSS, SUMSS, and RACS-low, and TGSS cross-matches. The median spectral indices for each survey comparison are shown as vertical lines on the top panel.}}
\end{figure*}

\begin{figure}
\centering
\begin{subfigure}[b]{1\linewidth}
\includegraphics[width=1\linewidth]{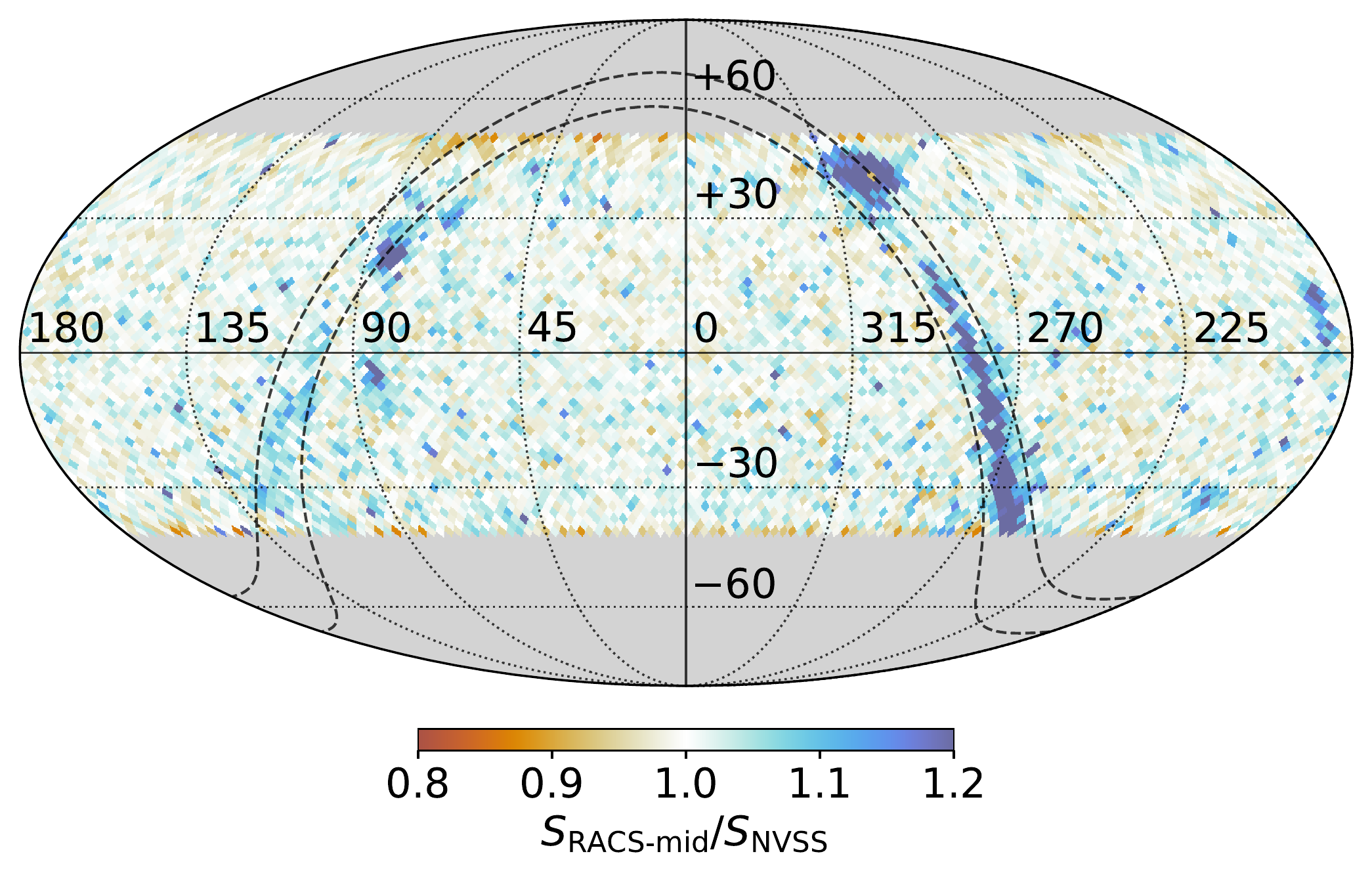}
\caption{\label{fig:fluxscale:sky:nvss} Full survey comparison NVSS.}
\end{subfigure}\\%
\begin{subfigure}[b]{1\linewidth}
\includegraphics[width=1\linewidth]{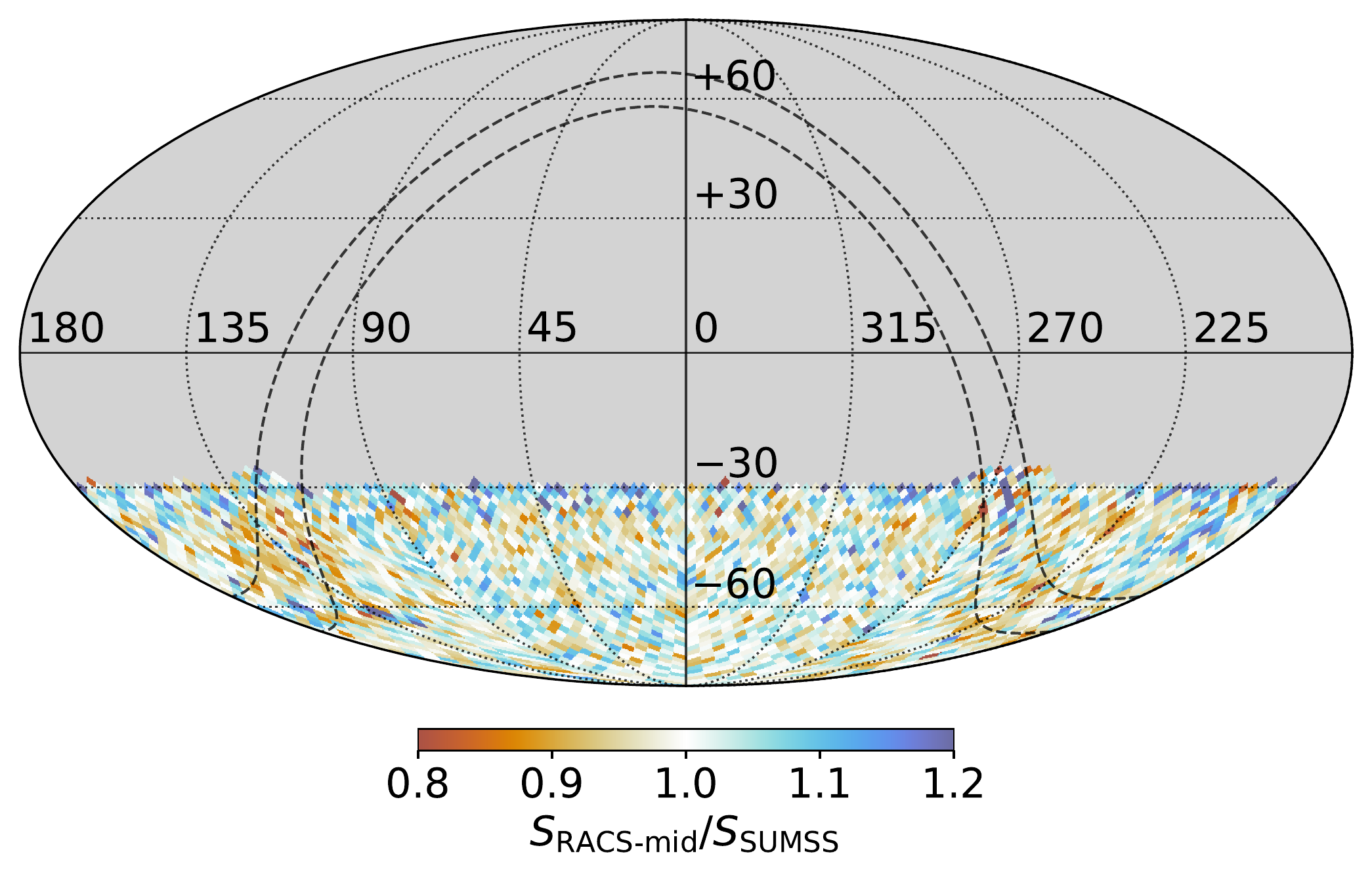}
\caption{\label{fig:fluxscale:sky:sumss} Full survey comparison with SUMSS and the MGPS-2.}
\end{subfigure}\\%
\begin{subfigure}[b]{1\linewidth}
\includegraphics[width=1\linewidth]{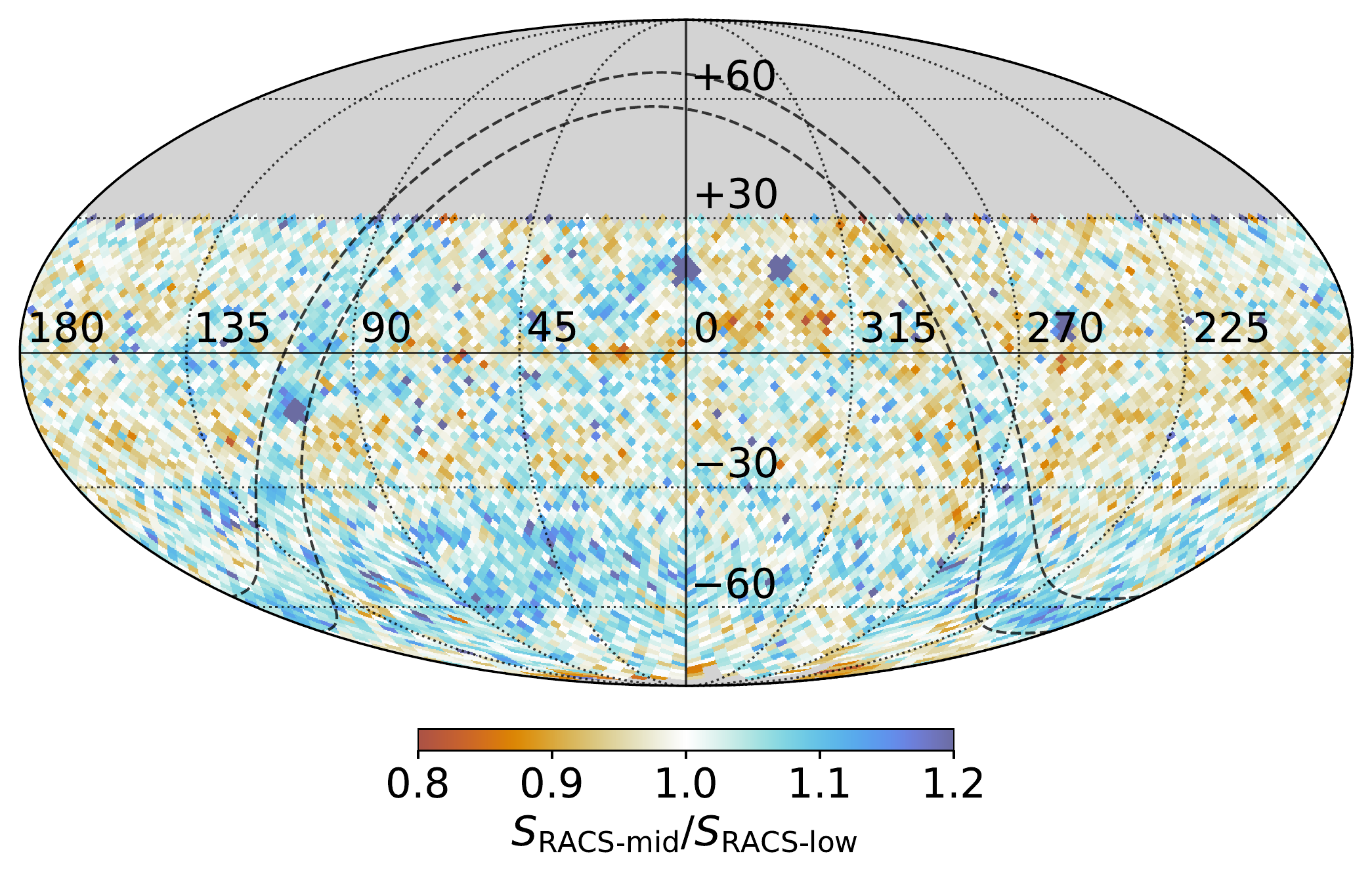}
\caption{\label{fig:fluxscale:sky:racs0} Full survey comparison with RACS-low.}
\end{subfigure}%
\caption{\label{fig:fluxscale:sky} \corrs{Equatorial representation of the brightness scale over the sky with comparison to NVSS \subref{fig:fluxscale:sky:nvss}, SUMSS and the MGPS-2 \subref{fig:fluxscale:sky:sumss}, and RACS-low \subref{fig:fluxscale:sky:racs0}, binned following Hierarchical Equal Area isoLatitude Pixelation \citep[HEALPix;][]{Gorski2005} binning with \texttt{nside = 32}, corresponding to bins of $\sim 3.4$\,deg$^2$. Galactic latitudes of $\pm 5^\circ$ are shown as dashed, black lines.}}
\end{figure}

\begin{figure}
    \centering
    \includegraphics[width=1\linewidth]{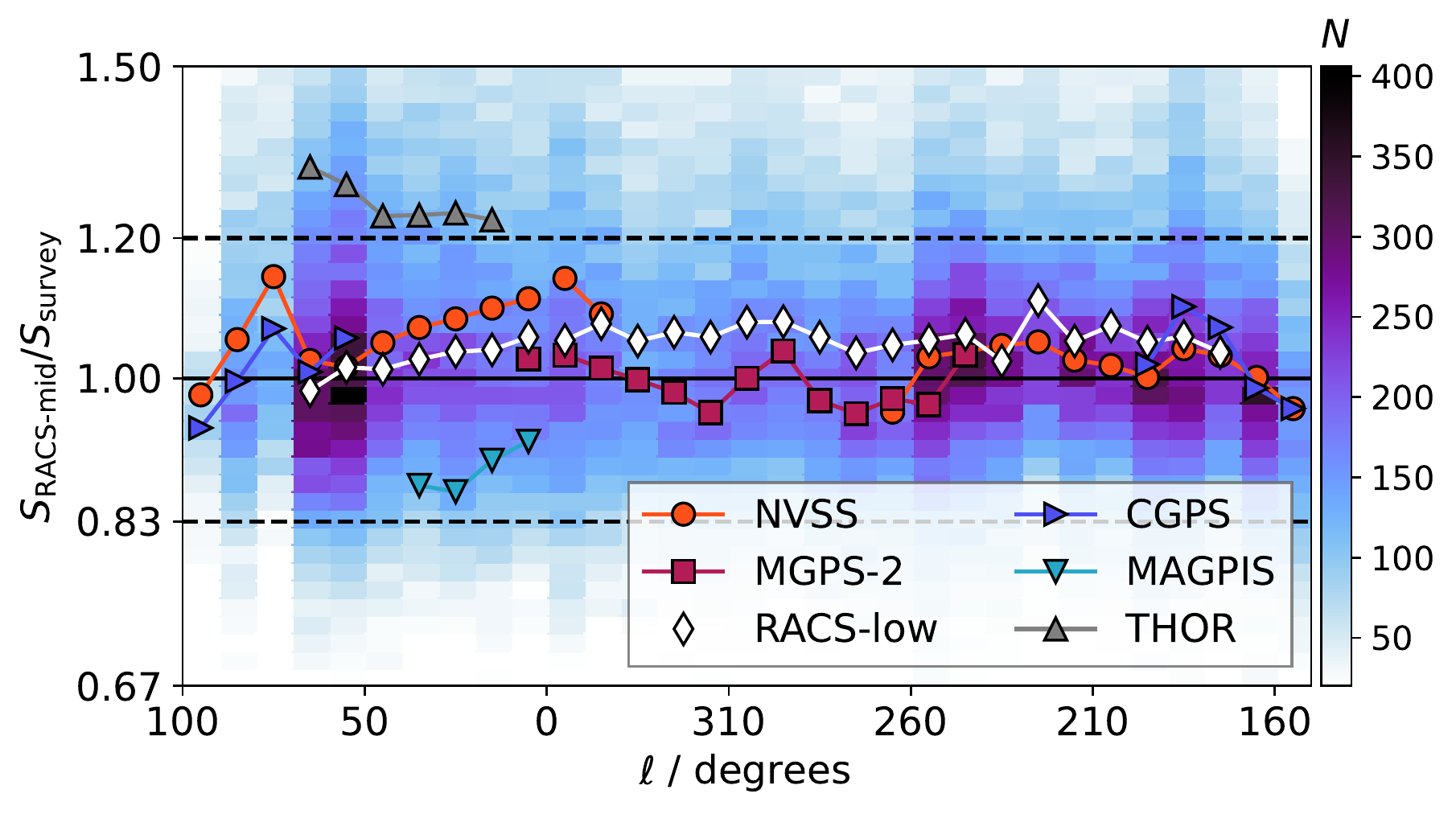}
    \caption{\label{fig:flux:gp} 2-D histogram of flux density ratios in the Galactic Plane within $|b|\leq 5^\circ$ from the six surveys that cover the region. Median values in 10\,deg bins in $l$ are provided for each survey separately.}
\end{figure}

\begin{figure*}
\centering
\begin{subfigure}[b]{1\linewidth}
\includegraphics[width=1\linewidth]{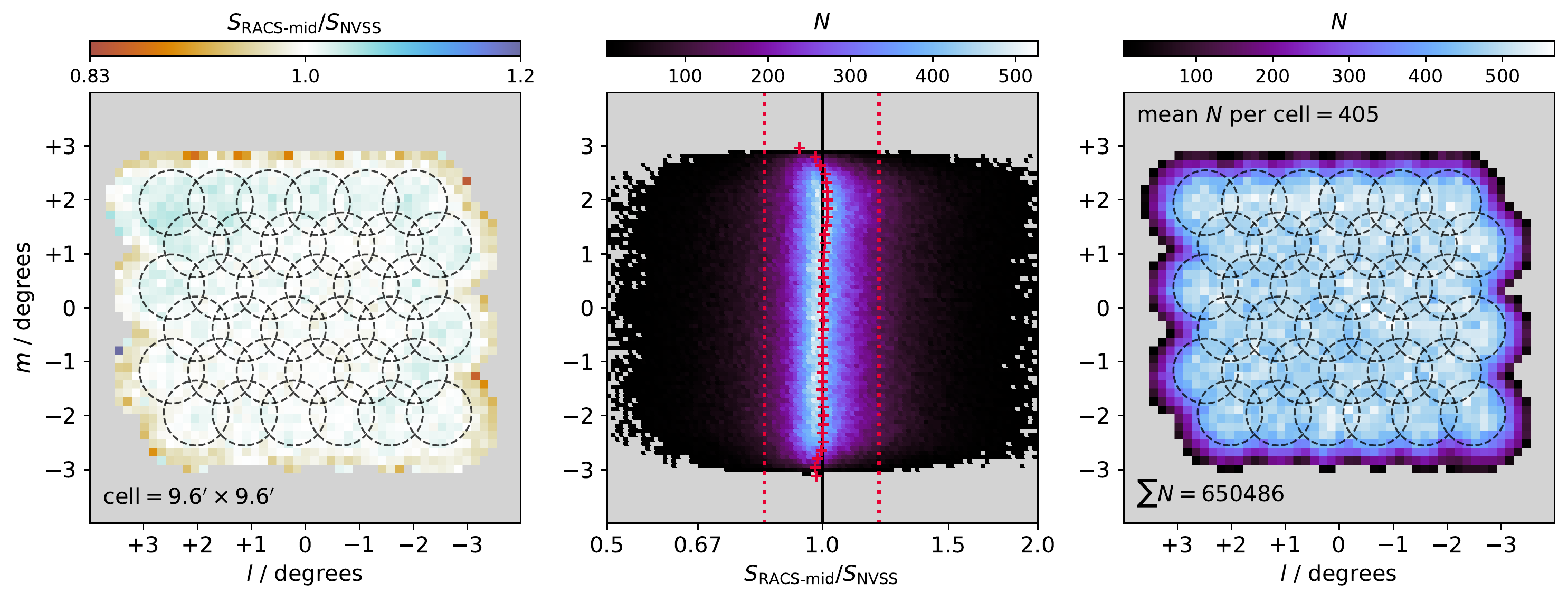}
\caption{\label{fig:tileflux:nvss} Full survey comparison with NVSS.}
\end{subfigure}\\%
\begin{subfigure}[b]{1\linewidth}
\includegraphics[width=1\linewidth]{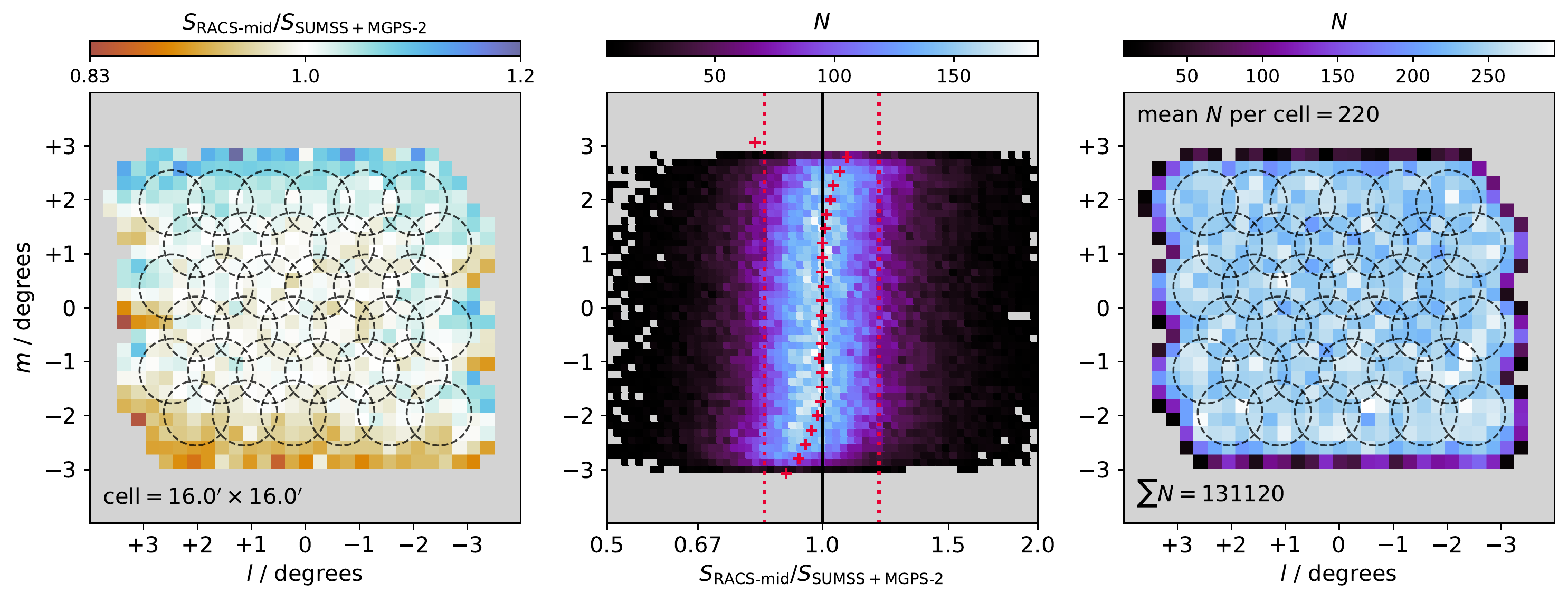}
\caption{\label{fig:tileflux:sumss} Full survey comparison with SUMSS and MGPS-2.}
\end{subfigure}\\%
\begin{subfigure}[b]{1\linewidth}
\includegraphics[width=1\linewidth]{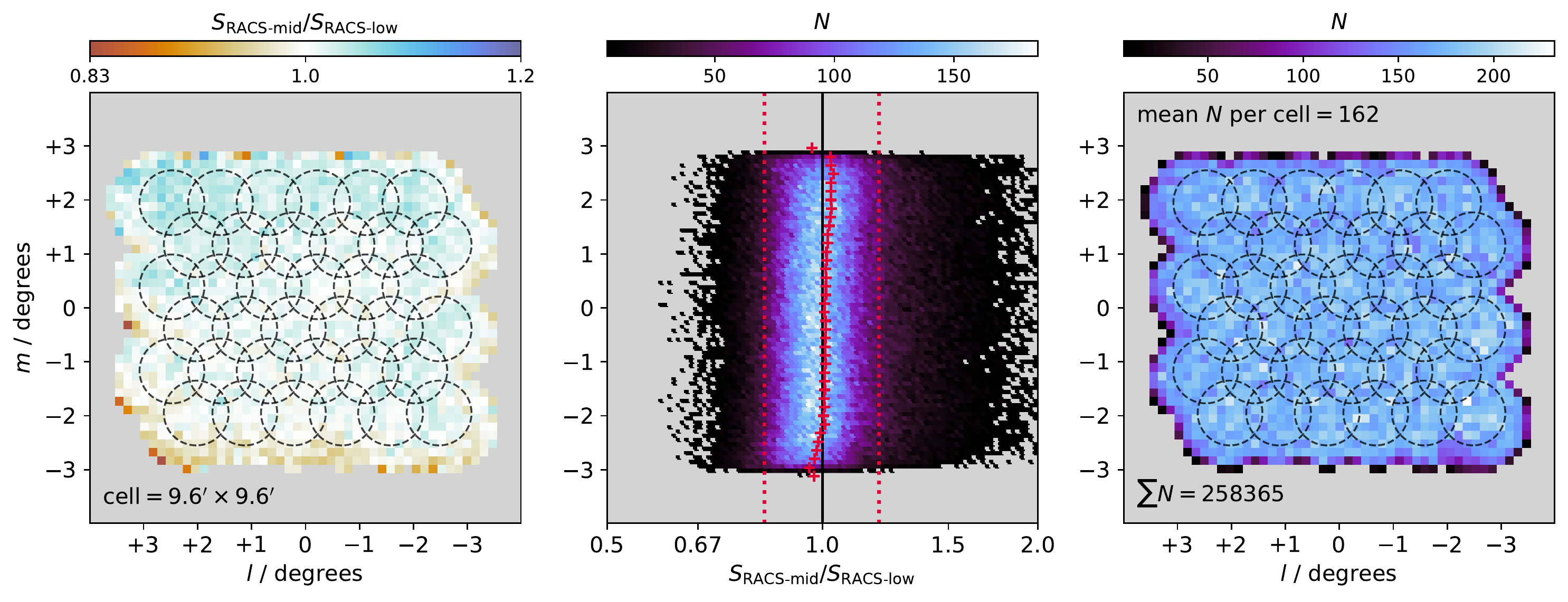}
\caption{\label{fig:tileflux:racs0} Full survey comparison with RACS-low.}
\end{subfigure}%
\caption{\label{fig:tileflux} \corrs{(\emph{Left panels.}) Median tile flux density ratio as a function of position across the observed tile for the full survey where overlap with NVSS \subref{fig:tileflux:nvss}, SUMSS (and MGPS-2) \subref{fig:tileflux:sumss}, and RACS-low \subref{fig:tileflux:racs0} exists. (\emph{Centre panels.}) Number density and flux density ratios in $m$ for the stacked tile. Dotted, red lines indicate ratios of 0.83 and 1.2, and the solid, black lines in dicate ratios of 1. Cross indicate median values across $l$ for each cell in $m$ from the left panels. (\emph{Right panels.}) Source counts in each cell in $l$ and $m$. The dashed, black circles indicate the idealised PAF primary beam positions with 0.6\,deg radius.}}
\end{figure*}

\begin{figure*}
\centering
\begin{subfigure}[b]{0.25\linewidth}
\includegraphics[width=1\linewidth]{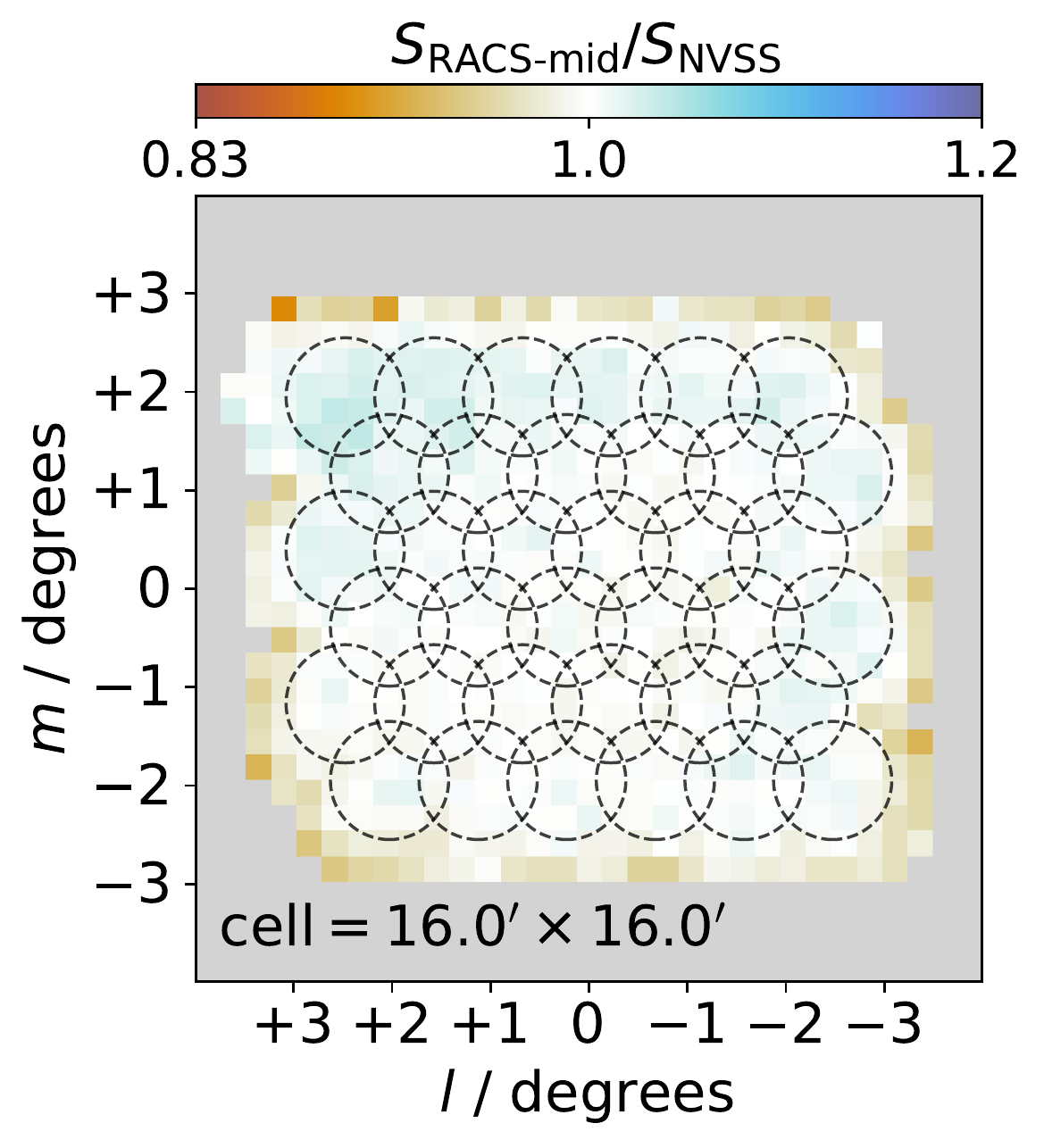}
\caption{\label{fig:tile:zernike:nvss} BWT-1--3\&5 (NVSS).}
\end{subfigure}%
\begin{subfigure}[b]{0.25\linewidth}
\includegraphics[width=1\linewidth]{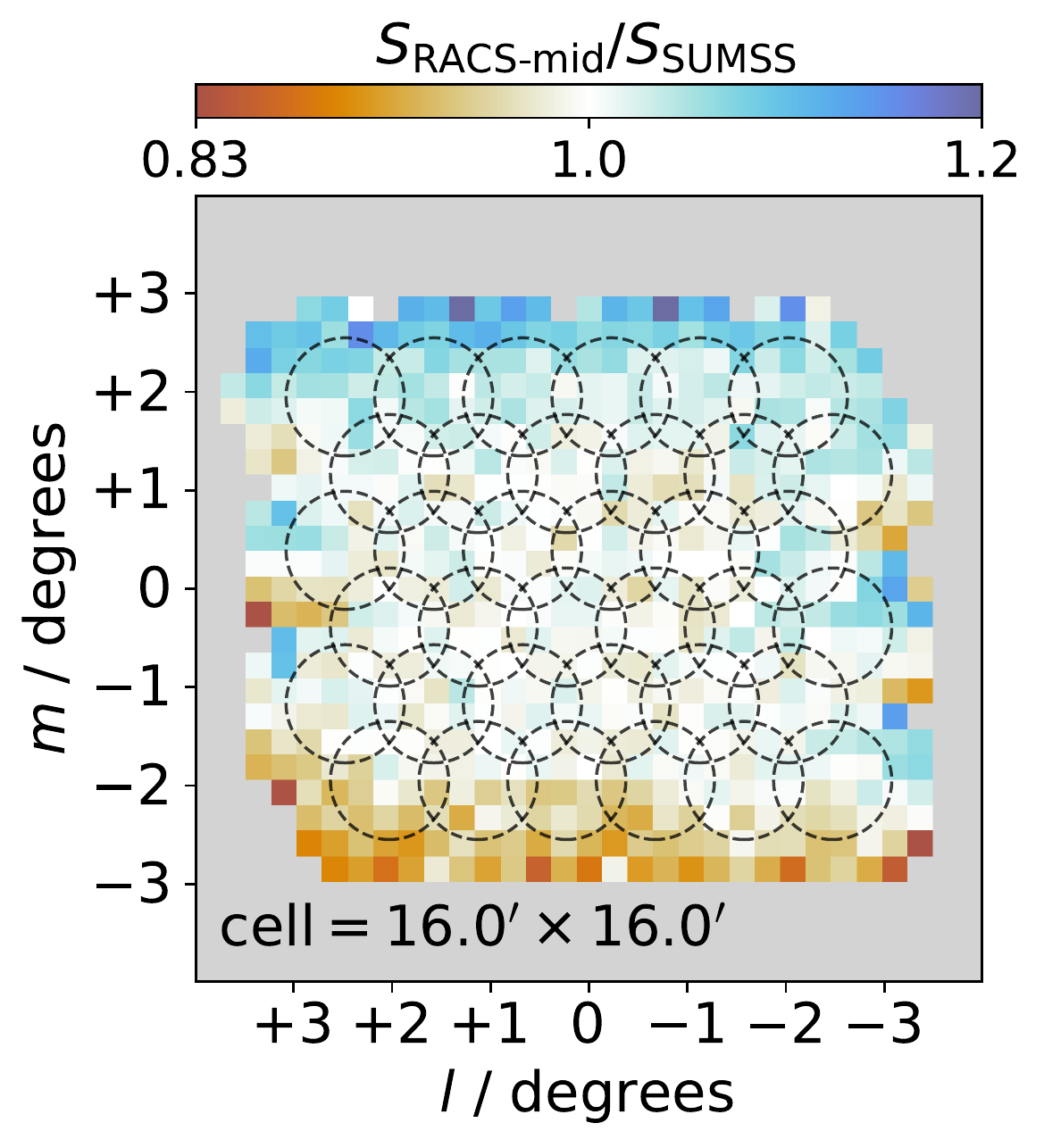}
\caption{\label{fig:tile:zernike:sumss} BWT-1--3\&5 (SUMSS).}
\end{subfigure}%
\begin{subfigure}[b]{0.25\linewidth}
\includegraphics[width=1\linewidth]{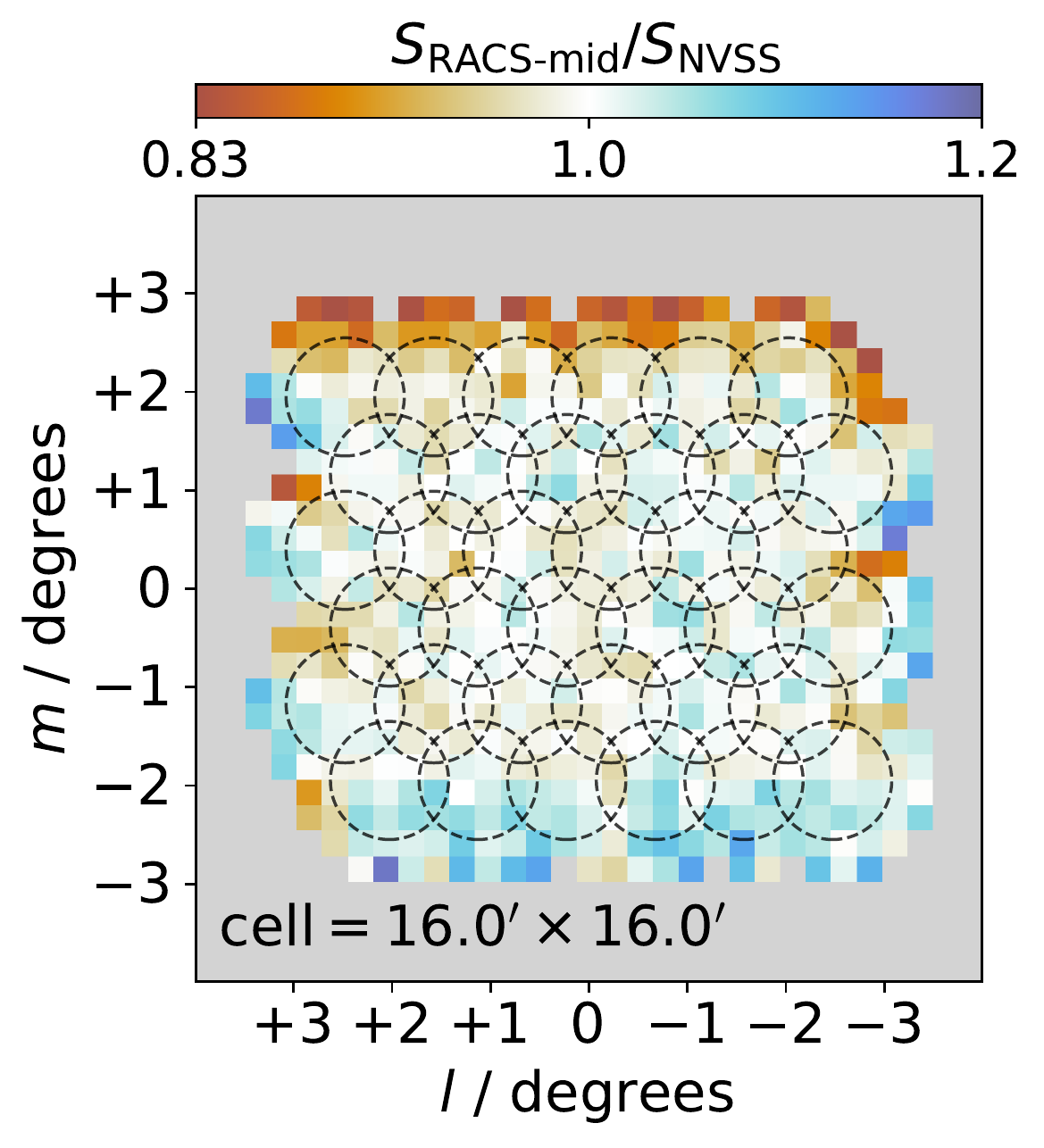}
\caption{\label{fig:tile:holography:nvss} BWT-4\&6--9 (NVSS).}
\end{subfigure}%
\caption{\label{fig:tile:split} \corrs{As in the left panels of Figure~\ref{fig:tileflux} except split into the following subsets: BWT-1--3\&5 cross-matched to NVSS \subref{fig:tile:zernike:nvss} and SUMSS \subref{fig:tile:zernike:sumss}, both using exclusively Zernike model primary beam responses, and BWT-4\&6--9 cross-matched to NVSS \subref{fig:tile:holography:nvss} using the holography approach.}}
\end{figure*}

\begin{figure}[t]
    \centering
    \includegraphics[width=1\linewidth]{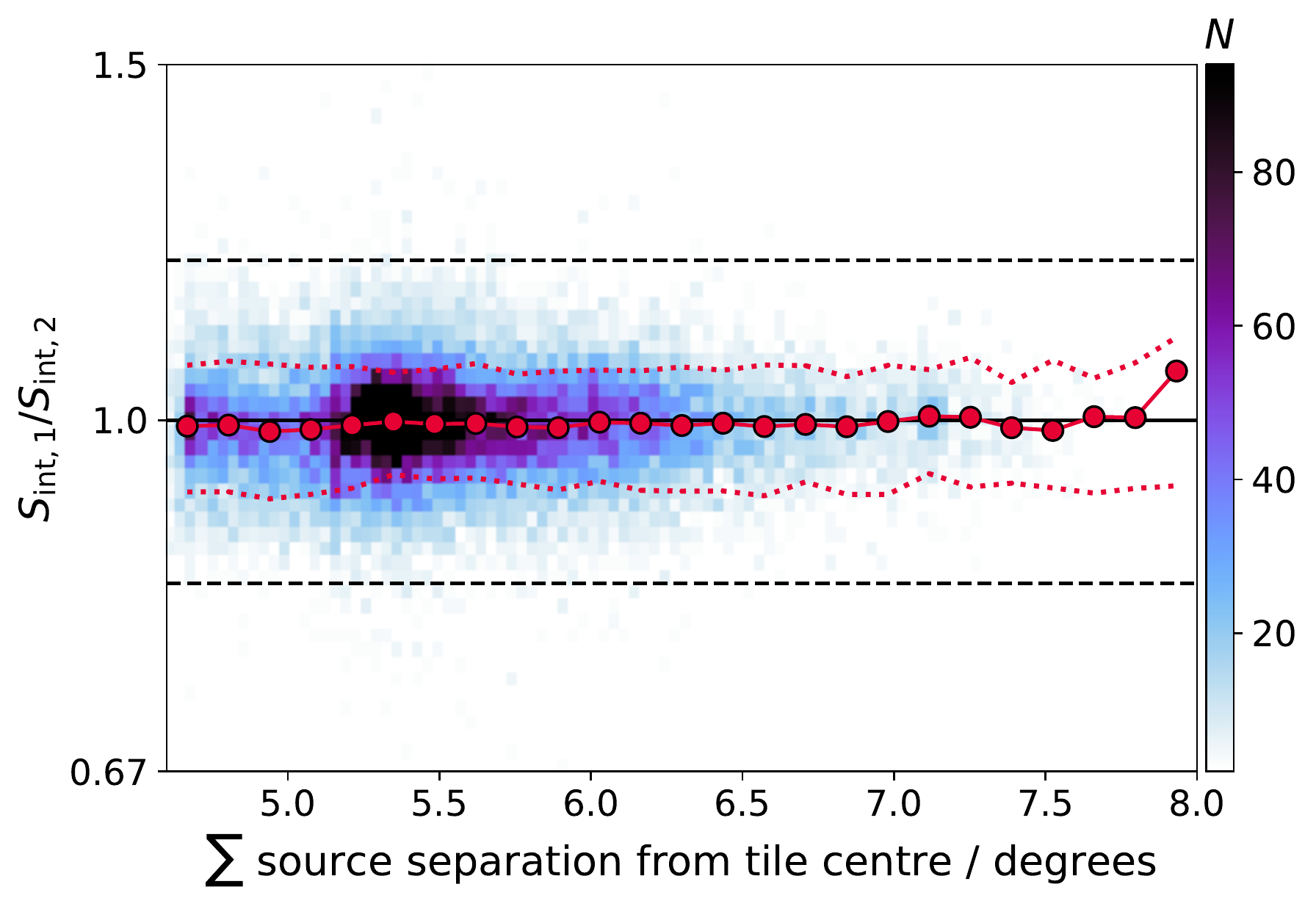}
    \caption{\label{fig:fluxoverlap} \corrs{2-D histogram of flux density ratios of $100\sigma_\text{rms}$ sources in overlapping tiles as a function of the summed separation from their respective tile centres. Medians are calculated in groups of four bins and are indicated by red circles. Corresponding 16$^\text{th}$ and 84$^\text{th}$ percentiles are shown as dashed, red lines. The dashed, black lines indicate ratios 0.83 and 1.2. Note the flux density ratios are shown in log scale.}}
 \end{figure}

\begin{figure}[t]
    \centering
    \includegraphics[width=1\linewidth]{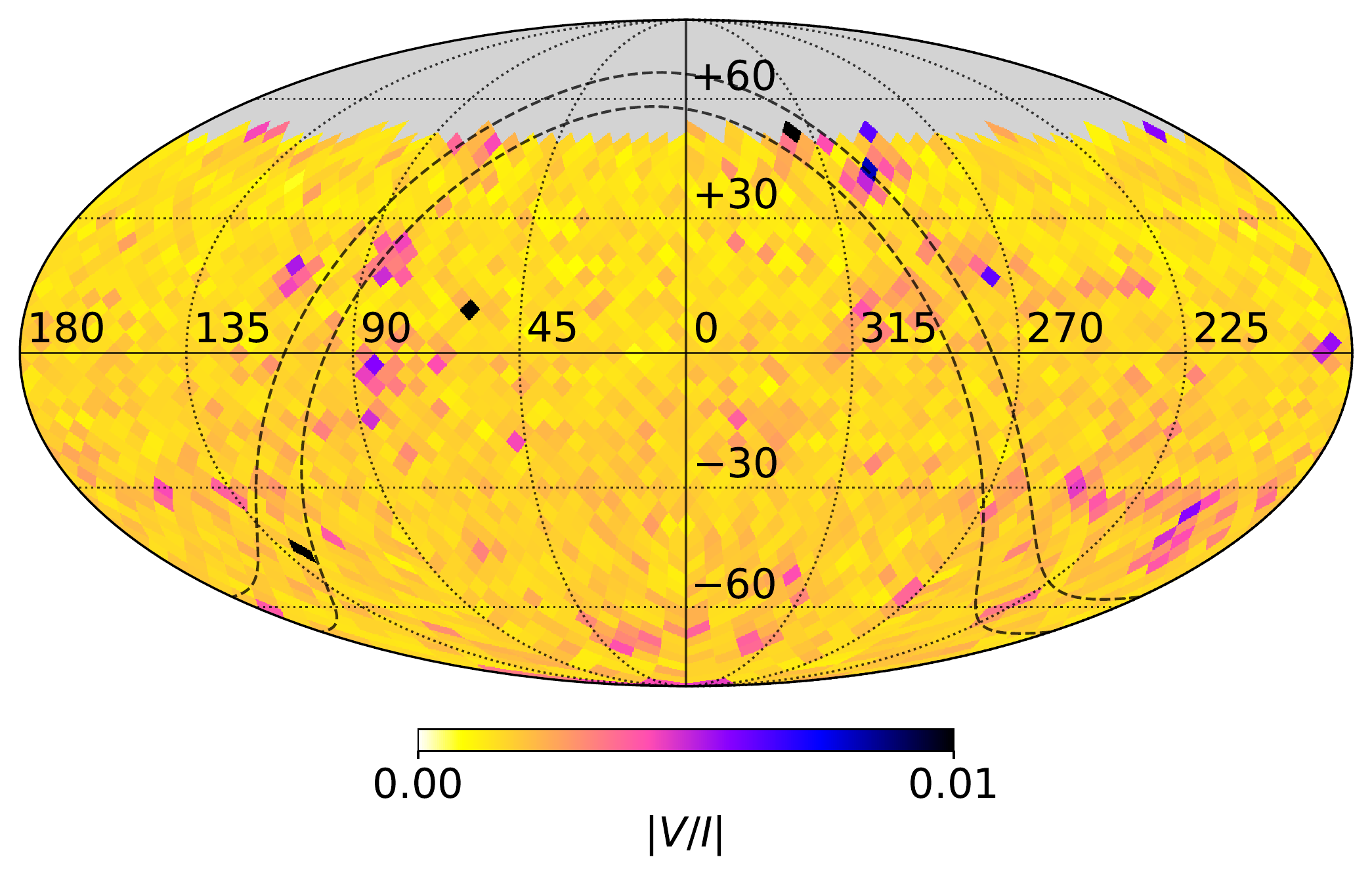}
    \caption{\label{fig:stokesv:allsky} Equatorial representation of the residual leakage, $|V/I|$, in $\sim 13.4$\,deg$^2$ HEALPix bins. Bin values are the \corrs{mean} $|V/I|$ per bin for \corrs{$S_I >500\sigma_{\text{rms},I}$} sources. Galactic latitudes of $|b| < 5^\circ$ are enclosed by the dashed, black lines. \CORRS{The colour scale is linear in the reported range.}}
\end{figure}

\begin{figure}[t]
    \centering
    \includegraphics[width=1\linewidth]{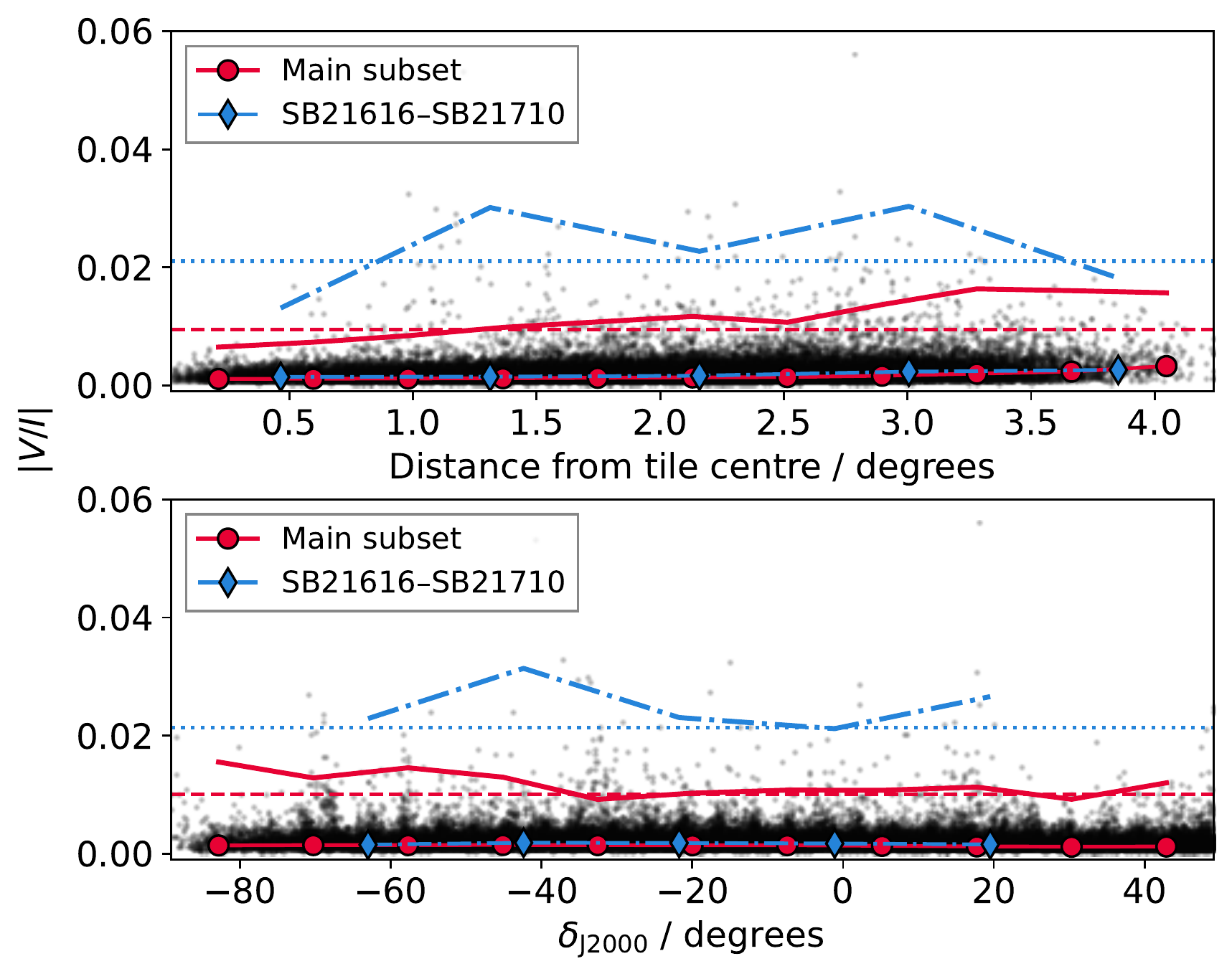}
      \caption{\label{fig:stokesv} Residual Stokes V leakage ($|V/I|$) as a function of distance from tile centres (\textit{top panel}) and declination (\textit{bottom panel}). \corrs{Medians are calculated in eleven equally spaced bins (for the main subset, and 5 bins for the smaller SB21616--SB21710 subset) for both tile centre separation and declination. The red, solid line indicates $3\sigma$ for each bin for the main subset, and the blue, dot-dash line shows the same for the SB21616-SB21710 subset. The red, dashed line shows the overall 3$\sigma$ for the main subset, and the blue, dashed line shows the same for the SB21616--SB21710 subset.}}
\end{figure}

\begin{figure}[t]
    \centering
    \includegraphics[width=1\linewidth]{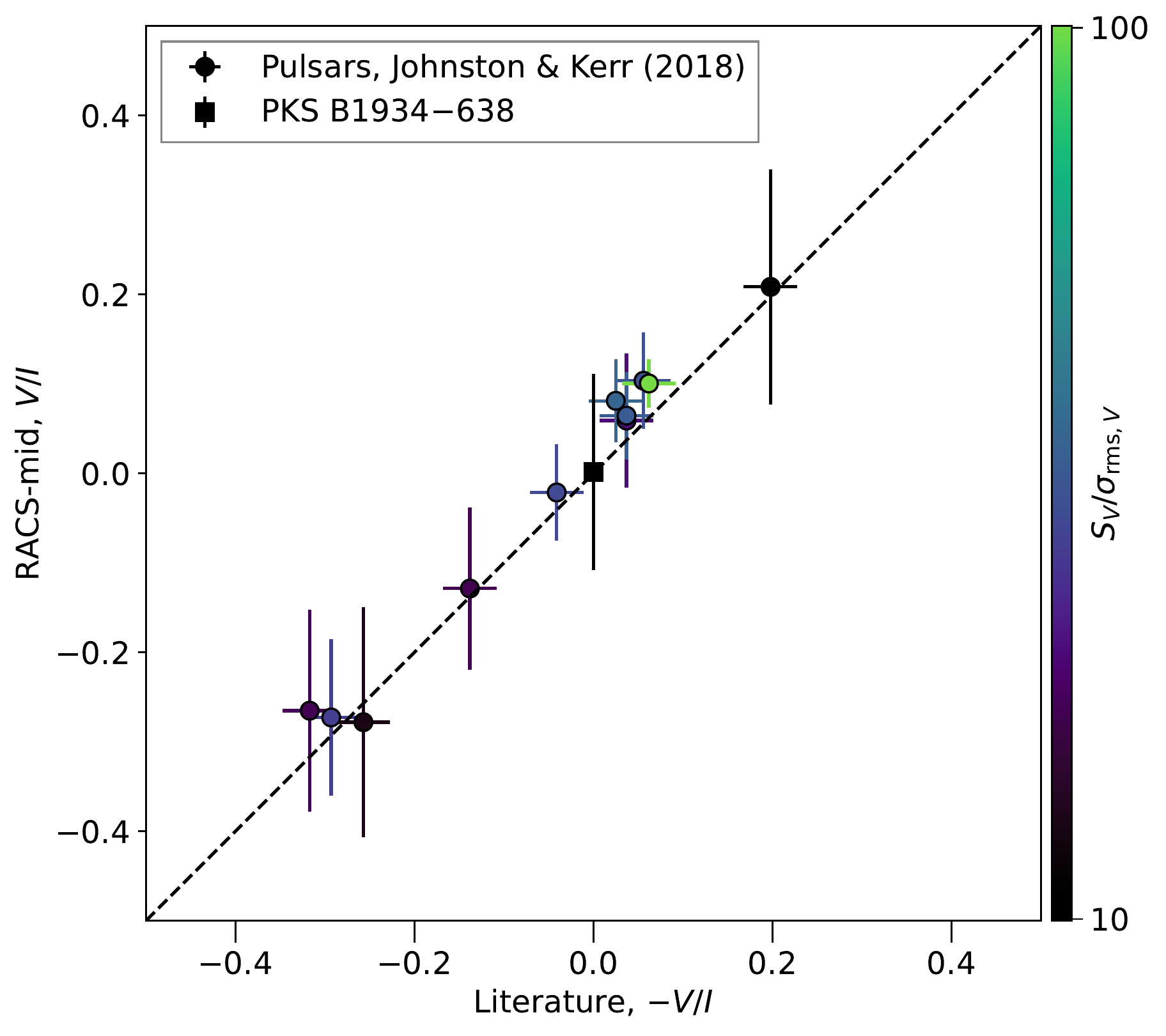}
    \caption{\label{fig:pulsars} \corrs{Comparison of the fractional circular polarization of 11 pulsars detected in both RACS-mid and by \citet{Johnston2018} alongside PKS~B1934$-$638. The sign of Stokes V is reversed for the pulsar catalogue measurement due to the different sign convention. The sources are coloured by their Stokes V SNR on a logarithmic scale.}}
\end{figure}

\corrs{Compared to the first epoch of RACS-low the overall stability of the instrument and some improvements to the observing strategy and data processing have resulted in a more consistent image quality across the RACS-mid survey. Quality metrics for the full survey based on the tile images---such as resolution (Section~\ref{sec:psf}), image noise (Section~\ref{sec:rms}), brightness scaling (Section~\ref{sec:fluxscale} and Section~\ref{sec:validation:stokesv}), and astrometry (Section~\ref{sec:validation:astrometry})---are discussed in detail in the following sections.}

\corrs{Figure~\ref{fig:example:large} shows an example image around $\delta_\text{J2000} \approx -10^\circ$ featuring extended radio galaxies in the galaxy cluster Abell~754 along with  point sources to provide a qualitative idea of the image fidelity for a representative extra-Galactic field. In Figure~\ref{fig:example:eso509} we show the radio galaxy associated with ESO~509$-$G108 as a comparison between RACS-mid and RACS-low. This extended radio galaxy features a well-detected compact core embedded within a  diffuse cocoon of emission with a $\sim$6\,arcmin angular extent. In Figure~\ref{fig:example:eso509} we  also show a comparison between RACS-mid and RACS-low when both are convolved to a common 25~arcsec angular resolution (as in \citetalias{racs2}), highlighting the difference in sensitivity to the diffuse, extended emission independent of a change in angular resolution. Additionally, in Figure~\ref{fig:example:j1034} we show another radio galaxy with both compact features and $\sim 1$--2\,arcmin diffuse components. This time the comparison is with NVSS and an image from the Faint Images of the Radio Sky at Twenty centimetres \citep[FIRST;][]{first,wbhg97,hwb15}, highlighting RACS-mid as a midpoint between NVSS and FIRST in terms of resolution, while retaining sensitivity to arcmin-scale structures.}

\subsection{Point-spread function}\label{sec:psf}

As each individual beam image is convolved to a common resolution for each SBID, \corrs{each individual tile image has} a \corrs{separate} position-independent PSF. This PSF does vary over the full survey region. \corrs{Figure~\ref{fig:psf} shows the positional variation for both the PSF major and minor axes over the sky. Pointings with multiple observations have their minimum PSF major/minor FWHM shown}. \corrs{Figure~\ref{fig:psfcoords} shows the PSF major and minor axes binned as a function of declination, showing both the median and minimum binned PSF FWHM. As in Figure~\ref{fig:psf}, binned values take the minimum values for duplicate observations.} The median PSF major axis is \medpsfmajor\,arcsec, but does range from $\minpsfmajor$--$\maxpsfmajor$\,arcsec, increasing at high and low declination where the low elevation observations result in a compressed projected $(u,v)$ plane. The PSF minor axis does not vary as much except at the SCP where it increases along with the major axis. In the northern-most declination strip, the PSF major axis reaches up to \maxpsfmajor\,arcsec where ASKAP's antennas are almost at their elevation limit of $15^\circ$ \citep{Hotan2021}. Conversely, the most southerly pointings only go down to $\sim 30^\circ$ in elevation. The choice to observe within $\pm 1$\,h of the meridian has limited the PSF major axis from varying as much as the first epoch of RACS-low. Additional variation beyond the observed declination (i.e.~elevation) dependence is largely due to the occasional offline antenna and otherwise flagged data.

\subsection{{Image noise and artefacts}}\label{sec:rms}

The full survey has a median \corrs{Stokes I} rms noise  of \rmserror\,\textmu Jy\,PSF$^{-1}$, calculated as the median from individual tile median values. \corrs{For Stokes V, the median noise is lower at \rmserrorv\,\textmu Jy\,PSF$^{-1}$.} This includes tile edges where noise increases significantly due to the primary beam de-attenuation. The median minimum rms \corrs{Stokes I} noise per tile is \rmsminerror\,\textmu Jy\,PSF$^{-1}$. The rms \corrs{Stokes I} noise across the full survey is shown in Figure~\ref{fig:fullskyrms}. There is an increase in noise near the Galactic Plane and LMC, bright (typically extended sources), and north of the celestial equator. While peeling (as described in Section~\ref{sec:peeling}) reduced the artefacts and noise around bright sources, the bright sources are not removed from beam images where they are within 1.2\,deg of a pointing centre. Therefore significant artefacts tend to remain within this aperture. This is particularly noticeable around Cygnus~A (labelled on Figure~\ref{fig:fullskyrms}); a particularly bright double radio source with reasonably small angular separation from the Galactic Plane. This creates further challenges in both peeling and general imaging in that region. Figure~\ref{fig:fullskyrms} also highlights some observations with a slightly reduced observation time which appear as tiles with higher non-local rms noise. Figure~\ref{fig:rmshist} shows the rms noise \corrs{for Stokes I and V} as a function of declination, highlighting the increase towards the equator and at the SCP. Extremes at the edges of the survey declination range are from both an increase in PSF size (see Section \ref{sec:psf}) and a decrease in $(u,v)$ coverage for the low-elevation pointings.

\corrs{While the noise properties are generally less variable than in RACS-low, RACS-mid does still suffer from artefacts around sources at the $\lesssim 1$\% level. While these artefacts are faint and decrease in magnitude away from sources, they can be seen with the colourmap scaling used in Figure~\ref{fig:example:large}. Such artefacts are kept at the $\lesssim 1$\% level throughout RACS-mid, whereas there is more variation in RACS-low.}

\subsection{{Stokes I} brightness scale}\label{sec:fluxscale}

\subsubsection{Absolute flux density scale}\label{sec:absflux}
As an initial check of the absolute flux density scale, we follow \citetalias{racs1} by comparing RACS-mid flux density measurements of a selection of bright calibrator sources modelled by \citet{Perley2017}. We also check the flux density of PKS~B1934$-$638 by comparison to the model reported by \citet{Reynolds1994}. For this purpose, we integrate the flux density above $2\sigma_\text{rms}$ directly from the maps within apertures that cover twice the angular extent reported by \citet{Perley2017}---assuming 0.1\,arcsec extent for PKS~B1934$-$638---convolved with the Gaussian restoring PSF of the images they are measured from. Figure~\ref{fig:pbscale} shows the RACS-mid flux density measurements against the model flux density. All source measurements lie within $\sim 6\%$ of model values except for extended sources. Typically the more extended the source is, the less agreement between measurement and model. This is particularly significant for Fornax~A and Taurus~A (enclosed by black diamonds in Figure~\ref{fig:pbscale}), though in these cases the additional 75\,m $(u,v)$ cut reduces the measurement in RACS-mid further (see Section~\ref{sec:gp} and Section~\ref{sec:angularscales}).

\subsubsection{The brightness scale across the sky}\label{sec:fluxscale:eg}

New source-lists are generated by \texttt{selavy} after the tile-dependent brightness scaling described in Section~\ref{sec:tilescale}. These per-SBID source-lists are used in standard pipeline processing for per-SBID validation and are also used for full-survey validation focused on flux densities/brightness scaling and astrometry (Section~\ref{sec:validation:astrometry}). Source-finding is performed to a limit of $5\sigma$ in the individual tile images. These source lists are cross-matched to various catalogues: \corrs{the NVSS; the Sydney University Molonglo Sky Survey \citep[SUMSS;][]{bls99,mmb+03}; the TIFR \footnote{Tata Insititute for Fundamental Research.} GMRT \footnote{Giant Metrewave Radio Telescope.} Sky Survey \citep[alternate data release 1;][]{ijmf16}; and RACS-low; and} sources in the third realisation of the International Celestial Reference Frame \citep[ICRF3;][]{Charlot2020}. Despite similarity in frequency to FIRST, we do not use FIRST for flux density comparisons as FIRST was observed over different frequency ranges for different parts of the survey due to the upgrade to the VLA \citep{hwb15}. Comparisons with NVSS, SUMSS, and RACS-low are useful in determining the quality (or consistency) of the final brightness scaling, particularly as NVSS and SUMSS catalogues have already been used extensively in deriving the primary beam response for a majority of the observations (Section~\ref{sec:holography}) and general brightness scaling per tile (Section~\ref{sec:tilescale}).

Figure~\ref{fig:tileflux:hist:alpha} shows the histogram of spectral indices derived from comparisons with SUMSS, RACS-low, and TGSS (excluding the Galactic Plane where source spectra may not reflect extragalactic source properties). The median spectral indices across the survey region are found to be $\alpha_\text{TGSS}^\text{RACS-mid} = -0.69_{-0.22}^{+0.25}$, $\alpha_\text{SUMSS}^\text{RACS-mid} = -0.81_{-0.42}^{+0.50}$, and $\alpha_\text{RACS-low}^\text{RACS-mid} = -0.88_{-0.28}^{+0.39}$, with \corrs{uncertainties} derived from the 16$^\text{th}$ and 84$^\text{th}$ percentiles of each distribution. For SUMSS, this is consistent with the spectral index used for SBID-dependent scaling earlier. For RACS-low, this is consistent to the spectral index between RACS-low and NVSS reported in \citetalias{racs2}. The TGSS comparison results in a spectral index representative of flatter spectra and is also consistent with similar comparisons in \citetalias{racs2} for RACS-low. \corrs{Similarly, the median spectral index when matching to TGSS is the same median in the NVSS--TGSS spectral index catalogue constructed by \citet{deGasperin2018}: $\alpha_\text{TGSS}^\text{NVSS} = -0.70_{-0.21}^{+0.26}$.} Direct comparisons of the flux densities for NVSS, SUMSS, and RACS-low over the full survey (excluding the Galactic Plane) are shown in Figure~\ref{fig:tileflux:hist:flux}. These comparisons assume $\alpha=-0.82$ for NVSS and SUMSS (as in Section~\ref{sec:tilescale}), and $\alpha=-0.88$ for RACS-low. As expected, due to SBID-dependent scaling, NVSS and SUMSS comparisons show a median ratio of 1: $S_\text{RACS-mid}/S_\text{NVSS} = 1.00^{+0.20}_{-0.14}$,  $S_\text{RACS-mid}/S_\text{SUMSS} = 1.00^{+0.27}_{-0.18}$, and $S_\text{RACS-mid}/S_\text{RACS-low} = 1.00^{+0.18}_{-0.11}$, with \corrs{uncertainties} estimated as for $\alpha$.

Figure~\ref{fig:fluxscale:sky} shows the median brightness scaling over the whole sky \corrs{with comparison to the NVSS, SUMSS with the second epoch Molonglo Galactic Plane Survey \citep[MGPS-2;][]{mmg+07}, and RACS-low}. While there is general agreement across the sky (by construction due to the use of NVSS and SUMSS in SBID-dependent brightness scaling), variation exists, particularly in the Galactic Plane (see Section~\ref{sec:fluxscale:gp}). The comparison with RACS-low [Figure~\ref{fig:fluxscale:sky:racs0}] also shows the same position-dependent variation with respect to NVSS and SUMSS revealed in \citetalias{racs1} and \citetalias{racs2}. \corrs{Likely the positional variation between RACS-mid and RACS-low is a result of uncorrected per-observation brightness scale variations reported in \citetalias{racs1}.} 

\subsubsection{The brightness scale in the Galactic Plane}\label{sec:fluxscale:gp}
In the Galactic Plane ($|b| \lesssim 5^\circ$), RACS-mid generally reports a higher flux density than expected compared to NVSS [Figure~\ref{fig:fluxscale:sky:nvss}], with $S_\text{RACS-mid}/S_\text{NVSS,GP} = 1.04_{-0.16}^{+0.28}$. A similar increase, though smaller in magnitude, is seen in comparison to RACS-low [Figure~\ref{fig:fluxscale:sky:racs0}] with $S_\text{RACS-mid}/S_\text{RACS-low} = 1.01_{-0.18}^{+0.33}$. While the Galactic Plane tiles are scaled in brightness according to the elevation-dependent fit shown in Figure~\ref{fig:fluxelevation}, a separate time-dependent effect remains uncorrected, as described in Section~\ref{sec:tilescale}. With reference to the NVSS, the regions around Cygnus A and between $l \gtrsim 350^\circ$ and $l \lesssim  55^\circ$ show an increase in brightness of order 15--25\% which is beyond what is seen in Figure~\ref{fig:timeflux} for non-Galactic Plane SBIDs. Artefacts from the bright, extended sources in the Galactic Plane (and Cygnus~A) may also influence the flux density measurements where they overlap with real sources.
 
In addition to the main all-sky catalogues noted in Section~\ref{sec:fluxscale:eg}, we also cross-match the RACS-mid source-lists with a collection of smaller-area Galactic Plane surveys: MGPS-2, THOR~\footnote{The H\textsc{i}/O\textsc{h}/Recombination line survey of the Milky Way.} \citep{Beuther2016:thor,Wang2020:thor}, MAGPIS~\footnote{Multi-Array Galactic Plane Imaging Survey.} \citep{White2005:magpis,Helfand2006:magpis}, and the CGPS~\footnote{Canadian Galactic Plane Survey.} \citep{Taylor2003:cgps} where overlap exists with RACS-mid.  The MGPS-2 comparison is shown alongside SUMSS in Figure~\ref{fig:fluxscale:sky:sumss} and does not show the same increase in brightness seen with either RACS-low or NVSS comparisons, with a marginally lower median $S_\text{RACS-mid}/S_\text{MGPS-2} = 0.98_{-0.21}^{+0.29}$. Figure~\ref{fig:flux:gp} shows the binned flux density ratios for sources within $|b|\leq 5^\circ$ from the Galactic Plane surveys as a function of $l$ along with NVSS and RACS-low. We note the overall median flux density ratios as $S_\text{RACS-mid}/S_\text{THOR} = 1.26_{-0.24}^{+0.30}$ (5231 sources), $S_\text{RACS-mid}/S_\text{MAGPIS} = 0.88_{-0.15}^{+0.20}$ (1481 sources), and $S_\text{RACS-mid}/S_\text{CGPS} = 1.02_{-0.20}^{+0.32}$ (14246 sources).

As only compact sources are used for comparison, it is unlikely RACS-mid recovers a larger fraction of flux density for the extended radio sources in the Galactic Plane. RACS-low showed a similar increase in flux density with respect to NVSS in the Galactic Plane, and it is clear from comparison with smaller Galactic Plane surveys there is large variation within this complex region of the sky. 

\subsubsection{Accuracy of the primary beam corrections}

To confirm the accuracy and consistency of the primary beam corrections \corrs{after mosaicking}, we also show the median flux density ratios as a function of position over the PAF footprint for the tile, binned in $(l,m)$. Figure~\ref{fig:tileflux} shows the comparison for NVSS [\ref{fig:tileflux:nvss}], SUMSS [\ref{fig:tileflux:sumss}], and RACS-low [\ref{fig:tileflux:racs0}] for the full RACS-mid survey (where overlaps exists with the relevant comparison survey). We find residual errors on the edges of the tiles due to errors in the individual beams which are reduced in the centre of the tile. Additional errors are seen, particularly for southern fields and those in BWT-4 using holographic primary beam correction. In Figure~\ref{fig:tile:split} we show the same median tile flux density scale split into three subsets: BWT-1--3\&5 cross-matched to NVSS [\ref{fig:tile:zernike:nvss}] and SUMSS [\ref{fig:tile:zernike:sumss}], both using exclusively Zernike model primary beam responses, and BWT-4\&6--9 cross-matched to NVSS [\ref{fig:tile:holography:nvss}] using the holography approach. This highlights that the edge effect seen in Figure~\ref{fig:tileflux:sumss} is present for both the Zernike and holography subsets, though the residual flux density ratio is flipped N-S. The error appears to increase away from the location where the primary beam correction is defined. As the Zernike models are defined with reference to the NVSS in the range $-40^\circ < \delta_\text{J2000} < +48^\circ$, they would be most applicable to the celestial equator. For holography, PKS~B0407$-$658 is used as the target source and $\delta_\text{J2000} \approx -65^\circ$ is the reference location. Due to the limited holography availability for RACS-mid, future investigation with RACS data is reserved for work with RACS-high and the second epoch of RACS-low, both of which have been observed alongside corresponding holographic measurements. In a future paper, when cataloguing, we will be linearly mosaicking nearby tiles as in \citetalias{racs2} which reduces this error for most tiles. With the improved data quality and consistency, RACS-mid and other recent ASKAP imaging is becoming sensitive to higher-order effects, such as antenna pointing errors, which are yet to be fully analysed and corrected.

\corrs{As a check of the internal consistency of the primary beam models after application during linear mosaicking, we compare the flux densities of sources detected in two tiles (in overlapping regions between tile mosaics).  Sources are selected by cross-matching individual source-lists and considering sources with cross-matches within half their reported size. We restrict this to compact ($S_\text{int} / S_\text{peak} < 1.2$) sources only. This process is similar to the per-beam comparisons shown in Section~\ref{sec:beamvalidation:i} except cross-matches occur between observations and we only consider sources once. Figure~\ref{fig:fluxoverlap} shows a 2-D histogram of the integrated flux density ratios ($S_\text{int,1} / S_\text{int,2}$) for $100\sigma_\text{rms}$ sources present in two tiles, plotted as a function of the sum of the source separations from their respective tile centres. We also show medians along with 16$^{\text{th}}$ and 84$^{\text{th}}$ percentiles in $\sim 8$\,arcmin bins. While some features appear in the histogram, these are due to larger densities of sources at specific separation distances and the reported summary statistics do not vary as a function of distance from the tile centre in these overlap regions. The overall median ratio is $0.99_{-0.07}^{+0.07}$. }

\subsubsection{Brightness scale uncertainty}
We find that despite tile edge effects, the uncertainty \corrs{in the brightness scale} for each BWT, as a function of declination, \corrs{and internally} is reasonably consistent. Therefore to derive a brightness scale uncertainty for RACS-mid, we focus on the bulk NVSS-derived flux density ratios for $100\sigma$ sources to reduce contribution from frequency differences and low-SNR measurement errors. As an estimate of this uncertainty, we use the maximum between the 16$^\text{th}$ and $84^\text{th}$ percentile of the $S_\text{RACS-mid} / S_\text{NVSS}$ ratio to avoid being overly conservative within the fairly well-modelled tiles (within the central $6\,\text{deg} \times 5\,\text{deg}$). This yields a brightness scale uncertainty of 6\%, consistent with scatter in measurements of bright calibrator sources (Figure~\ref{fig:pbscale}). The exceptions to this are the edges of the PAF footprint (within $\sim 0.5$\,deg of the footprint boundary). In this exterior region, we follow \citetalias{racs1} and estimate the uncertainty by cross-matching compact~\footnote{$0.8 < S_\text{int} / S_\text{peak} < 1.2$.} sources that are detected in adjacent tiles outside of the central $6\,\text{deg} \times 5\,\text{deg}$. This provides a median integrated flux density ratio of $1.00_{-0.12}^{+0.13}$. Adding the tile exterior uncertainty in quadrature with the NVSS-based uncertainty we estimate a brightness scale uncertainty, $\xi_\text{RACS-mid}$, of 
\begin{equation}
    \xi_\text{RACS-mid}(l,m) = \begin{cases}
        0.06 S_\text{RACS-mid}, & \text{if}~(|l| < 3^\circ, |m| < 2.5^\circ) \\
        0.14 S_\text{RACS-mid}, & \text{otherwise}
    \end{cases}
\end{equation}
as a function of $(l,m)$ across a given tile for all SBIDs. \corrs{These externally-estimated uncertainties are consistent with the internal accuracy of the primary beam discussed in the previous section and Section~\ref{sec:beamvalidation:i}.}

\subsection{{Stokes V}}\label{sec:validation:stokesv}

\subsubsection{{Stokes V residual leakage}}\label{sec:validation:stokesv:leakage}

\corrs{To assess the residual leakage of Stokes I in to V, we extract residual leakage using the \texttt{selavy} Stokes I source-lists.} In the Stokes V images we take the absolute brightest pixel within a 9-pixel box around the Stokes I position, avoiding sources near beam edges which may be reduced in size in either the Stokes I or V images. Residual leakage tables for each SBID are stored in the RACS database \corrs{as an additional source list for use in data validation}. Figure~\ref{fig:stokesv:allsky} shows the \corrs{mean} $|V/I|$ for sources with $S_I >500\sigma_{\text{rms},I}$ in HEALPix bins across the sky. 

Figure~\ref{fig:stokesv} shows the residual $|V/I|$ leakage as a function of distance from the tile centre and declination \corrs{ for the same sample of sources.} \corrs{In Figure~\ref{fig:stokesv} we also show the binned median $|V/I|$ split between the SB21616--SB21710 subset, which is corrected using a single-day leakage model (see Section~\ref{sec:stokesv}) and the remaining SBIDs. The median $|V/I|$ is $0.0014_{-0.0007}^{+0.0012}$ for the main subset and $0.0017_{-0.0009}^{+0.0022}$ for the SB21616--SB21710 subset. For the main subset, we estimate the residual leakage from $3\sigma$ of the $|V/I|$ distribution: $|V/I| \lesssim 0.009$. Figure~\ref{fig:stokesv} shows higher residual leakage in the SB21616--SB21710 subset, and we estimate the residual leakage for this subset separately as $|V/I| \lesssim 0.024$. These are shown on Figure~\ref{fig:stokesv}, along with the binned $3\sigma$ as a function of tile centre separation and declination. The higher residual leakage found for the SB21616--SB21710 subset explains the spurious sources noted Section~\ref{sec:beamvalidation:v}, including the sign-changing source NVSS~J205111$+$081859 which has $|V/I| \approx 0.016$ in the mosaic of SB21634.} 

\subsubsection{{Comparison of bright Stokes V sources with the literature}}

\corrs{Sources with high fractional circular polarization are generally associated with stars, including pulsars \citep[e.g.][]{Lenc2018,Pritchard2021}. We create a subset of the Stokes V sources from aforementioned leakage tables by extracting sources detected at both $10\sigma_{\text{rms,}I}$ and $10\sigma_{\text{rms,}V}$. From these 1\,011 sources, we inspect the top twenty with the highest fractional polarization. We find that five correspond to either extended/double radio sources with mismatched peak Stokes V peak flux density measurements that are normally removed in isolated/compact source cuts. The remaining fifteen are associated with stars (and pulsars) or are otherwise related to stellar systems. This includes the  magnetic chemically peculiar star CU~Vir detected with $V/I \sim 0.66$, which is known to have pulsed circularly polarized radio emission \citep[e.g.][]{Lo2012} and has been previously detected by RACS-low \citep{Pritchard2021}.}

\corrs{As many circularly polarized sources are variable, to assess the absolute flux scale and sign of the Stokes V data we also cross-match the $>10\sigma_{\text{rms},V}$ Stokes V sources with the pulsar catalogue reported by \citet{Johnston2018}. This catalogue is reported at 1\,400~MHz with measurements integrated over pulse profiles. With a 2~arcsec cross-match (with no proper motion correction), we find fourteen pulsars from the ATNF pulsar catalogue \footnote{\url{https://www.atnf.csiro.au/research/pulsar/psrcat/}.}, of which eleven are cross-matched with the \citet{Johnston2018} catalogue. We re-measure the Stokes I and V flux densities within apertures as in Section~\ref{sec:absflux} instead of relying on the peak flux measurement as in the leakage tables. Figure~\ref{fig:pulsars} shows the RACS-mid $V/I$ compared to the \citet{Johnston2018} $V/I$ after a sign flip (to account for a difference in sign convention for the pulsar catalogue) coloured by their Stokes V SNR. There is agreement with the signs for each pulsar with $|V/I|$ typically slightly larger in RACS-mid. The ratio of RACS-mid fractional polarization to the pulsar catalogue values has a median of $1.08_{-0.19}^{+0.70}$. We also show $V/I$ for PKS~B1934$-$638 for reference, assuming zero circular polarization, which is used for on-axis leakage correction prior to widefield corrections.}

\corrs{No sources from the validation files that satisfy $10\sigma_{\text{rms,}I}$ and $10\sigma_{\text{rms,}V}$ are cross-matched with the circularly polarized LOFAR Two-metre Sky Survey detections \citep[V-LoTSS;][]{Callingham2022} despite some overlap in the the respective survey areas. More in-depth comparisons will be made in future Stokes V cataloguing work.}

\subsection{Recovery of large angular scales}\label{sec:angularscales}

RACS-low could recover significant signal from angular scales up $\sim 10$\,arcmin, depending on brightness. RACS-mid, at higher frequency and with less bandwidth is overall less sensitive to large angular scales (see Section~\ref{sec:gp}). Figure~\ref{fig:example:eso509} shows an example radio galaxy with $\sim 6$\,arcmin extent in comparison to RACS-low, the NVSS, and the TGSS images. The higher angular resolution naturally `resolves out' some of the extended emission, though the  most significant contributing factor to lower sensitivity to large angular scales is the difference in $(u,v)$ coverage for the mid-band data (Figure~\ref{fig:uvcover}\label{figuse:uvcover:2}).

Additionally, Figure~\ref{fig:pbscale}, which shows the difference in measured and model flux density for bright calibrator sources, also shows some of the brightest extended radio sources in the sky. For reference they are displayed using a colour that reflects their angular size as reported by \citet{Perley2017}. Generally, the larger a source is, the less the RACS-mid flux density measurement ($S_\text{RACS-mid}$) is compared to the model ($S_\text{model}$). The reduction in $S_\text{RACS-mid}/S_\text{model}$ is particularly noticeable for Pictor~A (8\,arcmin extent with $S_\text{RACS-mid}/S_\text{model}=0.84$), Virgo~A (14\,arcmin and 0.82), Taurus~A (8\,arcmin and 0.46), and Fornax~A (50\,arcmin and 0.20). For Taurus~A and Fornax~A, the SBIDs they feature in have had an additional 75\,m ($u,v$) cut applied to the data, reducing sensitivity to these large diffuse sources further. The recovery of flux density of large scales will also reduce as a function of surface brightness, and care should be taken to not use the reduction of these calibrator source flux densities as a precise guide for other sources.

The choice to remove some short baselines for a subset of the observations has helped to control the artefacts around bright extended sources and reduce solar interference. Artefacts become more localised to the extended source, rather than spread through the whole tile image \corrs{(see Figure~\ref{fig:multiscale})}. This however means that there is even less sensitivity to these extended sources, though we caution use of RACS-mid for flux density recovery of sources $>5$\,arcmin in extent even without the $(u,v)$ cut. To improve quality in the Galactic Plane, it may be sensible follow the LOFAR Multifrequency Snapshot Sky Survey \citep{Heald2015:msss}, adding an extra 15\,min observation to each observed direction at an alternating hour angle to improve $(u,v)$ coverage with only minimal observing time overhead. While not planned currently, such improvements (or otherwise deeper observations) will likely be necessary to model the Galactic Plane, other regions of extended emission such as the Large Magellanic Cloud, and for specific bright extended sources (e.g.~Fornax~A) to obtain the requisite detail for use as a sky model for calibration of future ASKAP observations. Many such deep ASKAP observations already exist, though are for now restricted to the low band \citep[e.g.][]{Pennock2021,Riggi2021,Loi2022}. A similar limitation will be present for RACS-high.

\begin{figure*}
    \centering
    \begin{subfigure}[b]{0.5\linewidth}
    \includegraphics[width=1\linewidth]{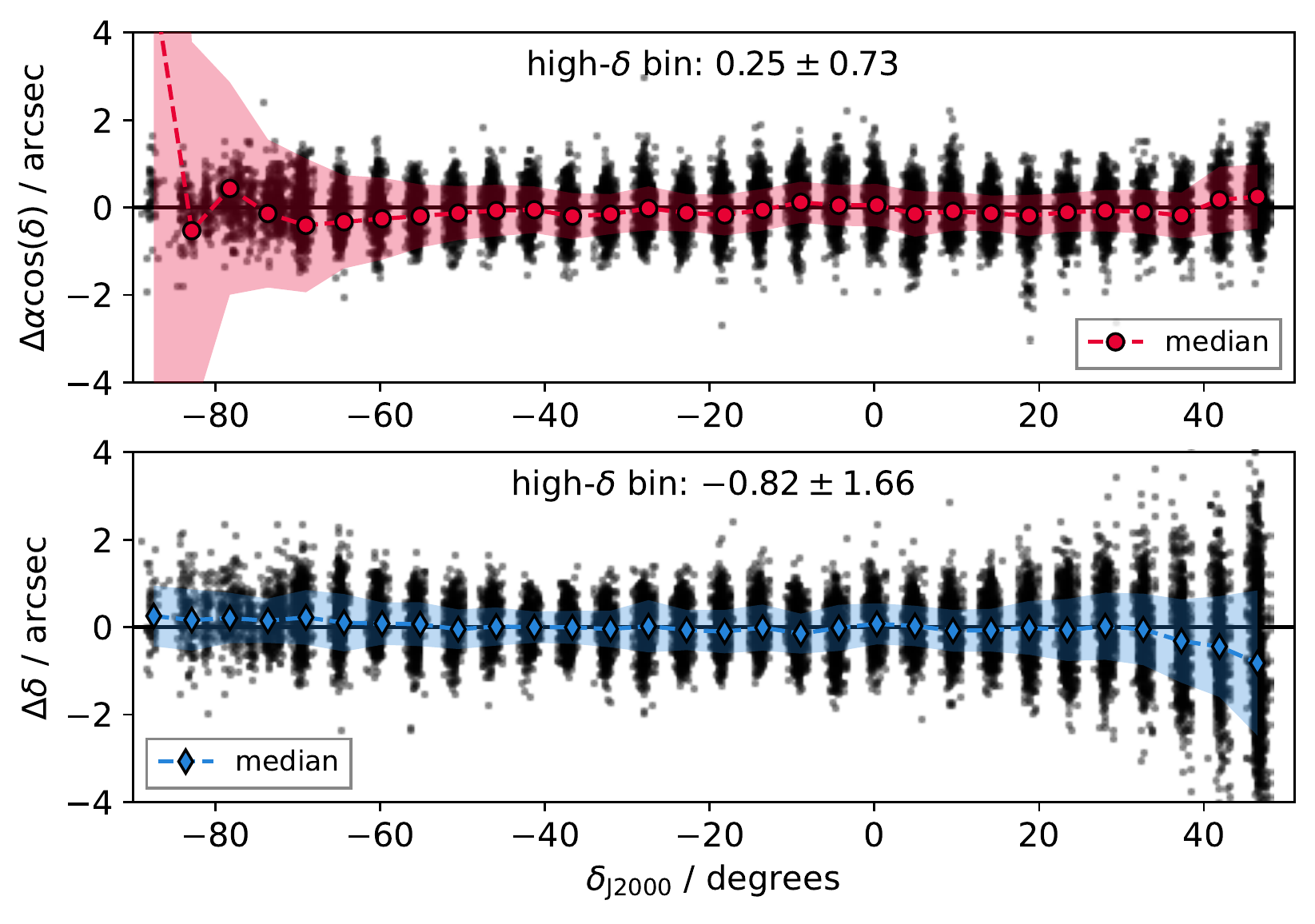}
    \caption{\label{fig:astrometry:beam2015:10sigma} Beam 20 $-$ beam 15, $10 < S_I / \sigma_\text{rms} < 100$.}
    \end{subfigure}%
    \begin{subfigure}[b]{0.5\linewidth}
    \includegraphics[width=1\linewidth]{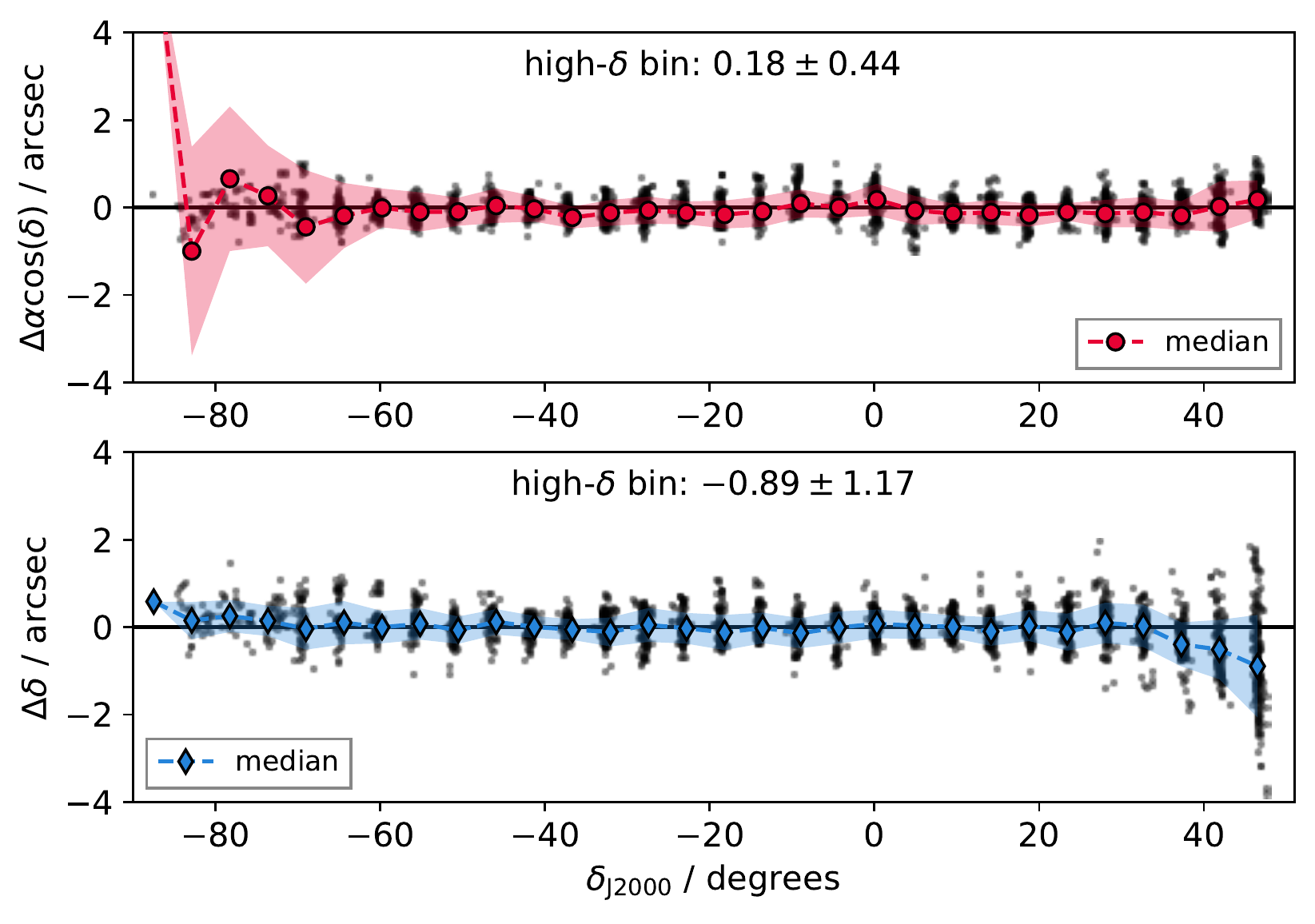}
    \caption{\label{fig:astrometry:beam2015:100sigma} Beam 20 $-$ beam 15, $S_I > 100\sigma_\text{rms}$.}
    \end{subfigure}\\%
    \begin{subfigure}[b]{0.5\linewidth}
    \includegraphics[width=1\linewidth]{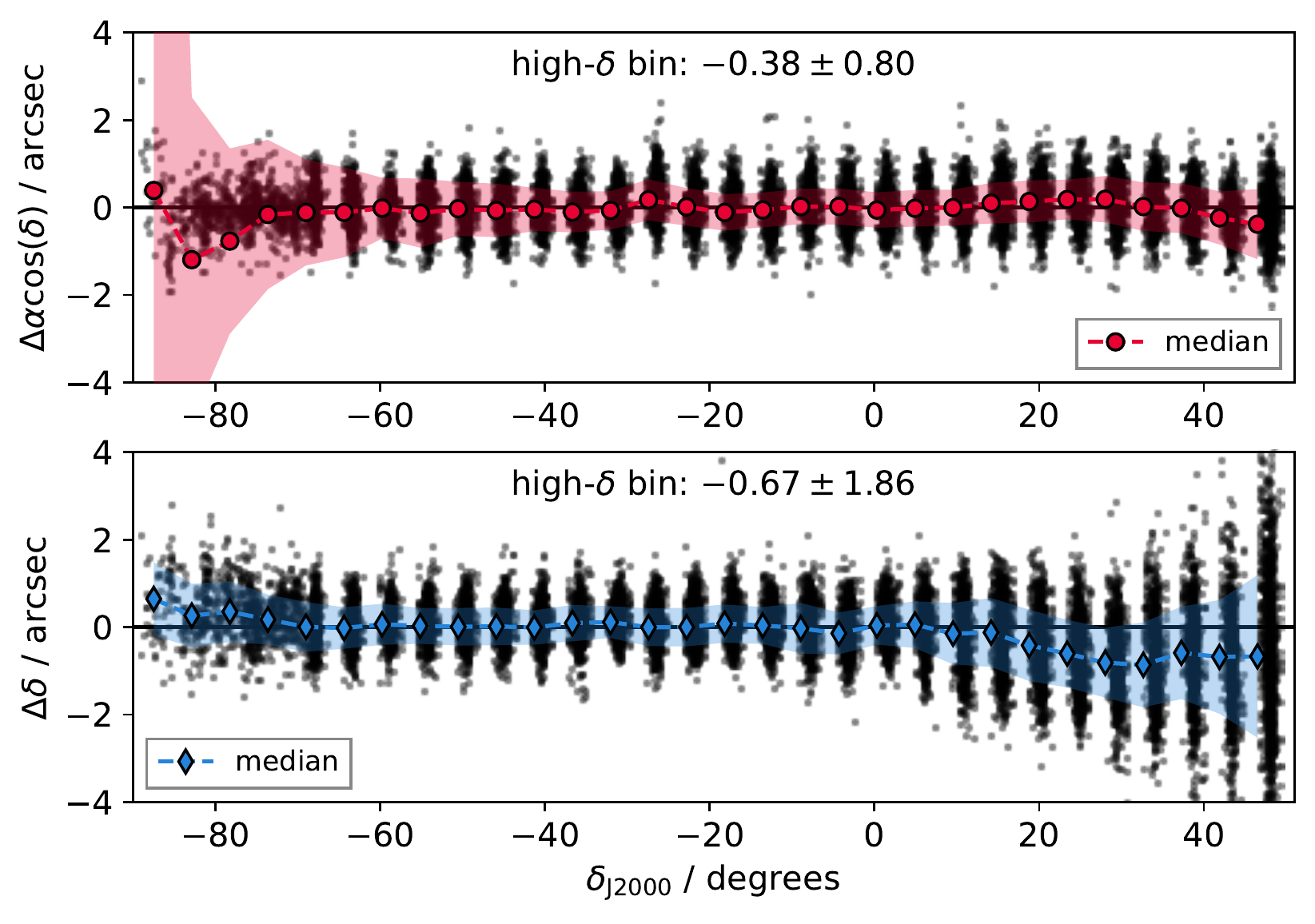}
    \caption{\label{fig:astrometry:beam3529:10sigma} Beam 35 $-$ beam 29, $10 < S_I / \sigma_\text{rms} < 100$.}
    \end{subfigure}%
    \begin{subfigure}[b]{0.5\linewidth}
    \includegraphics[width=1\linewidth]{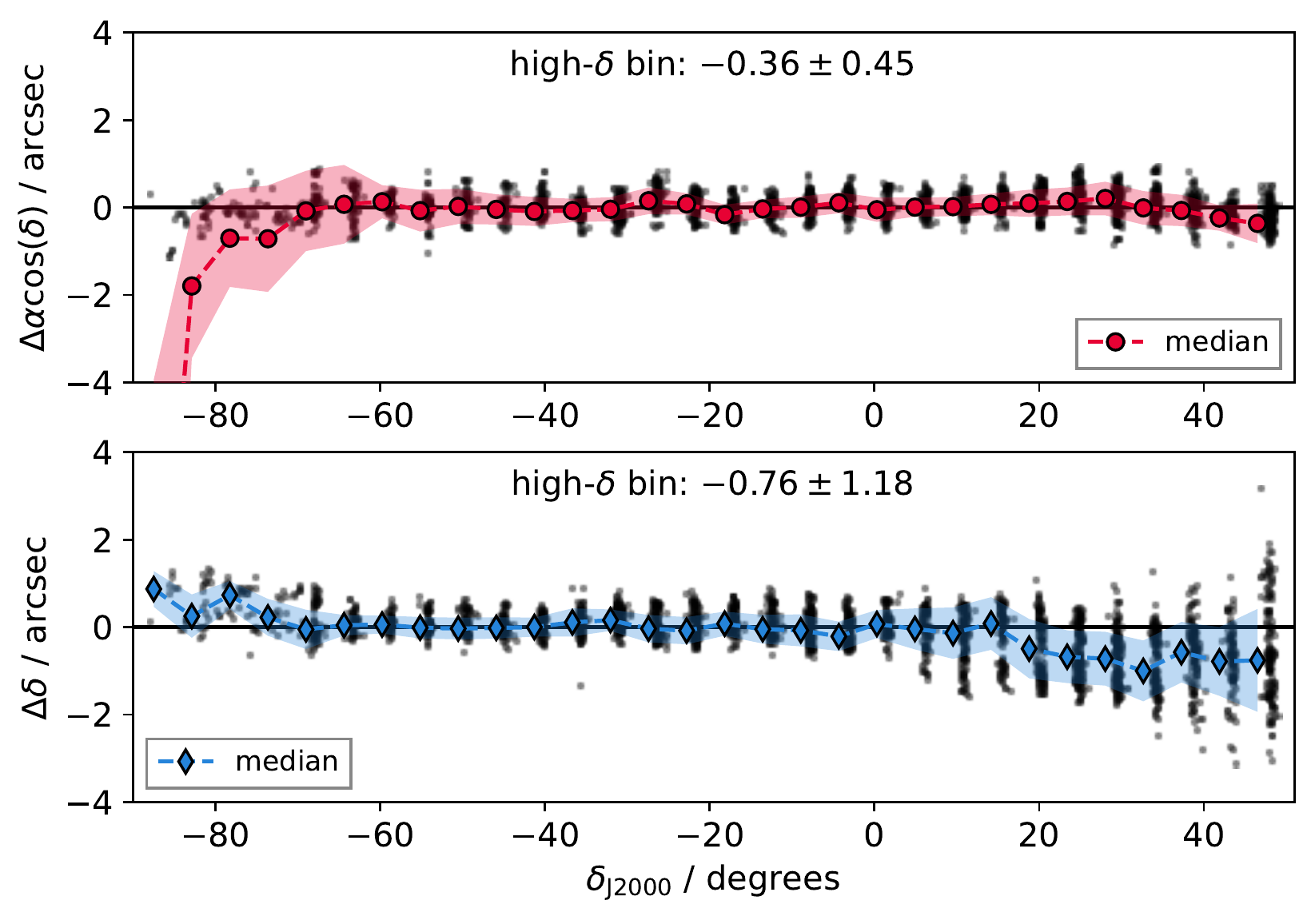}
    \caption{\label{fig:astrometry:beam3529:100sigma} Beam 35 $-$ beam 29, $S_I > 100\sigma_\text{rms}$.}
    \end{subfigure}\\%
    \begin{subfigure}[b]{0.5\linewidth}
    \includegraphics[width=1\linewidth]{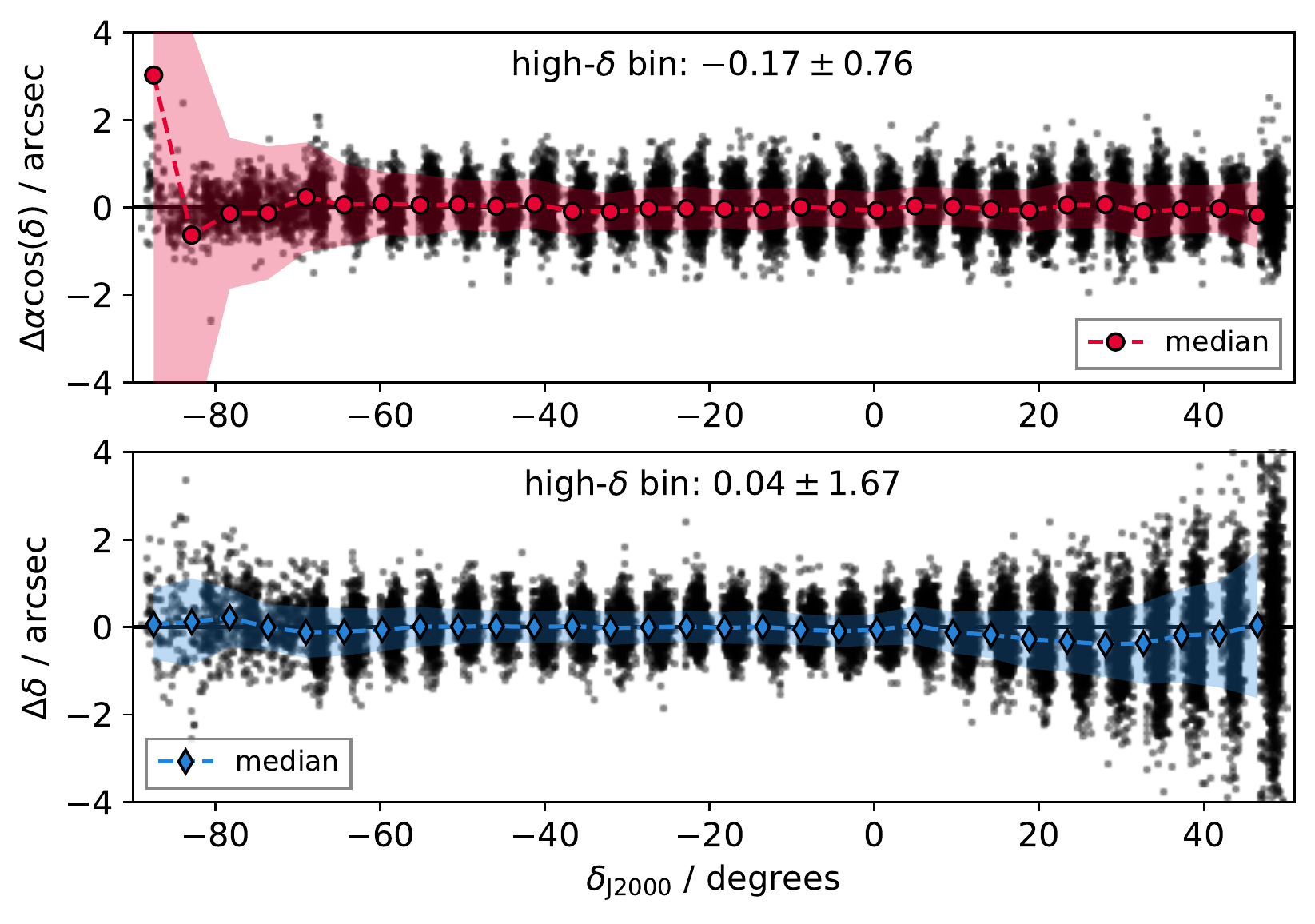}
    \caption{\label{fig:astrometry:beam3534:10sigma} Beam 35 $-$ beam 34, $10 < S_I / \sigma_\text{rms} < 100$.}
    \end{subfigure}%
    \begin{subfigure}[b]{0.5\linewidth}
    \includegraphics[width=1\linewidth]{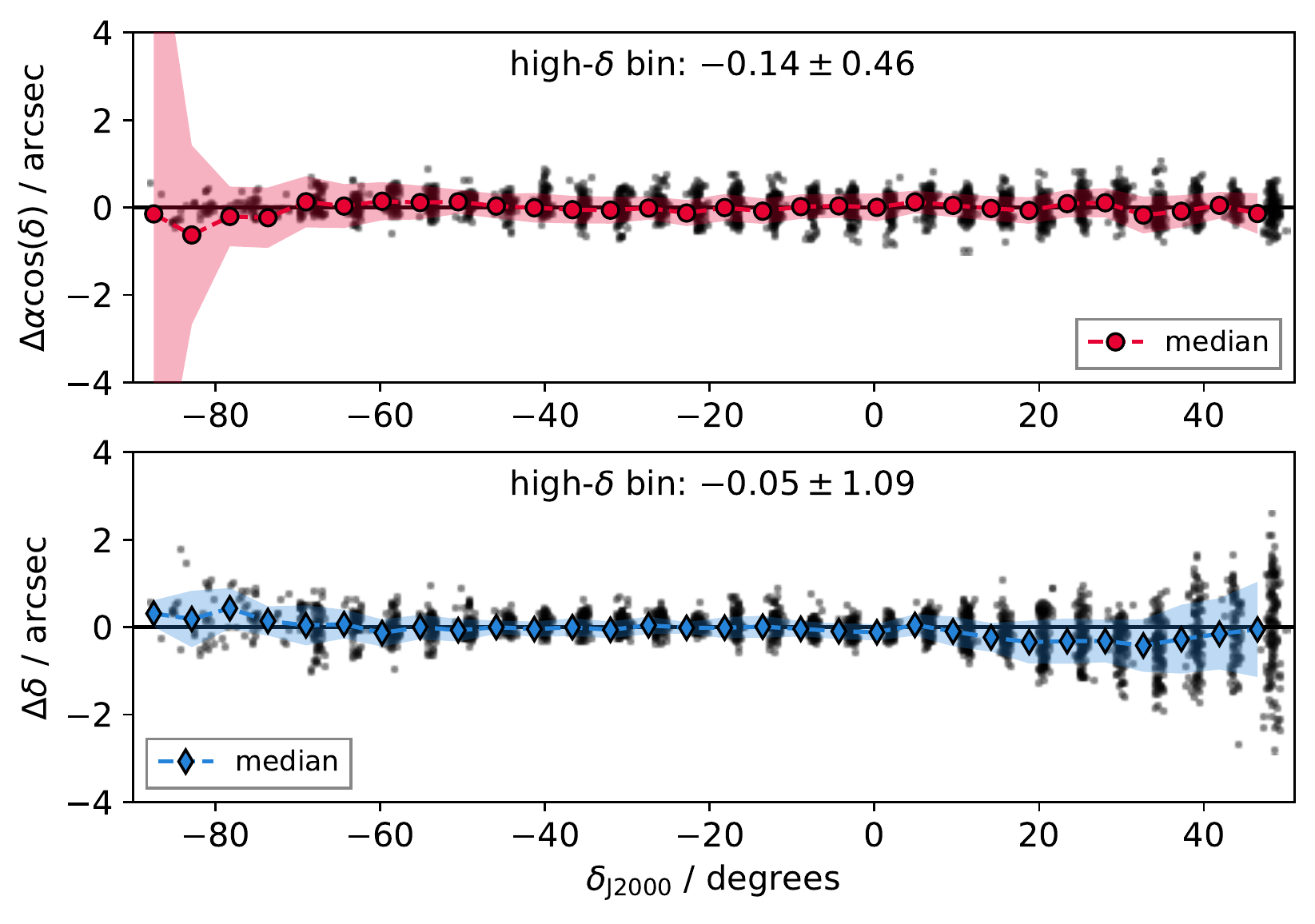}
    \caption{\label{fig:astrometry:beam3534:100sigma} Beam 35 $-$ beam 34, $S_I > 100\sigma_\text{rms}$.}
    \end{subfigure}%
    \caption{\label{fig:astrometry:beam2015} R.~A. ($\Delta\alpha \cos(\delta)$, \textit{top panels}, red) and declination ($\Delta\delta$, \textit{bottom panels}, blue) differences for sources detected in both adjacent beam pairs (in descending order $20 - 15$, $35 - 29$, and $35 - 34$)  for sources with $10 < S_I / \sigma_\text{rms} < 100$ [\subref{fig:astrometry:beam2015:10sigma}. \subref{fig:astrometry:beam3529:10sigma}, and \subref{fig:astrometry:beam3529:100sigma}  and $S_I > 100\sigma_\text{rms}$ [\subref{fig:astrometry:beam2015:100sigma}, \subref{fig:astrometry:beam3534:10sigma}, and \subref{fig:astrometry:beam3534:100sigma} over the full survey. The median in each declination bin is shown for reference and we note the median offset in the highest-declination bin. Bins are defined from the tile centres. The shaded regions corresponds to $\pm 1$ standard deviation. Note beams 20 and 15 are near the centre of the PAF footprint, while 35, 34, and 29 and in the top left corner (see Figure~\ref{fig:footprint}).}
\end{figure*}

\begin{figure*}
\centering
\begin{subfigure}[b]{0.5\linewidth}
\includegraphics[width=1\linewidth]{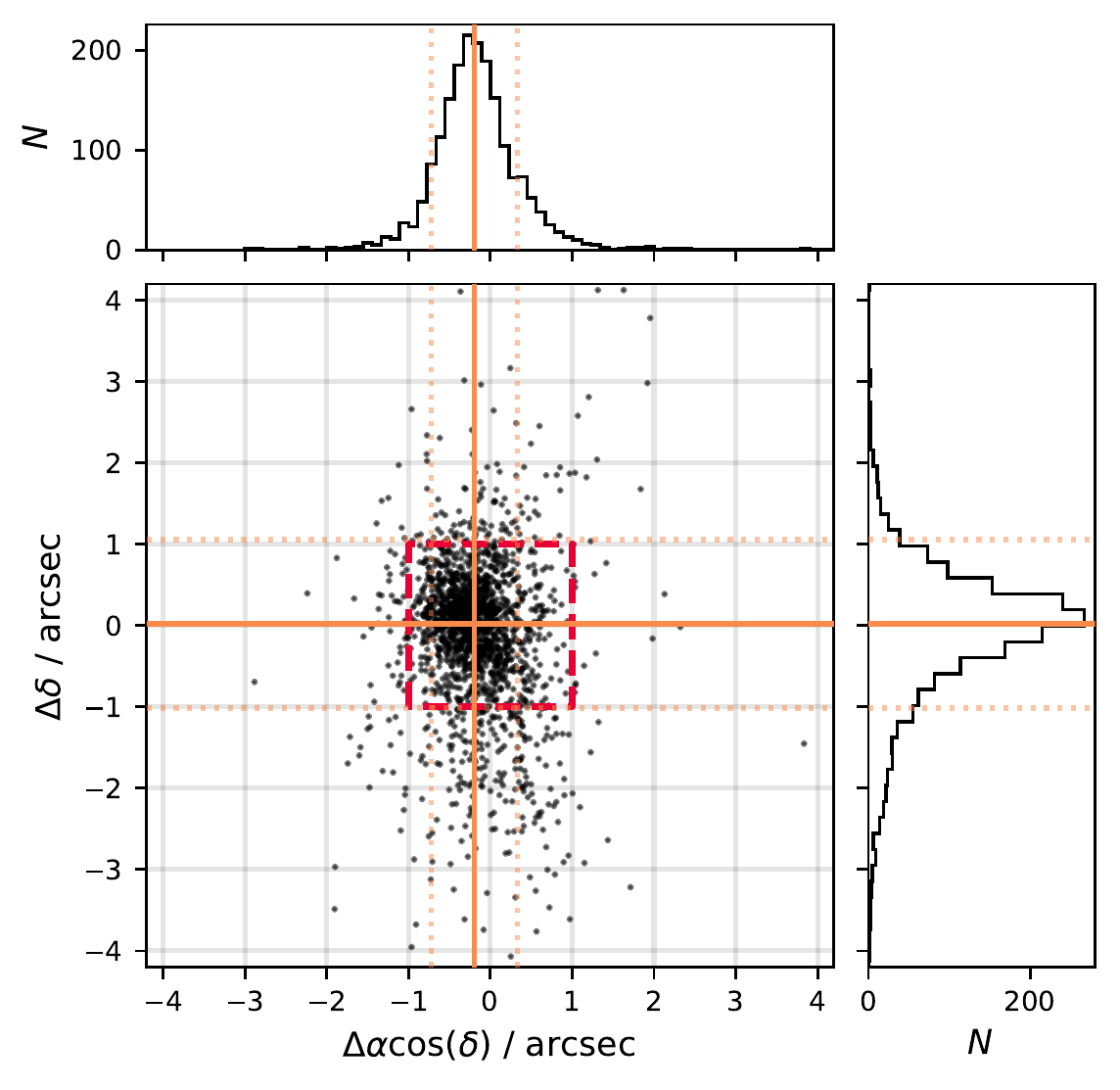}
\caption{\label{fig:astrometry:icrf} RACS-mid${}-{}$ICRF.}
\end{subfigure}\\%
\begin{subfigure}[b]{0.5\linewidth}
\includegraphics[width=1\linewidth]{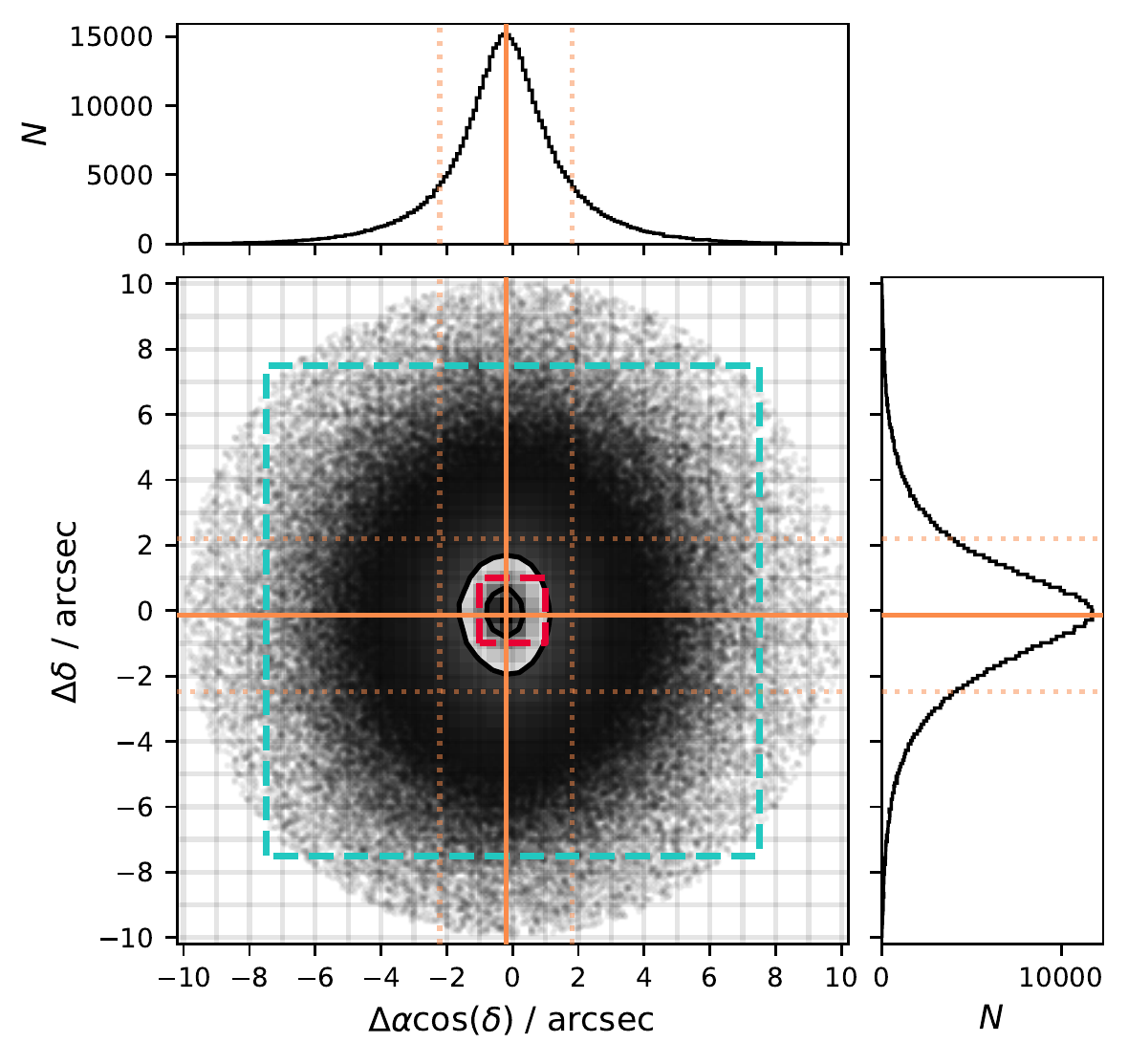}
\caption{\label{fig:astrometry:nvss} RACS-mid${}-{}$NVSS.}
\end{subfigure}%
\begin{subfigure}[b]{0.5\linewidth}
\includegraphics[width=1\linewidth]{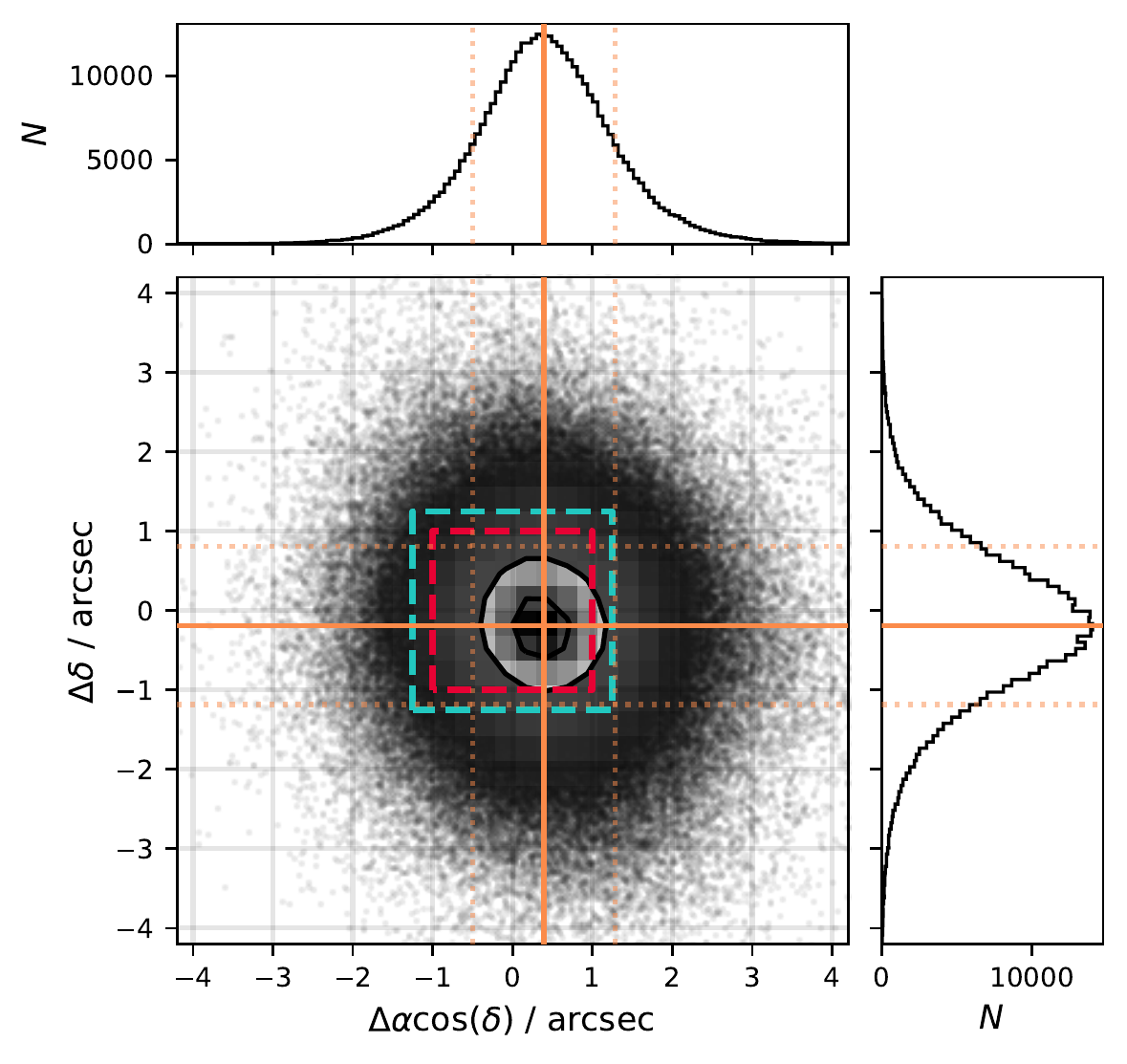}
\caption{\label{fig:astrometry:low} RACS-mid${}-{}$RACS-low \citepalias{racs2}.}
\end{subfigure}\\%
\caption{\label{fig:astrometry} Astrometric offsets, ($\Delta\alpha\cos{(\delta)},\Delta\delta$) for sources in RACS-mid compared to the ICRF3 \subref{fig:astrometry:icrf}, NVSS \subref{fig:astrometry:nvss}, and the RACS-low catalogue \subref{fig:astrometry:low}. Offsets are defined as RACS-mid position $-$ reference catalogue position. The small, dashed red boxes indicate the pixel size for RACS-mid, whereas the larger, dashed cyan boxes are the pixel size of the comparison survey. The solid, orange lines indicate medians in $\Delta\alpha \cos\left( \delta \right)$ and $\Delta\delta$. The dashed, orange lines indicate $\pm 1$ standard deviation to the distribution of offsets.}
\end{figure*}

\begin{figure*}[t]
    \centering
    \begin{subfigure}[b]{0.5\linewidth}
    \includegraphics[width=1\linewidth]{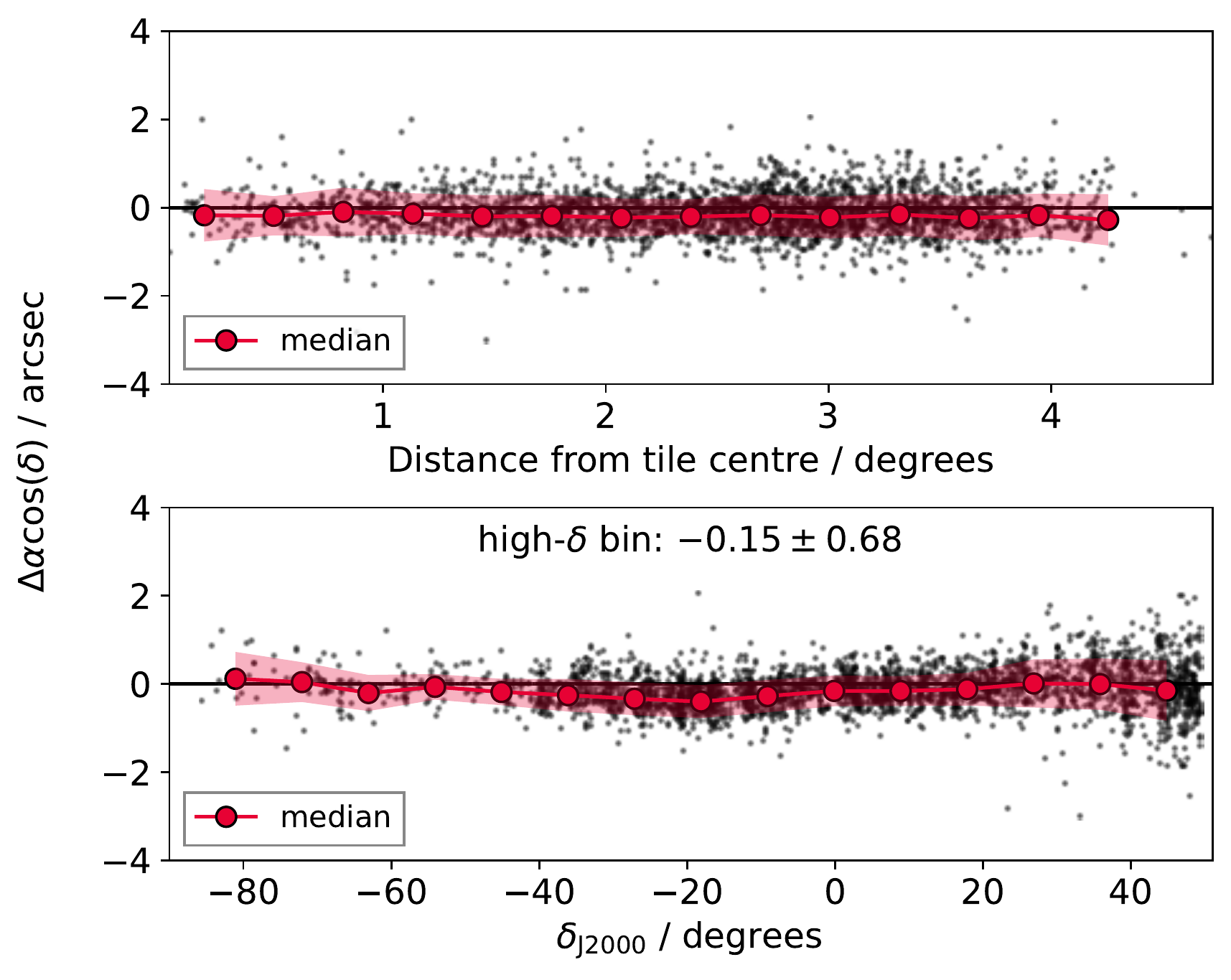}
    \caption{\label{fig:icrf:ra}}
    \end{subfigure}%
    \begin{subfigure}[b]{0.5\linewidth}
    \includegraphics[width=1\linewidth]{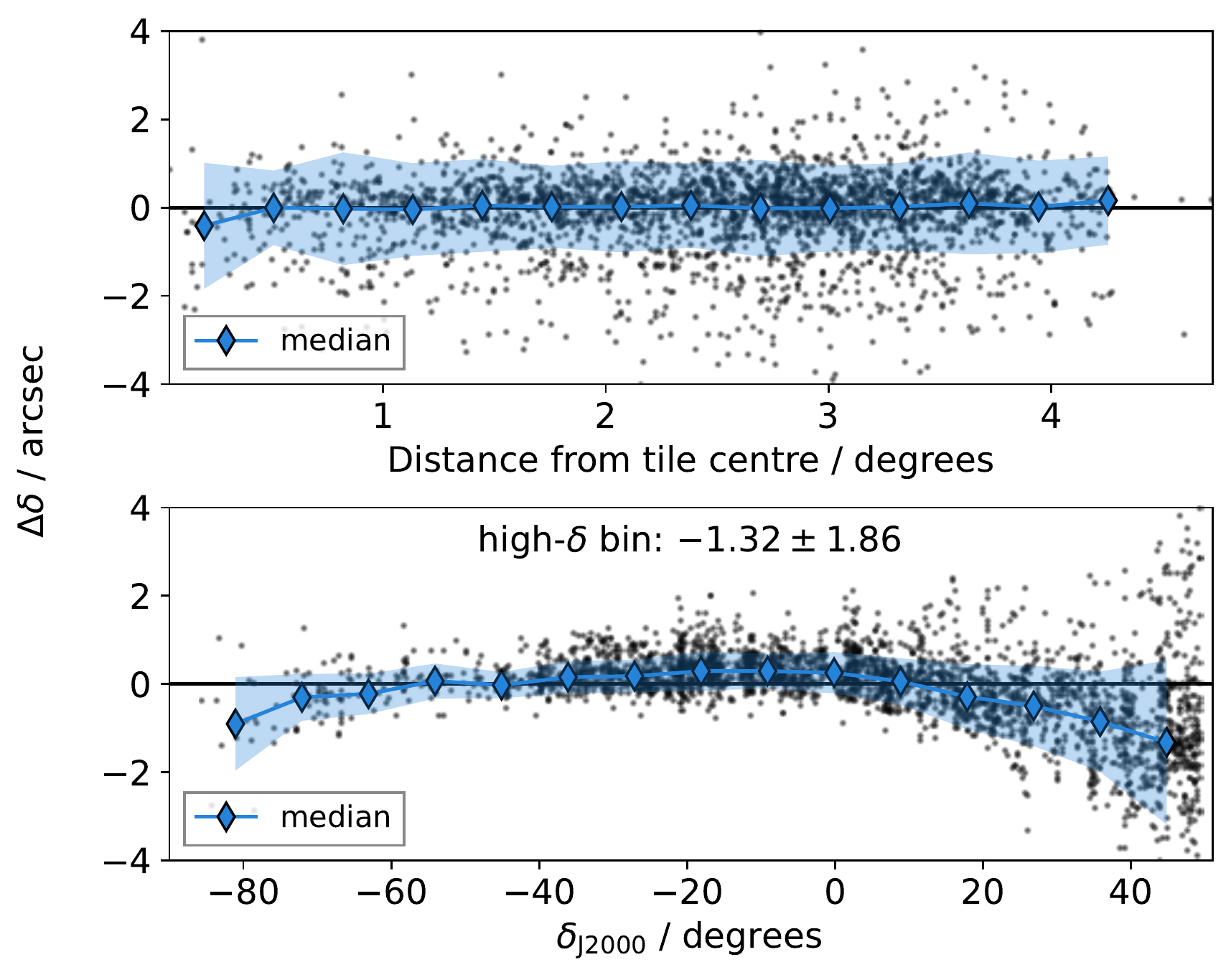}
    \caption{\label{fig:icrf:dec}}
    \end{subfigure}\\%
    \caption{\label{fig:icrf} $\Delta\alpha\cos\left(\delta\right)$ \subref{fig:icrf:ra} and $\Delta\delta$ \subref{fig:icrf:dec} for RACS-mid and the ICRF3 as a function of distance from the tile centre (\emph{top panels}) and declination (\emph{bottom panels}). Binned medians are shown along with shaded regions corresponding to $\pm 1$ standard deviation.}
\end{figure*}

\subsection{Astrometry}\label{sec:validation:astrometry}

\subsubsection{{Beam-to-beam consistency}}
Variation in astrometry between individual beams of order $\lesssim 2.5$\,arcsec ($\sim 1$\,pixel) was noted in \citetalias{racs1}. The cause of scatter between adjacent beams is largely due to self-calibration and will therefore affect all sources regardless of SNR. To investigate this for RACS-mid, we cross-matched compact, isolated sources in the individual apparent brightness beam images used for primary beam response modelling (Section~\ref{sec:holography:zernike}). This was done for beams 35 and 29, beams 35 and 34 (top left corner of the footprint, see Figure~\ref{fig:footprint}) and beams 20 and 15 (centre of the footprint) for all SBIDs in the survey. Figure~\ref{fig:astrometry:beam2015} shows $\Delta\alpha\cos\left(\delta\right)$ and $\Delta\delta$ for the cross-matched sources in beams 20 and 15, beams 35 and 29, and beam 35 and 34. We split the data into low- and high-SNR subsets, with $10 < S_I / \sigma_{\text{rms},I} < 100$ ($\sim 22\,000$ sources) and $S_I > 100\sigma_{\text{rms},I} $ ($\sim 2\,500$ sources) for each subset, respectively. Both the low- and high-SNR subset show similar features. Offsets in $\alpha_\text{J2000}$ become scattered towards low declination, in line with fewer sources per declination bin combined with a larger PSF minor axis (Figure~\ref{fig:psfcoords}). The PSF position angle is typically close to zero elsewhere aligning the minor axis closely with $\alpha_\text{J2000}$. \corrs{Scatter} in $\Delta\delta$ increases at high and low declination are largely due to the increase in PSF major axis. In the case of $\Delta\delta$, there is an additional bias in these low/high-declination offsets.  These offsets appear independent of source SNR.

\subsubsection{{External accuracy}}
To assess the \corrs{external} astrometric accuracy of RACS-mid, we matched the positions of bright ($>10\sigma$), isolated (no neighbours within 150\,arcsec), and compact ($0.8 < S_\text{int}/S_\text{peak} < 1.2$) sources to counterparts in other radio surveys. In Figure \ref{fig:astrometry} we show the astrometric offsets between 1\,875 matches to the ICRF3 [\ref{fig:astrometry:icrf}] with median offsets (with $1\sigma$ uncertainties) of $\Delta\alpha\cos{(\delta)} = -0.20 \pm 0.53$\,arcsec, $\Delta\delta = 0.02 \pm 1.04$\,arcsec; 509\,487 matches to NVSS [\ref{fig:astrometry:nvss}] with median offsets of $\Delta\alpha\cos{(\delta)} = -0.20 \pm 2.01$\,arcsec, $\Delta\delta = -0.14 \pm 2.34$\,arcsec; and 380\,066 matches to RACS-low [\ref{fig:astrometry:low}] with median offsets of $\Delta\alpha\cos{(\delta)} = 0.40 \pm 0.90$\,arcsec, $\Delta\delta = -0.19 \pm 1.00$\,arcsec.
The scatter in the cross-matched positional offsets is generally confined to the pixel scale of RACS-mid (2\,arcsec), except in the case of NVSS where the larger pixels of the NVSS images (15\,arcsec) increase the offset scatter further. 

Figure~\ref{fig:icrf} shows the ICRF3 offsets as a function of distance from the tile centre and declination. There is no change in offsets as a function of distance from the tile centre. The high-declination (corresponding to low elevation) sources feature $\Delta\delta$ offsets with increased scatter which can be partially attributed to the larger PSF major axis. The median $\Delta\delta$ offset at high declination drops below 0, with binned-median offset of $-1.32\pm1.86$\,arcsec in the highest declination bin. This is indicative of a systematic effect, but this remains within one standard deviation of the binned offsets. A similar offset is seen for low-declination sources, though with less certainty due to the smaller sky area covered and correspondingly fewer sources. \corrs{The high-declination offsets are similar to what is seen in the beam-to-beam comparisons, though in the low-declination case the offsets against the ICRF3 are reversed.}

As variation in astrometric precision changes per SBID and per beam, we do not offer any astrometric correction per image. There are insufficient ICRF3 sources per full mosaic to make such corrections for each individual beam image (typically 1--10 ICRF3 sources over a single tile mosaic). There are no other surveys that cover the full RACS-mid survey region at a similar (or better) resolution besides RACS-low, which has the same astrometric errors. Bulk offsets (i.e.~as a function of declination) may be corrected for in full-sky catalogues as part of the next data release.

\subsection{Observations of planets}\label{sec:planets}

\begin{figure*}[t]
    \centering
    \includegraphics[width=1\linewidth]{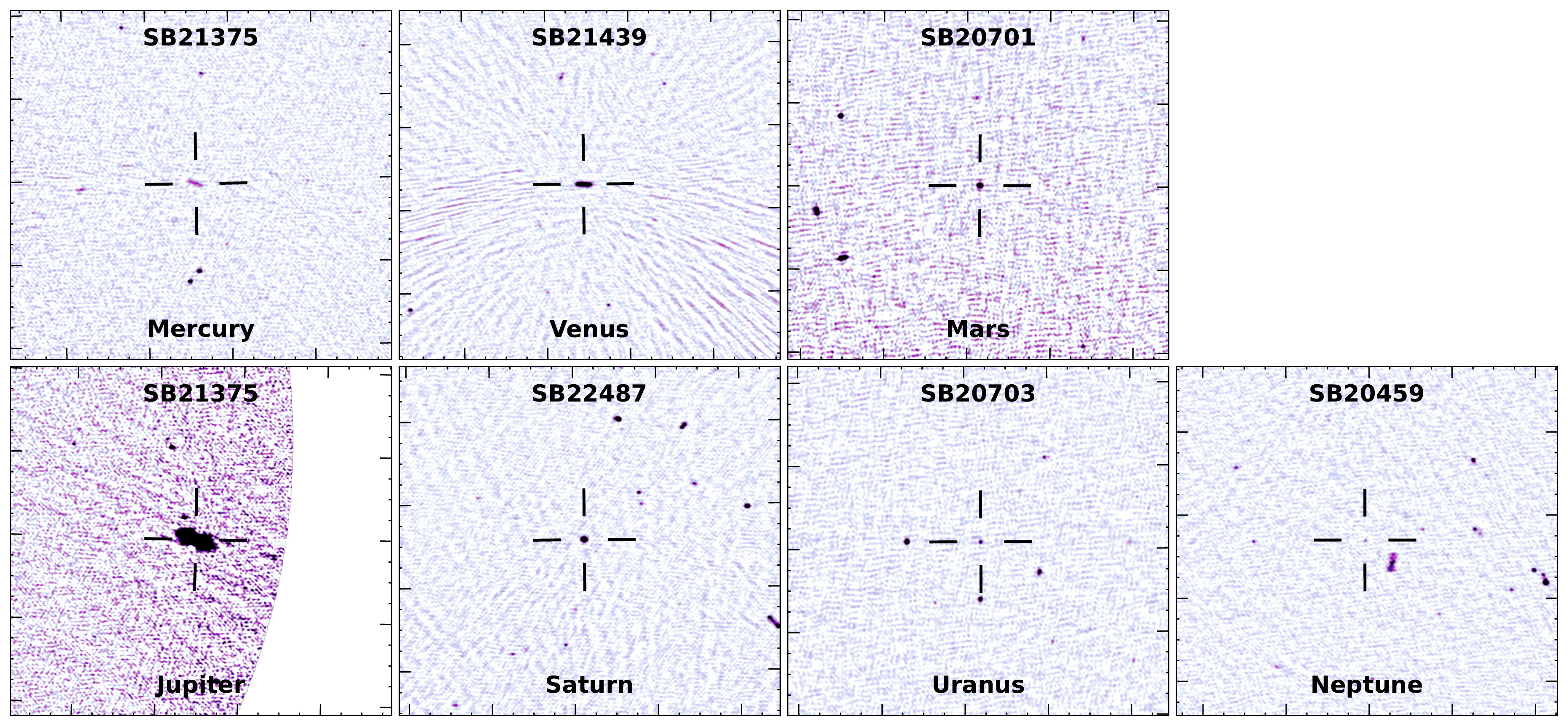}
    \caption{\label{fig:planets}Examples of the Solar system planets as they appear in the RACS-mid images.The cross-hairs are centered on the location of the planet and have total lengths of 6\,arcmin. Jupiter appears at the edge of a beam in the footprint. }
\end{figure*}

\begin{table}[t]
    \centering
    \caption{\label{tab:planets} Planets and the SBIDs in which they feature.}
    \begin{tabular}{c c}\toprule
    Planet & SBIDs \\\midrule
    Mercury & 21375,33098 \\
    Venus & 21439,21935,22000,22487 \\
    Mars & 20701,20769 \\
    Jupiter & 21375 \\
    Saturn & 22487 \\
    Neptune & 20459 \\
    Uranus & 20703 \\
    \bottomrule
    \end{tabular}

\end{table}

Over the course of RACS-mid observing runs all seven extra-terrestrial planets in our Solar System were observed in one or more SBIDs. Generally, these SBIDs have not been re-observed as the inclusion of the planet does not decrease overall image quality (unlike other Solar system objects like the Moon or Sun). Planets and the SBIDs in which they appear in are listed in Table~\ref{tab:planets}. \corrs{Figure~\ref{fig:planets} shows examples of each of the detected planets, and shows some low-level artefacts close to Venus due to movement across the sky during the observation}. The Sun does appear in primary beam sidelobes frequently, and solar interference on degree angular scales is a significant source of artefacts. Many SBIDs with clear solar interference are imaged with a minimum $(u,v)$ cut corresponding to 75\,m baselines as described in Section~\ref{sec:gp} which avoids the most significant artefacts. The automated observation scheduler SAURON takes into account a specified constraint for RACS such that the Sun and Moon are $\geq 10$\,deg from the pointing direction for any RACS-mid observation so they do not appear within the mainlobes (or first sidelobes) of any beams.

\section{Data products and availability}
As with the RACS-low data releases, RACS-mid data products are available through the CSIRO ASKAP Science Data Archive \citep[CASDA;][]{casda,Huynh2020}. Survey data products can be accessed through the online CASDA interface~\footnote{\url{https://data.csiro.au/domain/casdaObservation}.} under project AS110 with the Digital Object Identifier \url{https://doi.org/10.25919/6mr6-rd83}. Note that the first epoch of RACS-low (and future RACS data releases) share this project code. RACS-mid data can be searched alone by specifying an appropriate frequency range in the online interface. The main available data products include

\begin{itemize}
\item \texttt{image.*.taylor.0.restored.conv.fits} \\
Stokes I and V $0^{\text{th}}$-order Taylor term mosaics, after all individual beams are convolved to a common resolution. \CORRS{Stokes I and V images are primary beam corrected during the mosaicking process, and Stokes V images have had widefield leakage removed as described in Section~\ref{sec:stokesv}.}
\item \verb|image.*.taylor.0.fits| \\
Stokes I and V $0^{\text{th}}$-order Taylor term CLEAN component model image.
\item \verb|residual.*.taylor.0.fits| \\
Stokes I and V $0^{\text{th}}$-order Taylor term residual maps.
\item \verb|weights.*.taylor.0.fits| \\
Stokes I and V $0^{\text{th}}$-order Taylor term combined weight maps, useful for linear combinations of adjacent fields.
\item \texttt{*beam??*.ms.tar} \\
Archive of the calibrated and on-axis leakage-corrected MeasurementSet for a particular beam in the PAF footprint---[00, ..., 35]. These data have been flagged as per Section~\ref{sec:spectralcoverage} and only the remaining 144 channels are kept.
\end{itemize}

Additional metadata and validation files for each SBID are also made available, including \texttt{selavy}-generated source lists. While these pipeline output products are available, we suggest general users locate both images and catalogues from the next data release (Paper V, Duchesne et al., in prep) when available which are discussed in the proceeding section. \CORRS{As noted in Section~\ref{sec:holography:zernike}, $1^{\text{st}}$-order Taylor term data will be unreliable and should not be used in scientific analysis. Despite this, for completeness the $1^{\text{st}}$-order Taylor term images are included as ancillary data products on CASDA.}

\section{Future RACS data releases}\label{sec:futurework}

This paper and data release are limited to an initial set of data products for RACS-mid, including images covering the full survey region, calibrated and on-axis leakage-corrected visibility datasets, initial source-lists for images, along with associated metadata. Paper V will detail the cataloguing process for RACS-mid, similar to work done for the first epoch of RACS-low \citepalias{racs2}. This cataloguing process will include images formed as a linear mosaic of common resolution neighbouring tiles to produce an image of each pointing with full sensitivity to the tile edges. Source lists for these full-sensitivity mosaics will be made available at the same time and we aim to create a multi-resolution combined catalogue, with a PSF that largely varies as a function of declination, to maximise the resolution and source density across the survey region.

Following the second RACS-mid data release, we plan to provide RACS-high and the second epoch of RACS-low in a similar fashion. These two surveys have been fully observed and a first pass of processing/imaging has been performed. These initial images have helped with validation of the performance of regular weights-matched holographic primary beam measurements over a whole survey. This is in contrast to RACS-mid, which has matched holographic measurements for only a small fraction of the observations. The residual declination-dependent flux density scale error discussed in Section~\ref{sec:fluxscale} will be investigated further in those works, with consistent holography application making determination of underlying issues less complex. \CORRS{Further to the Stokes I brightness scale, future RACS releases will also aim to make Stokes I into V leakage corrections using holography data, which may improve upon the residual leakage reported for RACS-mid. Finally,} with the goal of creating a global sky model for ASKAP operations, a combined catalogue of the three bands will be also be produced \CORRS{which will contain model spectra for the catalogued sources.}

Alongside the Stokes I and V continuum data releases, there has been work to also provide spectro-polarimetric results using Stokes Q and U low band data as part of SPICE-RACS. The 30-field pilot for SPICE-RACS is described in Paper III (Thomson et al., submitted) using data from the first epoch of RACS-low. Work with SPICE-RACS will continue with the second epoch of RACS-low, providing a more consistent quality over the full surveyed region, before also including the mid and high bands.

\section{Summary}

An all-sky survey using ASKAP is being conducted in three bands across its operational frequency range. The first centred at 887.5\,MHz has already been released \citepalias{racs1,racs2} and the second centred at 1367.5\,MHz is released alongside this description paper. The 1367.5-MHz survey, named `RACS-mid', covers the sky up to $\delta_\text{J2000} = +49^\circ$, with a median rms noise of \rmserror\,\textmu Jy\,PSF$^{-1}$ \corrs{in Stokes I and \rmserrorv\,\textmu Jy\,PSF$^{-1}$ in Stokes V, with} a declination-dependent PSF with a major axis FWHM ranging from \minpsfmajor--\maxpsfmajor\,arcsec (and minor axis FWHM \minpsfminor--\maxpsfminor\,arcsec). Much of the observation and data-processing strategy is similar to the first data release, however we have highlighted a number of changes that have helped the overall quality of the RACS-mid data, including: \begin{itemize}
    \item Compact \texttt{closepack36} PAF footprint to reduce sensitivity ripple across individual observations and when mosaicking nearby fields,
    \item Observations scheduled within $\pm 1$\,h of the meridian, reducing variation to $(u,v)$ coverage between observations allowing significantly more consistent PSF and noise properties compared to RACS-low,
    \item Subtraction and/or peeling of select bright, off-axis sources, to further reduce artefacts and improve self-calibration of beam images near these sources,
    \item Primary beam correction with measured Stokes I beam responses, using holography and in-field measurements to derive models,
    \item Widefield Stokes V leakage correction using $V/I$ leakage models derived from  measured leakage of in-field sources resulting in \corrs{ $\lesssim 0.9$--$2.4$\%} residual leakage \CORRS{of Stokes I into the Stokes V images},
    \item Correction of time-dependent flux scale variations.
\end{itemize}
The data from the this survey are being made available to the wider astronomical community through CASDA, and as part of this data release, we supply primary beam-corrected Stokes I images, Stokes V images with widefield leakage corrections, and self-calibrated visibility datasets with on-axis leakage corrections applied. A future paper in this series will describe the resulting radio source catalogue at 1367.5\,MHz.

\begin{acknowledgement}
We thank the referee for their extensive comments that have helped to improve this manuscript. This scientific work uses data obtained from Ilyarrimanha Ilgari Bundara / the Murchison Radio-astronomy Observatory. We acknowledge the Wajarri Yamaji People as the Traditional Owners and native title holders of the Observatory site. The Australian SKA Pathfinder is part of the Australia Telescope National Facility (https://ror.org/05qajvd42) which is managed by CSIRO. Operation of ASKAP is funded by the Australian Government with support from the National Collaborative Research Infrastructure Strategy. ASKAP uses the resources of the Pawsey Supercomputing Centre. Establishment of ASKAP, Ilyarrimanha Ilgari Bundara / the Murchison Radio-astronomy Observatory and the Pawsey Supercomputing Centre are initiatives of the Australian Government, with support from the Government of Western Australia and the Science and Industry Endowment Fund. 

Numerous \texttt{python} and other software packages have been used during the production of RACS and this manuscript. Those not explicitly noted in other sections include: \texttt{aplpy} \citep{Robitaille2012}, \texttt{astropy} \citep{astropy:2018}, \texttt{matplotlib} \citep{Hunter2007}, \texttt{numpy} \citep{numpy}, \texttt{scipy} \citep{scipy}, and \texttt{cmasher} \citep{cmasher}. Some of the results in this paper have been derived using the \texttt{healpy} \citep{healpy} and HEALPix package. Some colourmaps have been selected from \texttt{cmasher} \citep{cmasher} while other colourmaps have been created using \texttt{gradient} (\url{https://github.com/eltos/gradient}). We make use of \texttt{ds9} and \citep{ds9}, \texttt{topcat} \citep{topcat} for visualisation, as well as the ``Aladin sky atlas'' developed at CDS, Strasbourg Observatory, France \citep{aladin1,aladin2}. We make use of \texttt{CASA} \citep{casa} including its modular \texttt{python} implementation \citep{Raba2019} and \texttt{casacore} (\url{https://github.com/casacore/casacore}) including \texttt{python-casacore} (\url{https://github.com/casacore/python-casacore}).
\end{acknowledgement}


\bibliography{references}

\appendix

\begin{figure}[t!]
\centering
\begin{subfigure}[b]{1\linewidth}
\includegraphics[width=1\linewidth]{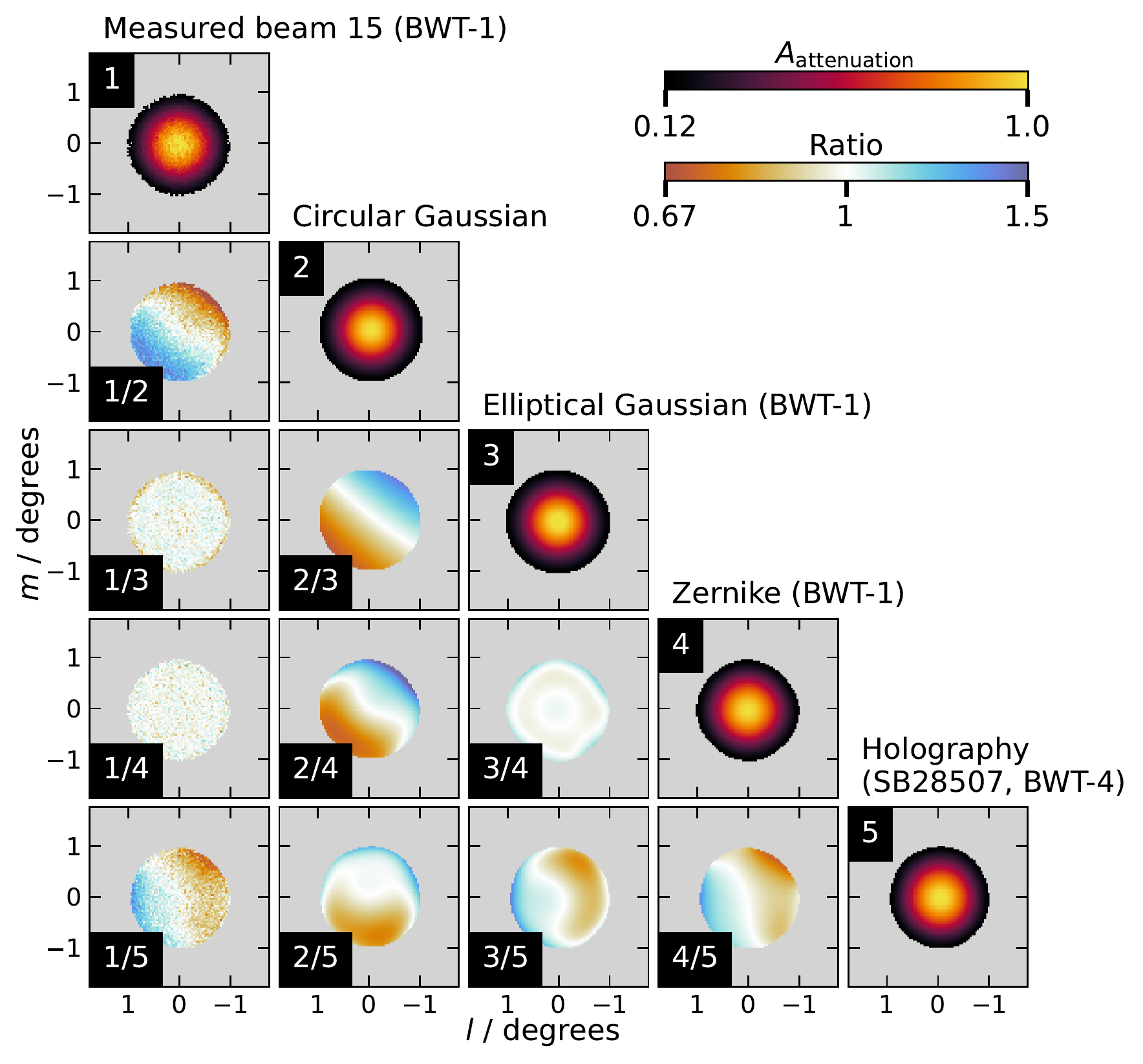}
\caption{\label{fig:compare:beam15} Beam 15.}
\end{subfigure}\\%
\begin{subfigure}[b]{1\linewidth}
\includegraphics[width=1\linewidth]{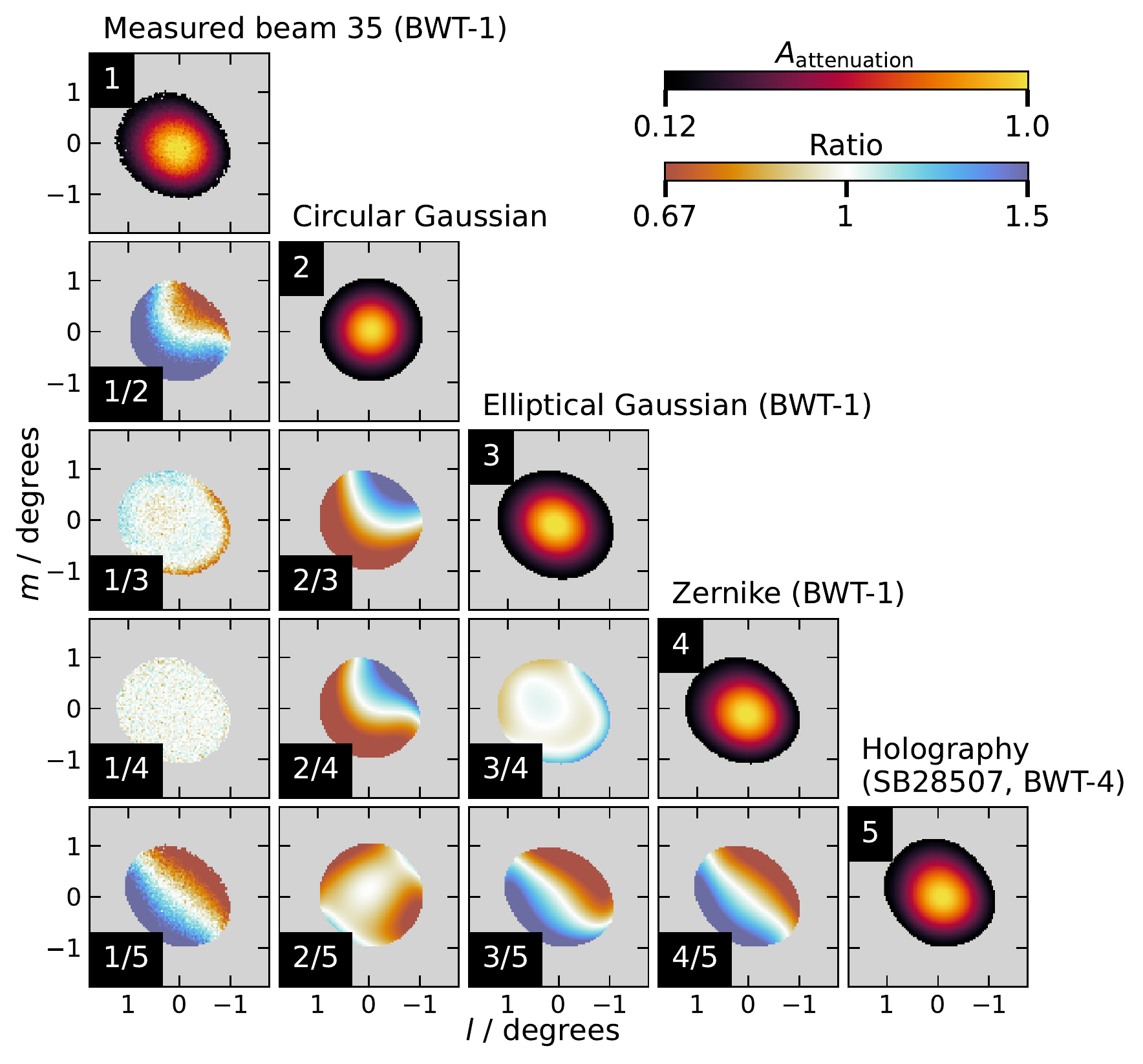}
\caption{\label{fig:compare:beam35} Beam 35.}
\end{subfigure}%
\caption{\label{fig:compare} \corrs{A comparison of the measured and model $A_\text{attenuation}$ for beam 15 \subref{fig:compare:beam15} and beam 35 \subref{fig:compare:beam35}.}}
\end{figure}

\section{Deviation from a 2-D Gaussian primary beam}\label{app:beams}

\corrs{Figure~\ref{fig:compare:beam15} shows the measured and model Stokes I response of beam 15 for BWT-1, $A_\text{attenuation}$. Beam 15 is located near the centre of the \texttt{closepack36} footprint. The measured response is shown in the top panel as a binned median, with $2.1 \times 2.1$\,arcmin$^2$ bins. The diagonal panels then show the normally assumed 2-D circular Gaussian model, a fitted 2-D elliptical Gaussian model, the best-fit Zernike model described in Section~\ref{sec:holography:zernike}, and the holographic measurements used for BWT-5. We also show the difference in the measured attenuation and models, highlighting not only a shift in peak position, but also the general deviation from the elliptical and purely circular Gaussian shapes. Figure~\ref{fig:compare:beam35} shows the same but for beam 35, which is located in the top-left corner of the \texttt{closepack36} footprint. Figure~\ref{fig:compare:beam35} shows the same measured response, Zernike model, and holography model that is shown in Figure~\ref{fig:icomparison}.}

\end{document}